\documentclass[a4paper,11pt]{article}
\pdfoutput=1 

\usepackage[utf8]{inputenc}

\usepackage{jcappub} 

\usepackage[T1]{fontenc} 

\usepackage{amsmath}
\usepackage{amssymb}
\usepackage{graphicx}
\usepackage{hyperref}
\usepackage{color}
\usepackage{commath}
\usepackage{url}
\usepackage{bm}
\usepackage{braket}
\usepackage{tikz}
\usepackage{url}
\usepackage{accents}
\usepackage{natbib}

\hypersetup{colorlinks=true,citecolor=blue, linkcolor=red, urlcolor=blue}

\newcommand{\inputnum}{1} 
\newcommand{\hiddennum}{4}  
\newcommand{\outputnum}{2}

\newcommand{\udt}[3]{#1^{#2}_{\phantom{#2}#3}}
\newcommand{\udut}[4]{#1^{#2\phantom{#3}#4}_{\phantom{#2}#3\phantom{#4}}}

\newcommand{\dut}[3]{#1_{#2}^{\phantom{#2}#3}}

\newcommand{\lc}[1]{\accentset{\circ}{#1}}

\title{\boldmath Neural Network Reconstruction of \texorpdfstring{$H'(z)$}{} and its application in Teleparallel Gravity}

\author[a,d,1]{Purba Mukherjee,\note{Corresponding author}} 
\author[b,c]{Jackson Levi Said,}
\author[b]{Jurgen Mifsud}

\affiliation[a]{Department of Physical Sciences, Indian Institute of Science Education and Research Kolkata, \vskip 0.05cm Mohanpur, West Bengal - 741246, India.}
\affiliation[b]{Institute of Space Sciences and Astronomy, University of Malta, Malta, MSD 2080.}
\affiliation[c]{Department of Physics, University of Malta, Malta, MSD 2080.}
\affiliation[d]{Physics and Applied Mathematics Unit, Indian Statistical Institute, Kolkata - 700108, India}

\emailAdd{purba16@gmail.com}
\emailAdd{jackson.said@um.edu.mt}
\emailAdd{jurgen.mifsud@um.edu.mt}

\abstract{In this work, we explore the possibility of using artificial neural networks to impose constraints on teleparallel gravity and its $f(T)$ extensions. We use the available Hubble parameter observations from cosmic chronometers and baryon acoustic oscillations from different galaxy surveys. We discuss the procedure for training a network model to reconstruct the Hubble diagram. Further, we describe the procedure to obtain $H'(z)$, the first order derivative of $H(z)$, using artificial neural networks which is a novel approach to this method of reconstruction. These analyses are complemented with further studies on the impact of two priors which we put on $H_0$ to assess their impact on the analysis, which are the local measurements by the SH0ES team ($H_0^{\text{R20}} = 73.2 \pm 1.3$~km~Mpc$^{-1}$~s$^{-1}$) and the updated TRGB calibration from the Carnegie Supernova Project ($H_0^{\text{TRGB}} = 69.8 \pm 1.9$~km~Mpc$^{-1}$~s$^{-1}$), respectively. Additionally, we investigate the validity of the concordance model, through some cosmological null tests with these reconstructed data sets. Finally, we reconstruct the allowed $f(T)$ functions for different combinations of the observational Hubble data sets. Results show that the $\Lambda$CDM model lies comfortably included at the 1$\sigma$ confidence level for all the examined cases.}


\begin{document}
\maketitle
\flushbottom

\section{Introduction} \label{sec:intro}

The standard model of cosmology, the $\Lambda$CDM model, is the most widely accepted model that accurately explains observations at astrophysical and cosmological scales \cite{Peebles:2002gy,Copeland:2006wr}. Here, the appearance of cold dark matter (CDM) acts on galaxies to sustain the rotational velocities of their constituents, as well as on larger scales to form the large-scale structure of the Universe \cite{Baudis:2016qwx,Bertone:2004pz}. On the other hand, the dark energy producing the accelerating expansion of the Universe \cite{Riess:1998cb,Perlmutter:1998np} is modeled by a cosmological constant ($\Lambda$) that appears in the description of gravitation together with the Einstein-Hilbert action description of general relativity (GR). With the inclusion of an inflationary field \cite{Guth:1980zm,Linde:1981mu}, this gives the concordance model that describes the evolution of the Universe from its initial conditions. However, $\Lambda$CDM has been plagued for decades by theoretical problems associated with the cosmological constant \cite{Weinberg:1988cp} as well as its UV completeness \cite{Addazi:2021xuf} as well as other issues \cite{CANTATA:2021ktz} such as the prospect of direct observations of CDM becoming ever more elusive \cite{LUX:2016ggv,Gaitskell:2004gd}. Most recently, the growing tension in measurements of the Hubble constant from different scales of observations has reached a potentially critical point \cite{DiValentino:2020vhf,DiValentino:2020zio,DiValentino:2020vvd,Staicova:2021ajb,DiValentino:2021izs,Perivolaropoulos:2021jda,DiValentino:2022oon}.

Recently, there has been increased reporting of the value of the Hubble constant $H_0$, partly due to the growing discrepancy between direct local observations of $H_0$ against those based on predictions coming from the cosmic microwave background (CMB) radiation using $\Lambda$CDM. The last report from the Planck Collaboration gives a low value of $H_0^{\rm P18} = 67.4 \pm 0.5 \,{\rm km\, s}^{-1} {\rm Mpc}^{-1}$ for the Hubble constant \cite{Aghanim:2018eyx} while the last ACT release gives similar values with ACT-DR4 giving $H_0^{\rm ACT-DR4} = 67.9 \pm 1.5 \,{\rm km\, s}^{-1} {\rm Mpc}^{-1}$ \cite{ACT:2020gnv}. These predictions use early Universe data in tandem with $\Lambda$CDM to produce best-fit values of the Hubble constant. This contrasts with direct measurements of $H_0$ from local sources, the highest of which come from Cepheid calibrated observations of Supernovae Type Ia (SN-Ia) by the SH0ES Team giving $H_0^{\rm R20} = 73.2 \pm 1.3 \,{\rm km\, s}^{-1} {\rm Mpc}^{-1}$ \cite{Riess:2020fzl}. Along a similar vein of measurement values, the H0LiCOW Collaboration \cite{Wong:2019kwg} reports a comparable Hubble constant $H_0^{\rm HW} = 73.3^{+1.7}_{-1.8} \,{\rm km\, s}^{-1} {\rm Mpc}^{-1}$ based on observations of the strong lensing from quasars. Another pivotal measurement of the Hubble constant is that based on the Tip of the Red Giant Branch (TRGB) calibration technique which has been reported to give a value $H_0^{\rm TRGB} = 69.8 \pm 1.9 \,{\rm km\, s}^{-1} {\rm Mpc}^{-1}$ \cite{Freedman:2020dne}. This Hubble constant value is more consistent with the early Universe based predicted values. Other measurements exist that hold the promise of offering new calculations of $H_0$ that do not rely on electromagnetic observations such as the novel approach of gravitational wave standard sirens \cite{Abbott:2017xzu} but the precision of such methods is not competitive with standard approaches as of yet. In this context, we endeavour to extend the literature in the direction of producing nonparametric estimates of the Hubble parameter together with its derivative which are core to a number of important themes of research such as cosmography \cite{Bargiacchi:2021fow,Capozziello:2019cav,Bamba:2012cp} and modified gravity \cite{Cai:2019bdh,Ren:2022aeo,Bernardo:2021qhu,Briffa:2020qli,LeviSaid:2021yat}.

There have been a variety of responses to the Hubble tension issue such as modifications to the behaviour of early Universe dark energy to new modifications to the matter sector, particularly neutrino physics, as well as renewed modifications to the gravitational sector \cite{Addazi:2021xuf}. One such way to go beyond GR is to consider teleparallel gravity (TG) where the curvature associated with the Levi-Civita tensor is exchanged with the torsion connected with the teleparallel connection \cite{Aldrovandi:2013wha,Bahamonde:2021gfp,Krssak:2018ywd,Cai:2015emx}. This change in connection implies that all measures of curvature will identically vanish, such as the Ricci scalar $R=0$. However, this does not mean that the regular, Levi-Civita connection, Ricci scalar vanishes ($\lc{R} \neq 0$ - We use over-circles to represent quantities calculated with the Levi-Civita connection). Along this line of thought, TG can produce a torsion scalar $T$ which is equivalent to the Ricci scalar (up to a boundary term). The action based on a linear formulation of the torsion scalar is called the \textit{Teleparallel equivalent of General Relativity} (TEGR) and produces equations of motion that are dynamically equivalent to GR.

Taking the same rationale $f(\lc{R})$ gravity \cite{Sotiriou:2008rp,Faraoni:2008mf,Capozziello:2011et}, TEGR can be directly generalized to $f(T)$ gravity \cite{Ferraro:2006jd,Ferraro:2008ey,Bengochea:2008gz,Linder:2010py,Chen:2010va,Bahamonde:2019zea}. Unlike $f(\lc{R})$ gravity, $f(T)$ gravity turns out to produce generally second order equations of motion which means that it depends only on the Hubble parameter $H(z)$ and its first derivative making it more amenable to reconstruction approaches. Due to this fact alone, there has been a lot of work in the literature on using supervised learning approaches to use reconstructions of the Hubble diagram to reconstruct the $f(T)$ gravity functional in a nonparametric way. This has mainly taken the form of using Gaussian processes (GP) \cite{10.5555/1162254} which is based on training a covariance function to reconstruct the Hubble diagram together with uncertainties at each point such as in Refs.~\cite{Busti:2014aoa,Busti:2014dua,Seikel:2013fda,Bernardo:2021mfs,Yahya:2013xma,2012JCAP...06..036S,Shafieloo:2012ht,Benisty:2020kdt,Mukherjee:2021epjc,Mukherjee:2022pdu}. In Refs.~\cite{Briffa:2020qli,Ren:2022aeo,Cai:2019bdh} the GP approach was applied to the $f(T)$ functional with reconstructions of both the mean and uncertainties of the functional form, which was later extended to include growth data in Ref.~\cite{LeviSaid:2021yat}. The approach has also been applied to other settings which contain second order equations of motion \cite{Bernardo:2021qhu,Bernardo:2021cxi}. However, other interesting approaches exist in the literature such as those proposed in Refs.~\cite{Montiel:2014fpa,2011A&A...527A..49I,Shafieloo:2005nd,Porqueres:2016kfv,Escamilla-Rivera:2021rbe}.

GP has a number of drawbacks such as overfitting and a possible over-reliance on choices in the covariance function. An alternative approach is to consider artificial neural networks (ANN) \cite{10.2307/j.ctt4cgbdj} which is a competing nonparametric approach by which the observational data can be approximated. Here, artificial neurons are modelled on their biological equivalent, and organized into layers in such a way to collectively respond to input signals (in this case redshift values) by outputting appropriate cosmological parameters (i.e. the corresponding Hubble values together with their uncertainties) \cite{aggarwal2018neural,Wang:2020sxl,Gomez-Vargas:2021zyl}. By optimizing the number of neurons and the layers that organize them, the various data sets and priors on the Hubble parameter can be used to reconstruct the Hubble diagram such as in Ref.~\cite{Dialektopoulos:2021wde}. However, in order to reconstruct the second order field equations, the reconstruction must be extended to the derivatives of the Hubble parameter, i.e. $H'(z)$. GP is designed to produce these derivatives organically but the generality of ANNs means that this is not as forthcoming to achieve. In this work, we use a Monte Carlo (MC) routine implementation together with ANNs to produce this derivative parameter. This then leads to an approach by which we can produce reconstruction fits for the $f(T)$ gravity functional form along with the uncertainties. We take this approach rather than a direct calculation of the derivative of $H'(z)$ since this avoid the correlation of uncertainties that such an approach would entail. This we find a clear calculation not only of the mean values of $H'(z)$ but also of the associated uncertainties at every data point. ANNs have been used in other areas of cosmology such as analyzing the power spectrum of the CMB \cite{Auld:2007qz,Auld:2006pm,2012MNRAS.421..169G} or studying the nature of dark energy \cite{Escamilla-Rivera:2019hqt} and the large scale structure of the Universe \cite{Aragon-Calvo:2018kfw,Ntampaka:2019ole,Ribli:2019wtw,Fluri:2019qtp,Fluri:2018hoy}. Here, we offer a new direction by which ANNs may further impact our understanding of gravity beyond GR.

In this paper, we probe $f(T)$ gravity models using Hubble data in conjunction with ANNs, which is organized as follows. The following section introduces the cosmological dynamics for both TEGR and its $f (T )$ gravity generalisation. Sec.~\ref{sec:ANN_intro} provides an introduction to ANNs and briefly reviews the observational Hubble data sets. In Sec.~\ref{H-ann-sect}, we describe the procedure to train our ANN to reconstruct the $H(z)$, followed by the reconstruction of $H'(z)$ using ANNs and undertake two null tests for the concordance model of cosmology. We further reconstruct the allowed $f(T)$ functions for different combinations of the observational Hubble data sets in Sec.~\ref{fT-recon}, from which any preferred deviation from the $\Lambda$CDM behaviour will become evident. Finally, we summarise our core conclusions in Sec.~\ref{conclusion}.

\section{\texorpdfstring{$f(T)$}{} cosmology}\label{sec:f_T_intro}

TG is based on the replacement of the Levi-Civita connection $\lc{\Gamma}_{\mu\nu}^{\sigma}$ (we recall that over-circles denote quantities determined by the Levi-Civita connection) that is used in curvature-based gravitational theories with the teleparallel connection $\Gamma^{\sigma}_{\mu\nu}$ which is curvature-less and continues to satisfy metricity \cite{Aldrovandi:2013wha,Cai:2015emx,Krssak:2018ywd,Bahamonde:2021gfp}. This is the basis on which teleparallel theories are constructed. A natural consequence of this is that the teleparallel Riemann tensor vanishes (not the regular Levi-Civita definition), so a new architecture of tensor measures of gravity is needed.

\subsection{Teleparallel gravity and its \texorpdfstring{$f(T)$}{} extension}

TG is best described using a tetrad formalism ($\udt{e}{a}{\mu}$) on which the metric tensor ($g_{\mu\nu}$) is derived \cite{Hayashi:1979qx,nakahara2003geometry,ortin2004gravity}. Tetrads connect local Minkowski spacetime coordinates (Latin indices) with coordinates on the general manifold (Greek indices) \cite{Aldrovandi:2013wha}. Thus, the tetrads relate tangent spaces with the general manifold through 
\begin{equation}\label{metric_tetrad_rel}
    g_{\mu\nu} = \udt{e}{a}{\mu}\udt{e}{b}{\nu}\eta_{ab}\,,\hspace{2cm}\eta_{ab} = \dut{e}{a}{\mu}\dut{e}{b}{\nu}g_{\mu\nu}\,,
\end{equation}
where the inverse tetrads $\dut{e}{a}{\mu}$ must also satisfy the orthogonality conditions
\begin{equation}
    \udt{e}{a}{\mu}\dut{e}{b}{\mu} = \delta^a_b\,,\hspace{2cm}\udt{e}{a}{\mu}\dut{e}{a}{\nu} = \delta^{\nu}_{\mu}\,,
\end{equation}
for consistency. The teleparallel connection can then be defined as \cite{Krssak:2018ywd,Bahamonde:2021gfp}
\begin{equation}
    \Gamma^{\sigma}_{\mu\nu}:= \dut{e}{a}{\mu}\partial_{\mu}\udt{e}{a}{\nu} + \dut{e}{a}{\sigma}\udt{\omega}{a}{b\mu}\udt{e}{b}{\nu}\,,
\end{equation}
where $\udt{\omega}{a}{b\mu}$ denotes the spin connection which appears as a flat connection in the TG context. The role of the spin connection is to preserve the local Lorentz invariance of the theory \cite{Krssak:2015oua}. In GR, spin connection components also appear but they are hidden in the internal structure of the theory \cite{chandrasekhar1998mathematical, Misner:1973prb}. Together, the tetrad-spin connection pair make up the fundamental variables of the theory in TG. They combine to produce the teleparallel connection.

The Riemann tensor gives a fundamental measure of curvature in GR ($\udt{\lc{R}}{\beta}{\mu\nu\alpha} \neq 0$). In TG, the teleparallel Riemann tensor identically vanishes ($\udt{R}{\beta}{\mu\nu\alpha} = 0$) since the teleparallel connection is curvature-less. Thus, we define a torsion tensor \cite{Krssak:2018ywd,Cai:2015emx}
\begin{equation}
    \udt{T}{\sigma}{\mu\nu} := 2\Gamma^{\sigma}_{[\mu\nu]}\,,
\end{equation}
where square brackets denote the anti-symmetry operator, and where torsion is the result of anti-symmetry \cite{Aldrovandi:2013wha}. Moreover, the torsion tensor is invariant under both local Lorentz and diffeomorphic transformations. The torsion tensor can be used to define a torsion scalar \cite{Krssak:2018ywd,Cai:2015emx,Aldrovandi:2013wha,Bahamonde:2021gfp}
\begin{equation}\label{eq:torsion_scalar_def}
    T:=\frac{1}{4}\udt{T}{\alpha}{\mu\nu}\dut{T}{\alpha}{\mu\nu} + \frac{1}{2}\udt{T}{\alpha}{\mu\nu}\udt{T}{\nu\mu}{\alpha} - \udt{T}{\alpha}{\mu\alpha}\udt{T}{\beta\mu}{\beta}\,,
\end{equation}
which results by demanding that an action based solely on the linear torsion scalar produces the same equations of motion as the Einstein-Hilbert action (up to a total divergence term).

By recalling that the teleparallel Ricci scalar vanishes identically, $R\equiv 0$, we can relate the regular curvature-based Ricci scalar with the torsion scalar through \cite{Bahamonde:2021gfp}
\begin{equation}\label{LC_TG_conn}
    R=\lc{R} + T - B = 0\,.
\end{equation}
where $B$ represents a total divergence term and is defined as
\begin{equation}\label{eq:boundary_term_def}
    B = \frac{2}{e}\partial_{\rho}\left(e\udut{T}{\mu}{\mu}{\rho}\right)\,,
\end{equation}
where $e=\det\left(\udt{e}{a}{\mu}\right)=\sqrt{-g}$ is the determinant of the tetrad. The action based solely on the torsion scalar is called the teleparallel equivalent of general relativity (TEGR). The relationship between the Ricci and torsion scalars in Eq.~(\ref{LC_TG_conn}) alone guarantees that GR and TEGR produce identical equations of motion, and so are dynamically equivalent. Hence, we can write the TEGR action as 
\begin{equation}
    \mathcal{S}_{\rm TEGR} = -\frac{1}{2\kappa^2}\int {\rm d}^4 x\; eT + \int {\rm d}^4 x\; e \mathcal{L}_{\rm m}\,,
\end{equation}
where $\kappa^2=8\pi G$ is the gravitational coupling and $\mathcal{L}_{\rm m}$ is the matter Lagrangian density.

Following the reasoning as in other avenues to modifying GR, we can consider direct generalizations of TEGR by taking arbitrary functional forms of the torsion scalar. Similar to $f(\lc{R})$ gravity \cite{DeFelice:2010aj,Capozziello:2011et}, TEGR can be straightforwardly generalized to an $f(T)$ gravity framework \cite{Ferraro:2006jd,Ferraro:2008ey,Bengochea:2008gz,Linder:2010py,Chen:2010va} through the action 
\begin{equation}\label{f_T_Lagrangian}
    \mathcal{S}_{\tilde{f}(T)} =  \frac{1}{2\kappa^2}\int {\rm d}^4 x\; e \tilde{f}(T) + \int {\rm d}^4 x\; e \mathcal{L}_{\rm m}\,,
\end{equation}
which interestingly produces second order equations of motion, and limits to TEGR for the case when $\tilde{f}(T)=-T$ and $\Lambda$CDM when $\tilde{f}(T)=-T+\Lambda$. The $f(T)$ gravity shares a number of interesting properties with GR such as sharing the same polarization modes \cite{Bamba:2013ooa,Farrugia:2018gyz,Cai:2018rzd,Abedi:2017jqx,Chen:2019ftv}, and also being Gauss-Ostrogradsky ghost-free (since it remains second order) \cite{Krssak:2018ywd,ortin2004gravity}. In our work, we map the functional to 
\begin{equation}
    \tilde{f}(T) \rightarrow -T + f(T)\,,
\end{equation}
so that the functional component appears as an extension to the TEGR Lagrangian.

\subsection{\texorpdfstring{$f(T)$}{f(T)} cosmology}

The spatially flat homogeneous and isotropic Friedmann–Lema\^{i}tre–Robertson–Walker metric is represented by
\begin{equation}\label{FLRW_metric}
    \mathrm{d}s^2=-\mathrm{d}t^2+a^2(t) \left(\mathrm{d}x^2+\mathrm{d}y^2+\mathrm{d}z^2\right)\,,
\end{equation}
which can be produced by the tetrad choice
\begin{equation}
    \udt{e}{a}{\mu}={\rm diag}\left(1,\,a(t),\,a(t),\,a(t)\right)\,,
\end{equation}
where $a(t)$ is the scale factor. Interestingly, this choice of tetrad is compatible with a vanishing spin connection ($\udt{\omega}{a}{b\mu} = 0$), also called the Weitzenb\"{o}ck gauge \cite{Krssak:2015oua,Tamanini:2012hg}. Taking the torsion scalar definition in Eq.~(\ref{eq:torsion_scalar_def}) results in 
\begin{equation}\label{Tor_sca_flrw}
    T=6H^2\,,
\end{equation}
where the boundary term will be $B=6\left(3H^2+\dot{H}\right)$, which straightforwardly gives the expected standard Ricci scalar for the flat FLRW setting, i.e. $\lc{R}=-T+B=6\left(\dot{H}+2H^2\right)$. 
The equations of motion for this choice of spacetime then turns out to be described by
\begin{align}
    3H^2 &= \kappa^2 \left(\rho_{\rm m}+\rho_{\rm eff}\right)\,,\label{Friedmann_eq}\\
    3H^2 + 2\dot{H} &= -\kappa^2\left(p_{\rm m}+p_{\rm eff}\right)\,,\label{Friedmann_eq2}
\end{align}
where $\rho_{\rm m}$ and $p_{\rm m}$ represent the energy density and pressure of the matter content respectively, while the $f(T)$ gravity can be interpreted as an effective fluid with components
\begin{align}
    \rho_{\rm{eff}} &:= \frac{1}{2\kappa^2}\left(2Tf_T - f\right)\,,\\
    p_{\rm{eff}} &:= -\frac{1}{\kappa^2}\left[2\dot{H}\left(f_T+2Tf_{TT}\right)\right] - \rho_{\rm{eff}}\,,
\end{align}
where $f_T$ and $f_{TT}$ are first and second derivatives of the $f(T)$ functional with respect to the torsion scalar $T$. Here, a perfect fluid setup is being employed for the matter sector. The effective fluid also turns out to satisfy the conservation equation
\begin{equation}
    \dot{\rho}_{{\rm eff}} + 3H\left(\rho_{{\rm eff}}+p_{{\rm eff}}\right) = 0\,,
\end{equation}
and can be utilized to define the effective equation of state (EoS) as \cite{Bahamonde:2016cul,Escamilla-Rivera:2019ulu}
\begin{equation}\label{EoS_func}
    \omega_{\rm{eff}}  := \frac{p_{\rm{eff}}}{\rho_{\rm{eff}}} = -1 +\left(1+\omega_m\right)\frac{\left(T+f-2Tf_T\right)\left(f_T+2Tf_{TT}\right)}{\left(-1+f_T+2Tf_{TT}\right)\left(-f+2Tf_T\right)}\,.
\end{equation}
Therefore, the Friedmann equations~\eqref{Friedmann_eq}--\eqref{Friedmann_eq2} can be rewritten as, 
\begin{align}
    H^2 - \frac{T}{3} f_T + \frac{f}{6} &= \frac{\kappa^2}{3}\rho_m \,,\label{eq:Friedmann_1}\\
    \dot{H}\left(1 - f_T - 2Tf_{TT}\right) &= -\frac{\kappa^2}{2} \left(\rho_m + p_m \right)\label{eq:Friedmann_2}\,.
\end{align}
It also turns out that the $\Lambda$CDM scenario is recovered for the case when $f(T)=\Lambda$.

\section{Methodology} \label{sec:ANN_intro}

\begin{figure}[h]
	\begin{center}
		
		\begin{tikzpicture}
		
		\foreach \i in {1,...,\inputnum}
		{
			\node[circle, 
			minimum size = 6mm,
			fill=blue!30] (Input-\i) at (0,-\i) {};
		}

		\foreach \i in {1,2,3}
		{
			\node[circle, 
			minimum size = 6mm,
			fill=orange!50,
			yshift=(\hiddennum-\inputnum)*5 mm
			] (Hidden-\i) at (2.5,-\i) {};
		}
		
		\node(dots) at (2.5,-1.9){\vdots};
		
		\node[circle, 
		minimum size = 6mm,
		fill=orange!50,
		yshift=(4-\inputnum)*5 mm
		] (Hidden-5) at (2.5,-4) {};

		\foreach \i in {1,...,\hiddennum}
		{
			\node[circle, 
			minimum size = 6mm,
			fill=orange!50,
			yshift=(\hiddennum-\inputnum)*5 mm
			] (Hidden-\i) at (2.5,-\i) {};
		}

		\foreach \i in {1,...,\outputnum}
		{
			\node[circle, 
			minimum size = 6mm,
			fill=purple!50,
			yshift=(\outputnum-\inputnum)*5 mm
			] (Output-\i) at (5,-\i) {};
		}
		
		\foreach \i in {1,...,\inputnum}
		{
			\foreach \j in {1,...,\hiddennum}
			{
				\draw[->, shorten >=1pt] (Input-\i) -- (Hidden-\j);	
			}
		}
		
		\foreach \i in {1,...,\hiddennum}
		{
			\foreach \j in {1,...,\outputnum}
			{
				\draw[->, shorten >=1pt] (Hidden-\i) -- (Output-\j);
			}
		}
		
		\foreach \i in {1,...,\inputnum}
		{            
			\draw[<-, shorten >=1pt] (Input-\i) -- ++(-1,0)
			node[left]{$z$};
		}
		
		\draw[->, shorten >=1pt] (Output-1) -- ++(1,0)
		node[right]{$H(z)$};
		
		\draw[->, shorten >=1pt] (Output-2) -- ++(1,0)
		node[right]{$\sigma_{H}(z)$};

		\end{tikzpicture}
		\vskip 0.2cm
		Input Layer~~~~~~~~~~~Hidden Layer~~~~~~~Output Layer 
		
	\end{center}
	\caption{A general structure of the adopted ANN, where the input is the redshift $z$, and the outputs are the corresponding $H(z)$ values and the associated uncertainties at each redshift $\sigma_{H(z)}$.} \label{ann}
\end{figure}
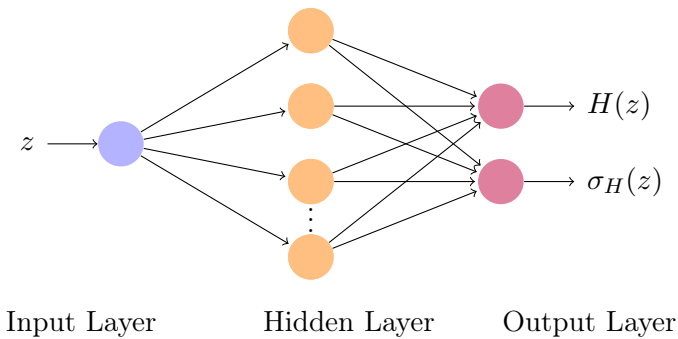

An outline of the adopted ANN technique is discussed briefly in this section. Fig.~\ref{ann} shows the general structure of a simple neural network for the Hubble data and associated uncertainties. It is composed of an input layer that is connected to a hidden layer (or a series of successive hidden layers in general) and an output layer. The input of the neural network is the redshift $z$, while the output is the corresponding Hubble parameter $H(z)$ and its respective uncertainty $\sigma_H(z)$ at that redshift. These layers are connected via nodes, known as \textit{neurons}. A connection of these interconnected neurons forms a network, called the \textit{neural network}. In the training process, the parameters of the neural network will be determined via a learning process using the observational Hubble data sets. 

A neural network involves the application of a linear transformation (composed of linear weights and biases) and a nonlinear activation on the input layer. The inferred results are propagated to the succeeding layers until a linear transformation is applied to the output layer. In this way, any input signal traverses the entire network in a structured manner. 

A wide variety of standard activation functions is available in the literature. In this work, make use of the Exponential Linear Unit (ELU)  \cite{2015arXiv151107289C}, given as
\begin{equation}
    f(x) = \begin{cases}
    x & ~~~~~x>0\,, \\
    \alpha \left( e^{x} -1 \right) & ~~~~~x\leq 0\,.
    \end{cases}
\end{equation}

Here $\alpha$ is a positive hyperparameter that controls the value to which an ELU saturates for negative net inputs, which we have set to unity. We have utilized the \texttt{PyTorch}\footnote{\url{https://pytorch.org/docs/master/index.html}} based code, Reconstruct Functions with ANN (\texttt{ReFANN\footnote{\url{https://github.com/Guo-Jian-Wang/refann}}}) \cite{Wang:2019vxv} for non-parametric reconstruction of $H(z)$ in this work. For reconstructing the Hubble diagram, the hyperparameters for any network model are the number of layers and neurons.

In order to optimize the hyperparameters, we minimize the difference between the predicted result $\hat{H}$ and the ground truth $H$, known as the mean absolute error loss (L1) function, during the training process of the ANN via the Adam \cite{2014arXiv1412.6980K} optimization algorithm. The loss function quantifies the degree to which the input data is modelled by the output reconstruction. Two other loss functions are the mean squared error (MSE) loss function that minimises the squared differences between $\hat{H}$ and $H$, along with the smooth L1 (SL1) loss function which uses a squared term if the absolute error falls below unity and absolute term otherwise. Since the L1 loss function is characterized by the lowest risk statistic with respect to the MSE and SL1 loss function networks, we have considered the L1 loss function in our work.

Besides, we have adopted a single hidden layer to structure the ANN, keeping in mind the lack of complexity when working individually with the Hubble data as already mentioned in the recent literature \cite{Dialektopoulos:2021wde,Wang:2019vxv,Wang:2020hmn,Wang:2020sxl}. Therefore,  the optimal network model is determined by finding the optimal number of neurons associated with this single hidden layer for the L1 loss function. The network is trained after $10^5$ iterations, to assure that the loss function no longer decreases. The initial learning rate is set to 0.01 which goes on decreasing with the number of iterations, and the training batch size is set to half of the number of available $H(z)$ measurements. 

We have utilized the latest 32 cosmic chronometer (CC) $H(z)$ measurements \cite{Jimenez:2003iv,Simon:2004tf,Stern:2009ep,Moresco:2012jh,Zhang:2012mp,Moresco:2015cya,Moresco:2016mzx,Ratsimbazafy:2017vga,Borghi:2022apj}, covering the redshift range up to $z \sim 2$. These measurements do not assume any particular cosmological model \cite{Jimenez:2001gg}, and contain both systematic and calibration errors as reported in the literature. Furthermore, we take into account the latest compilation of 18 baryon acoustic oscillation (BAO) $H(z)$ measurements \cite{Zhao:2018gvb,Gaztanaga:2008xz,Blake:2012pj,Samushia:2012iq,Xu:2012fw,BOSS:2014hwf,BOSS:2013igd,BOSS:2016wmc,Bourboux:2017cbm} from different galaxy surveys like Sloan Digital Sky Survey (SDSS), the Baryon Oscillation Spectroscopic Survey (BOSS) and the extended Baryon Oscillation Spectroscopic Survey (eBOSS). While BAO measurements are not entirely model-independent, particularly due to the assumption of a fiducial radius of the comoving sound horizon $r_d = 147.78$ Mpc \cite{Aghanim:2018eyx}. Nonetheless, they help in drawing perspective to the growing tension in the value of $H_0$ in that they offer expansion rate points derived from the large scale structure of the Universe. 

We are now aware of the rising tension between the local measurements of $H_0$ \cite{Riess:2019cxk, Riess:2020fzl, Riess:2020sih, Riess:2021jrx, Freedman:2019jwv, Freedman:2020dne, Freedman:2021ahq}, and the inferred values of $H_0$ via an extrapolation of data on the early universe \cite{Ade:2015xua, Aghanim:2018eyx}. In this work, we consider the most precise Cepheid calibration result of $H_0^{\text{R20}} = 73.2 \pm 1.3$ km Mpc$^{-1}$ s$^{-1}$ \cite{Riess:2020fzl} by the SH0ES team (hereafter referred to as R20) along with $H_0^\text{TRGB} = 69.8 \pm 1.9$ km Mpc$^{-1}$ s$^{-1}$ \cite{Freedman:2020dne} from the Carnegie Supernova Project which has been recently inferred via the Tip of the Red Giant Branch (TRGB) calibration technique. We shall investigate the impacts of these $H_0$ values, as priors, on the neural network reconstruction. It is well-known that the most precise early-time determination of $H_0^{\text{P18}} = 67.4 \pm 0.5$~km~Mpc$^{-1}$~s$^{-1}$) from the Planck survey \cite{Aghanim:2018eyx} is dependent on the adopted cosmological model. Hence, we have ignored using it in our work. In our analysis, we assume Gaussian prior distributions with the mean and variances corresponding to the central and 1$\sigma$ reported values of each prior above.

\section{Simulation and training with Hubble data \label{H-ann-sect}}

We first consider the generation of the $H(z)$ mock data and the ANN training process in Sec.~\ref{H-recon}, which will be used for structuring the number of layers and neurons of the ANN (more information on this can be found in Appendix \ref{app:mock_data}). The ANN will be used to reconstruct the Hubble diagram for various combinations of Hubble data and priors. In Sec.~\ref{dH-recon} we further apply a MC routine on multiple realizations of the Hubble diagram. This compounding effect of MC with ANNs is then applied to obtain the Hubble derivative $H'(z)$. We also perform some diagnostic tests on the reconstructed results to assess their behaviour against the concordance model in Sec.~\ref{null-test}.

\subsection{Reconstruction of \texorpdfstring{$H(z)$}{} \label{H-recon}}

For reconstructing the observational Hubble parameter, the network model is optimized using the mock $H(z)$ data set, which is simulated in the context of a spatially–flat $\Lambda$CDM model using
\begin{equation}
    H(z) = H_0^{\text{fid}} \sqrt{ \Omega_{m0}^{\text{fid}} ( 1 + z )^3 + 1 - \Omega_{m0}^{\text{fid}}}\,,
\end{equation}
with the ﬁducial $H_0^{\text{fid}} = 70$ km Mpc$^{-1}$ s$^{-1}$ and $\Omega_{m0}^{\text{fid}}=0.3$,  respectively. It should also be mentioned that our final results will be independent of these fiducial values since the actual ANN training is performed on real data rather than the mock generated data, as we are using this model to structure the network rather than actually train it. Note that for training the network and we consider the same number of mock data as the number of observational data available.

\begin{figure}[t!]
	\begin{center}
		\includegraphics[angle=0, width=0.325\textwidth]{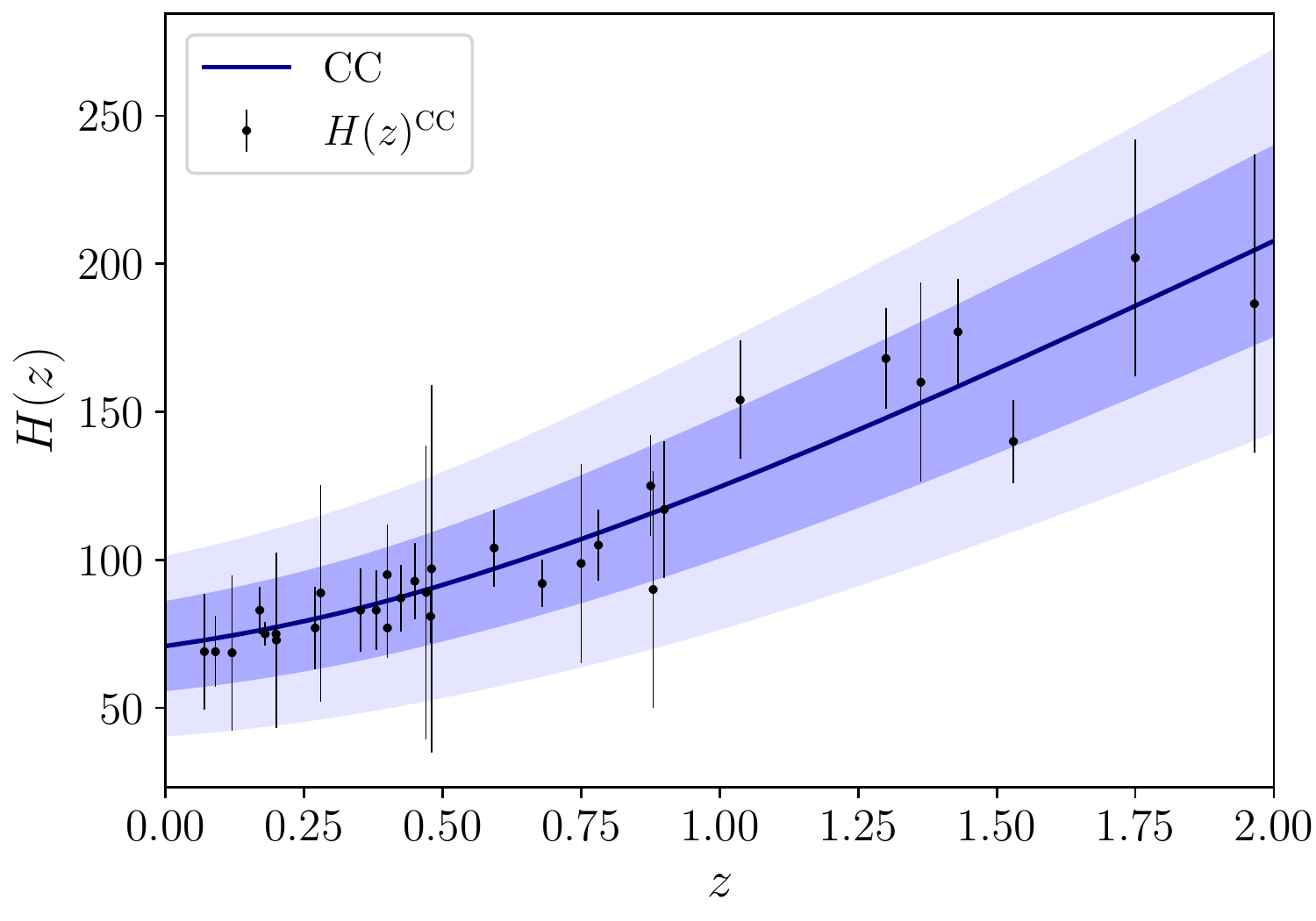}
		\includegraphics[angle=0, width=0.325\textwidth]{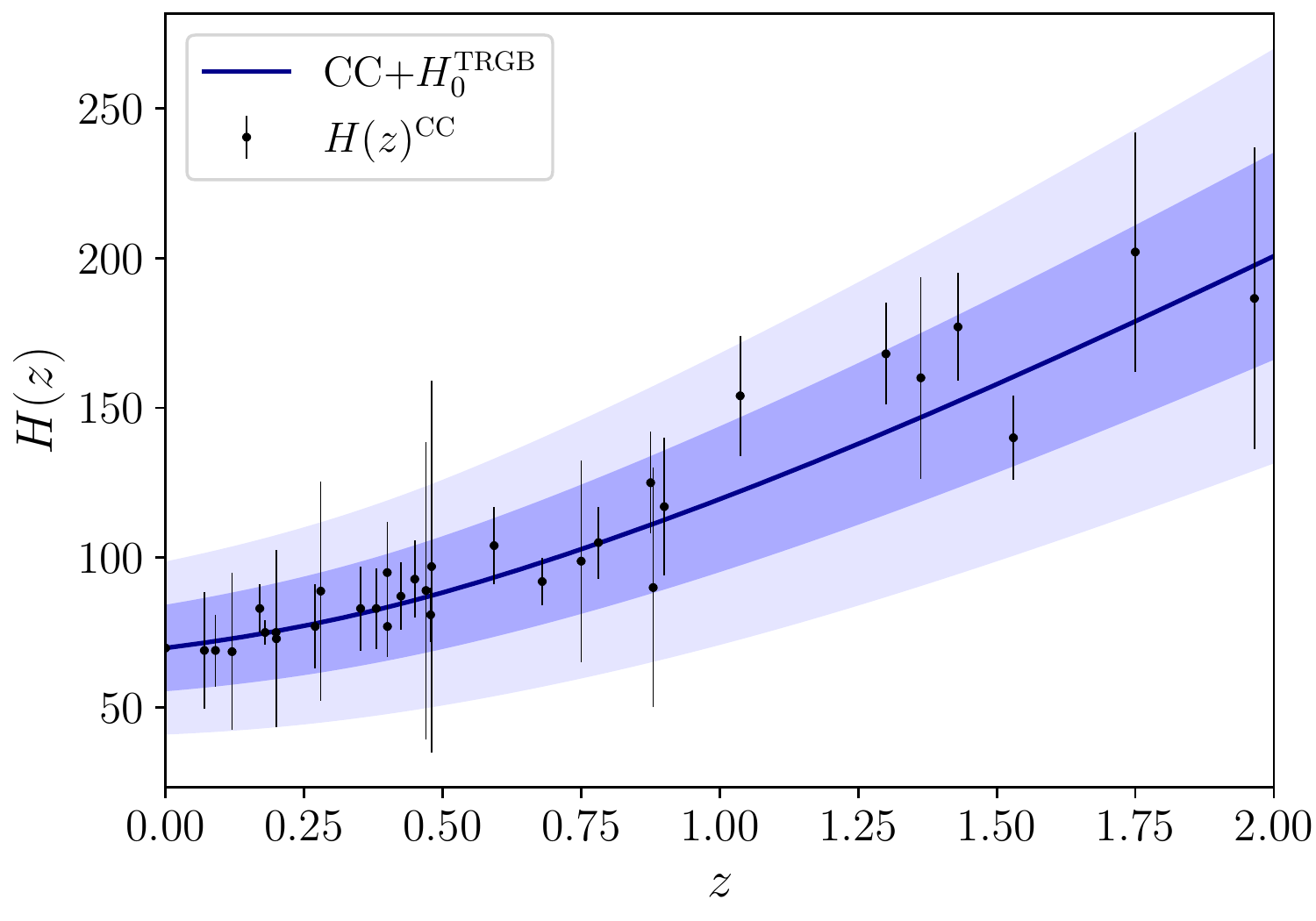} 		
		\includegraphics[angle=0, width=0.325\textwidth]{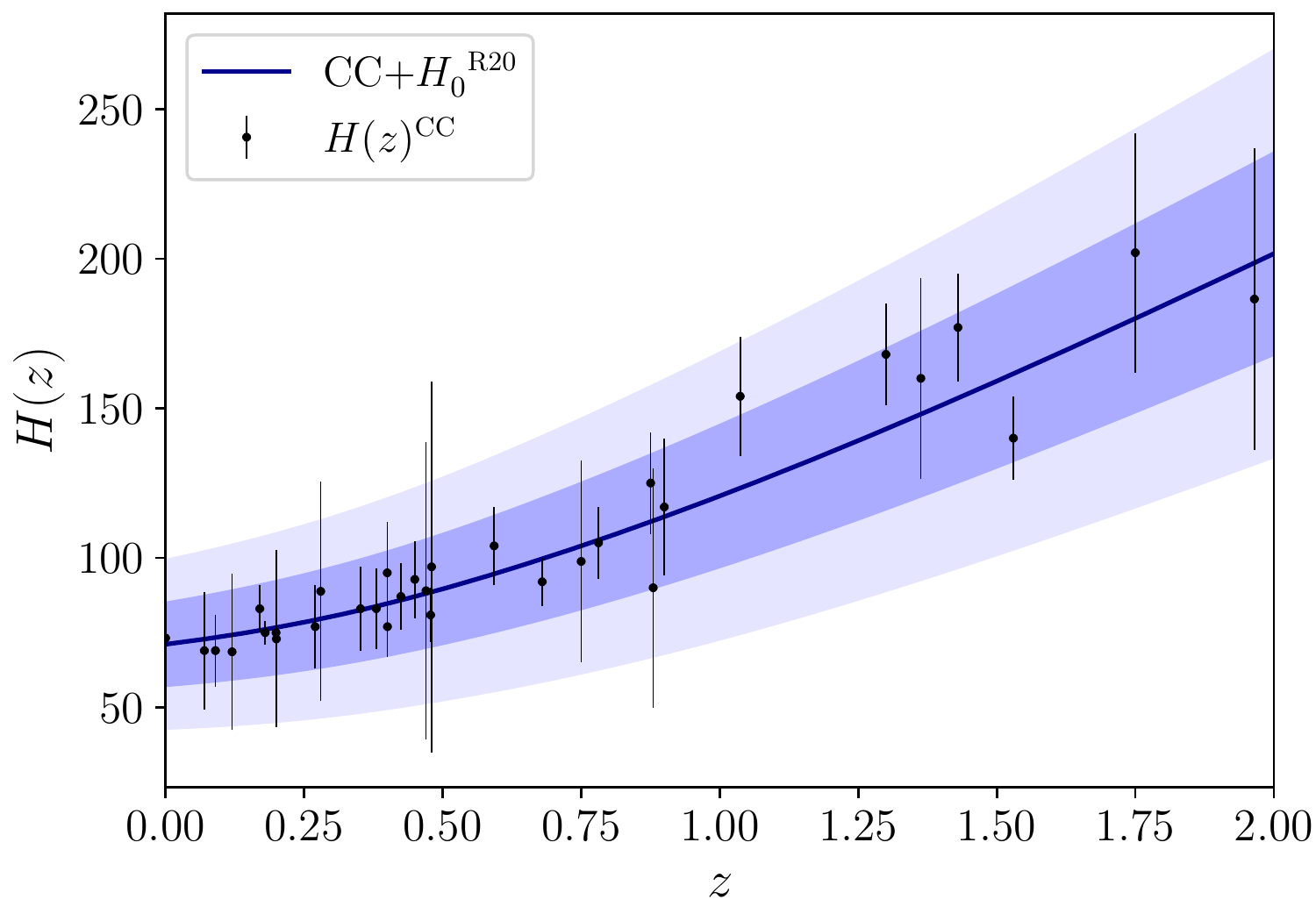} 		\\
		\includegraphics[angle=0, width=0.325\textwidth]{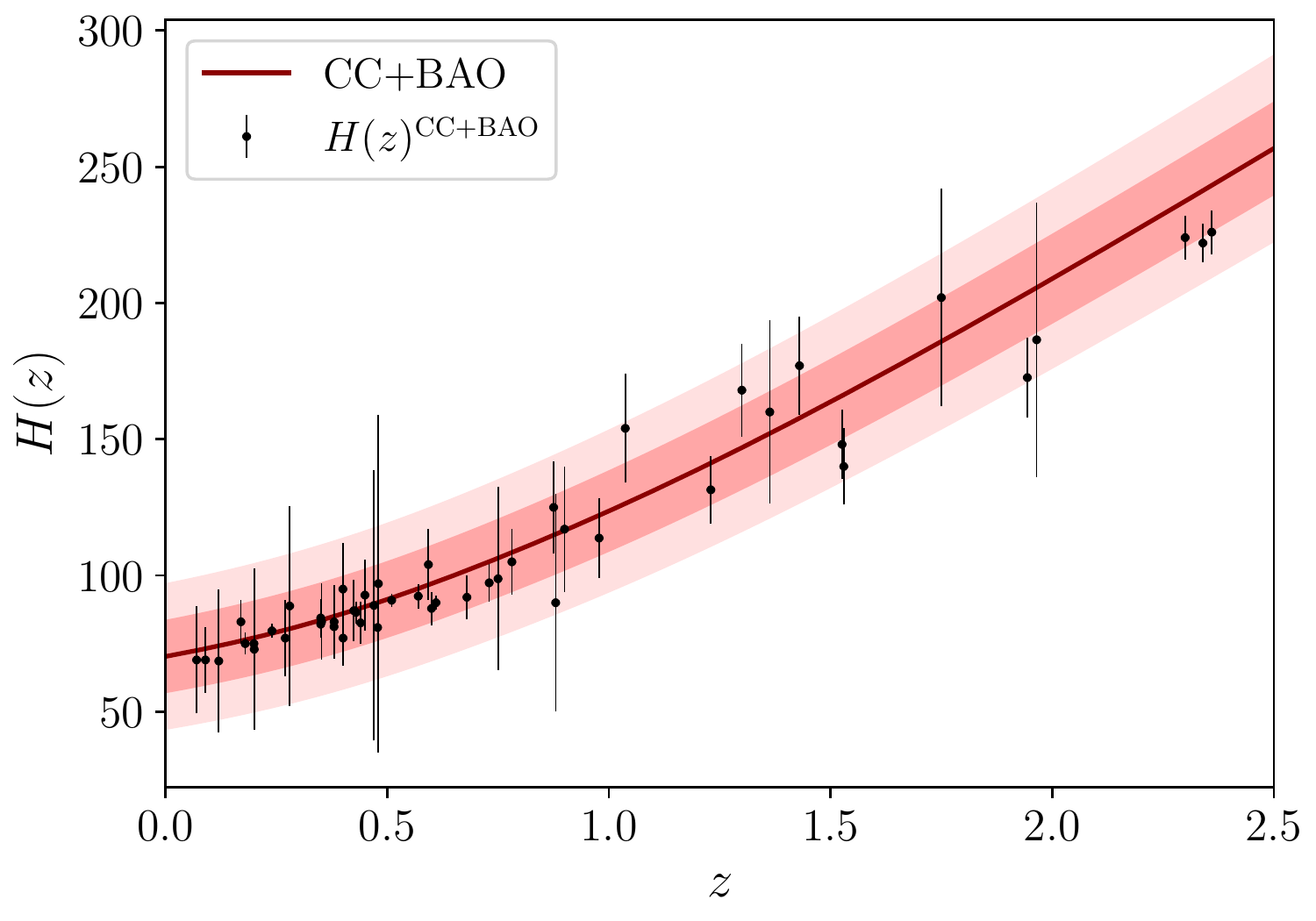}
		\includegraphics[angle=0, width=0.325\textwidth]{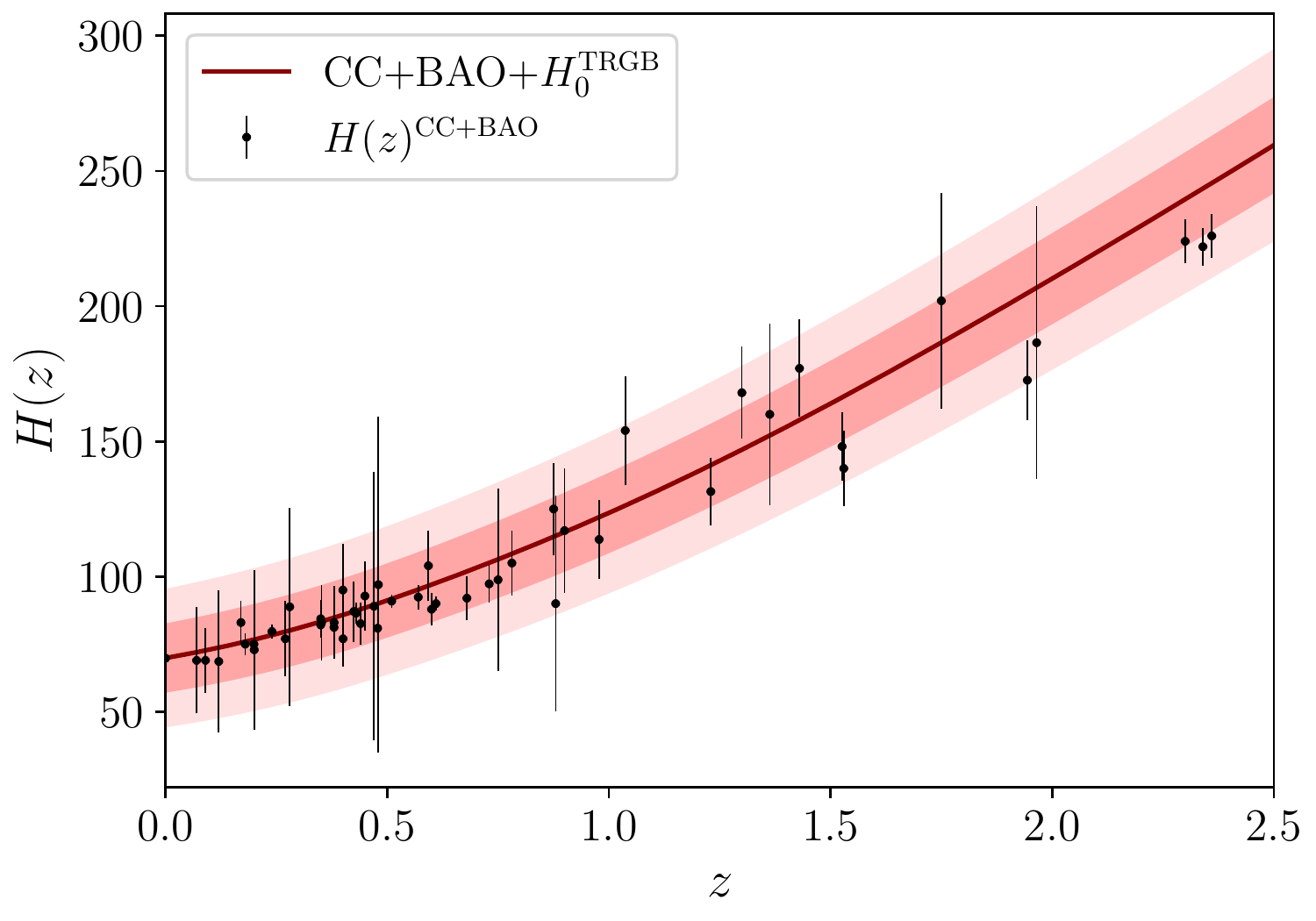} 		
		\includegraphics[angle=0, width=0.325\textwidth]{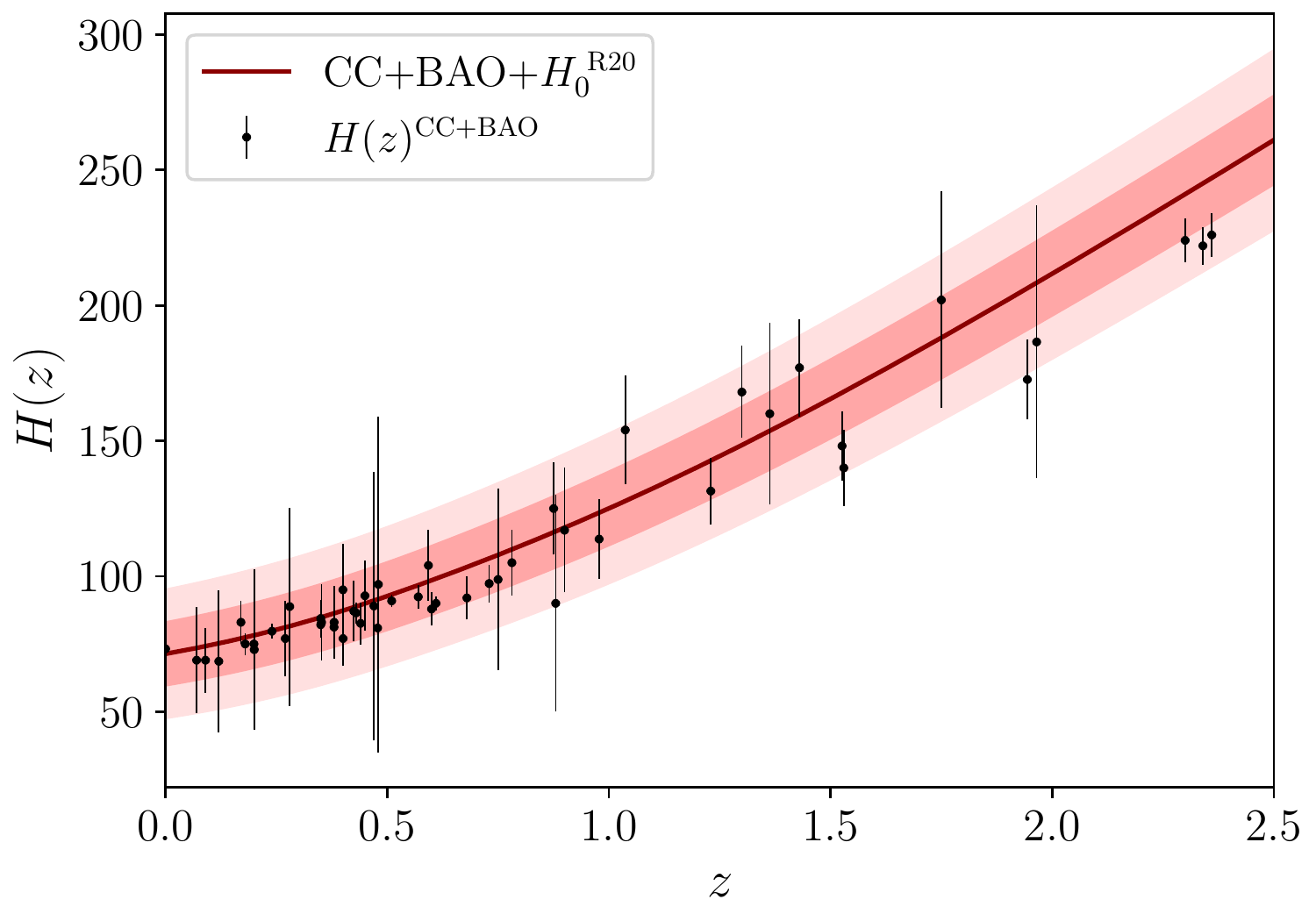} 		
	\end{center}
	\caption{{\small Plots for the reconstructed $H(z)$ using an artificial neural network. The solid line represents the mean reconstructed $H(z)$ curve. The associated 1$\sigma$ and 2$\sigma$ confidence regions are shown in lighter shades.}}
	\label{H-plot}
\end{figure}

\begin{table}[t!]
	\caption{{\small Table showing the reconstructed mean values of $H_0$ along with the associated 1$\sigma$ uncertainties using neural networks.}}
	\begin{center}
		\resizebox{0.70\textwidth}{!}{\renewcommand{\arraystretch}{1.5} \setlength{\tabcolsep}{50 pt} \centering  
			\begin{tabular}{l c }
				\hline \hline
				\textbf{Datasets} &  $H_0$ [in km Mpc$^{-1}$ s$^{-1}$]  \\ 
				\hline 
				CC &  $70.48 \pm 15.24$ \\ 
				CC+BAO & $70.27 \pm 13.47$  \\ 
				\hline
				CC+$H_0^\text{TRGB} $&  $69.85 \pm 14.45$ \\ 
				CC+BAO+$H_0^\text{TRGB} $&  $69.86 \pm 12.82$ \\ 
				\hline
				CC+$H_0^\text{R20} $&  $71.11 \pm 14.30$ \\ 
				CC+BAO+$H_0^\text{R20} $&  $71.36 \pm 12.07$  \\ 
				\hline 
				\hline 
			\end{tabular} 
		}
	\end{center}
	\label{H0_table}
\end{table}

We have done this exercise for the CC $H(z)$ as well as the joint CC and BAO $H(z)$ compilation (hereafter referred to as CC+BAO) consisting of 32 and 50 measurements, respectively. The redshift distribution of the observational $H(z)$ was assumed to follow a Gamma distribution
\begin{equation}
    p(z;\alpha,\lambda)=\frac{\lambda^\alpha}{\Gamma(\alpha)}z^{\alpha-1}e^{-\lambda z}\,, \label{gamma}
\end{equation}
where the free parameters $\alpha$ and $\lambda$ are fitted with the observational data. We made this choice since the data took this distribution overall, but other distribution options may also be appropriate. Given that the distribution was used to generate the mock data, this choice does not have an enormous impact on the eventual trained ANN, and thus the final reconstructed evolution profile.

In order to generate a mock $H(z)$ data, we take into account this fitted distribution of redshift and the uncertainties associated with  observational Hubble data. As the uncertainties tend to increase with $z$, following the prescription given in Ma \& Zhang~\cite{Ma:2010mr}, we assume a linear model for $\sigma_{H}(z)$, to which we fit an arbitrary first-degree polynomial in $z$. For the CC $H(z)$ data set, the mean fitting function is found to be $\sigma_H^0(z) = 15.15 + 9.84z$, while the symmetric upper and lower error bands are specified by $\sigma_H^+(z) = 25.56 + 16.64z$ and $\sigma_H^-(z) = 4.75 - 3.04z$ ensuring that majority of data lies in the area between them. Similarly, for the CC+BAO data set, $\sigma_H^0(z) = 14.29 + 2.77z$, $\sigma_H^+(z) = 20.47 + 9.18z$ and $\sigma_H^-(z) = 8.10 - 3.64z$, respectively.

The uncertainties associated to these mock $H(z)$ data sets are randomly generated assuming that the errors $\tilde{\sigma}_H^{}(z)$ follows the Gaussian distribution $\mathcal{N}(\sigma_H^0(z),\,\varepsilon_H^{}(z))$, where $\varepsilon_H^{}(z)=(\sigma_H^+(z)-\sigma_H^-(z))/4$, such that $\tilde{\sigma}_H^{}(z)$ falls in the area with a probability of 95\%. Therefore, every simulated $H_\mathrm{sim}^{}(z_i)$ at redshift $z_i$, is computed via $H_\mathrm{sim}^{}(z_i)=H_\mathrm{fid}^{}(z_i)+\Delta H_i$, with the associated uncertainty of $\tilde{\sigma}_H^{}(z_i)$, where $\Delta H_i$ is determined via $\mathcal{N}(0,\,\tilde{\sigma}_H^{}(z_i))$. 

Using these simulated $H(z)$ samples, eight network models are trained with $2^n$ number of neurons, where $7 \leq n \leq 14$. These sets of trained networks can then be used to select the optimal network structures on which to train the real CC and CC+BAO data sets by minimizing the risk \cite{Wasserman:2001ng}, defined as
\begin{equation}
    \mathrm{risk}=\sum_{i=1}^N\left[H(z_i)-\bar{H}(z_i)\right]^2+\sum_{i=1}^N\sigma^2\left(H(z_i)\right)\,.
\end{equation}
Here $N$ is the number of $H(z)$ data, and $\bar{H}(z)$ denotes the fiducial value of $H(z)$. For the CC $H(z)$ data set, we have $N=32$, whereas for the joint CC+BAO $H(z)$ data set, $N=50$. Thus, the minimum of the risk function represents the optimal number of neurons of an ANN structure where by the ANN outputs mimic the mock data set to the highest degree.

With these optimal network models, we make predictions by feeding the sequence of redshifts from the real data sets to the input layer. Consequently, we obtain a series of Hubble parameters and the associated uncertainties, which constitute the reconstructed $H(z)$ functions for the respective data sets. 

We further analyse the effect of the two $H_0$ prior values on the reconstruction of $H(z)$. For this exercise, we include the TRGB and R20 $H_0$ measurements and proceed with the full analysis as discussed above. Therefore, a total of 6 combinations are studied for reconstructing $H(z)$ with ANN, namely CC, CC+$H_0^\text{TRGB}$,  CC+$H_0^\text{R20}$, CC+BAO, CC+BAO+$H_0^\text{TRGB}$ and CC+BAO+$H_0^\text{R20}$ respectively. 

Plots for the reconstructed $H(z)$ functions are shown in Fig.~\ref{H-plot}. The reconstructed values of $H_0$ for the respective data set combinations are given in Table~\ref{H0_table}. We find that the mean values of the reconstructed $H_0$ are minimally affected by the inclusion of $H_0$ priors. However, the constraints are in excellent agreement with one another. Moreover, the reconstructed $H(z)$ functions are very similar to each other and are nearly independent of the $H_0$ prior choices, as reported in Ref.~\cite{Dialektopoulos:2021wde} and \cite{Wang:2019vxv}.

\subsection{Reconstruction of \texorpdfstring{$H'(z)$}{} \label{dH-recon}}

\begin{figure}[t!]
	\begin{center}
		\includegraphics[angle=0, width=0.325\textwidth]{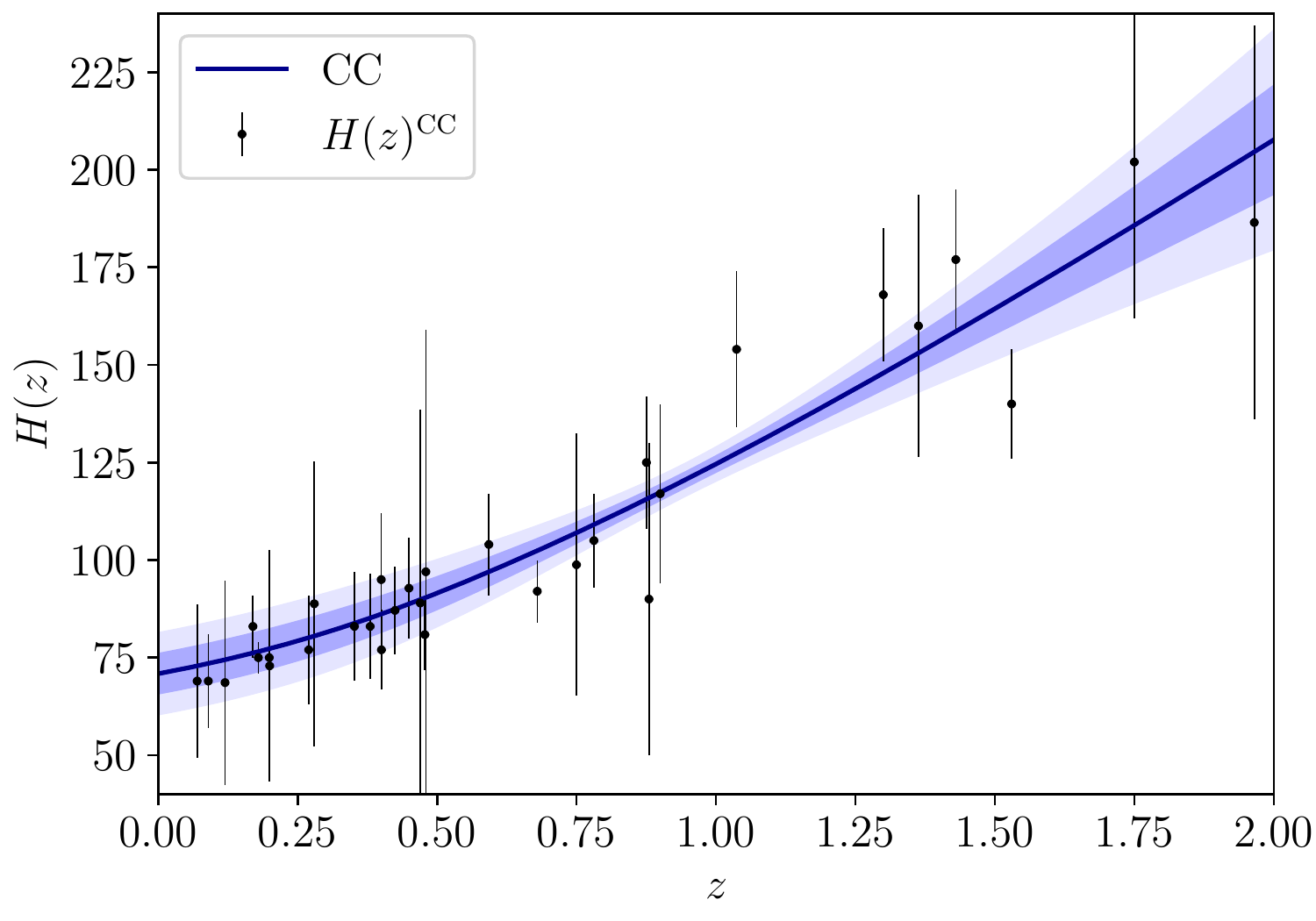}
		\includegraphics[angle=0, width=0.325\textwidth]{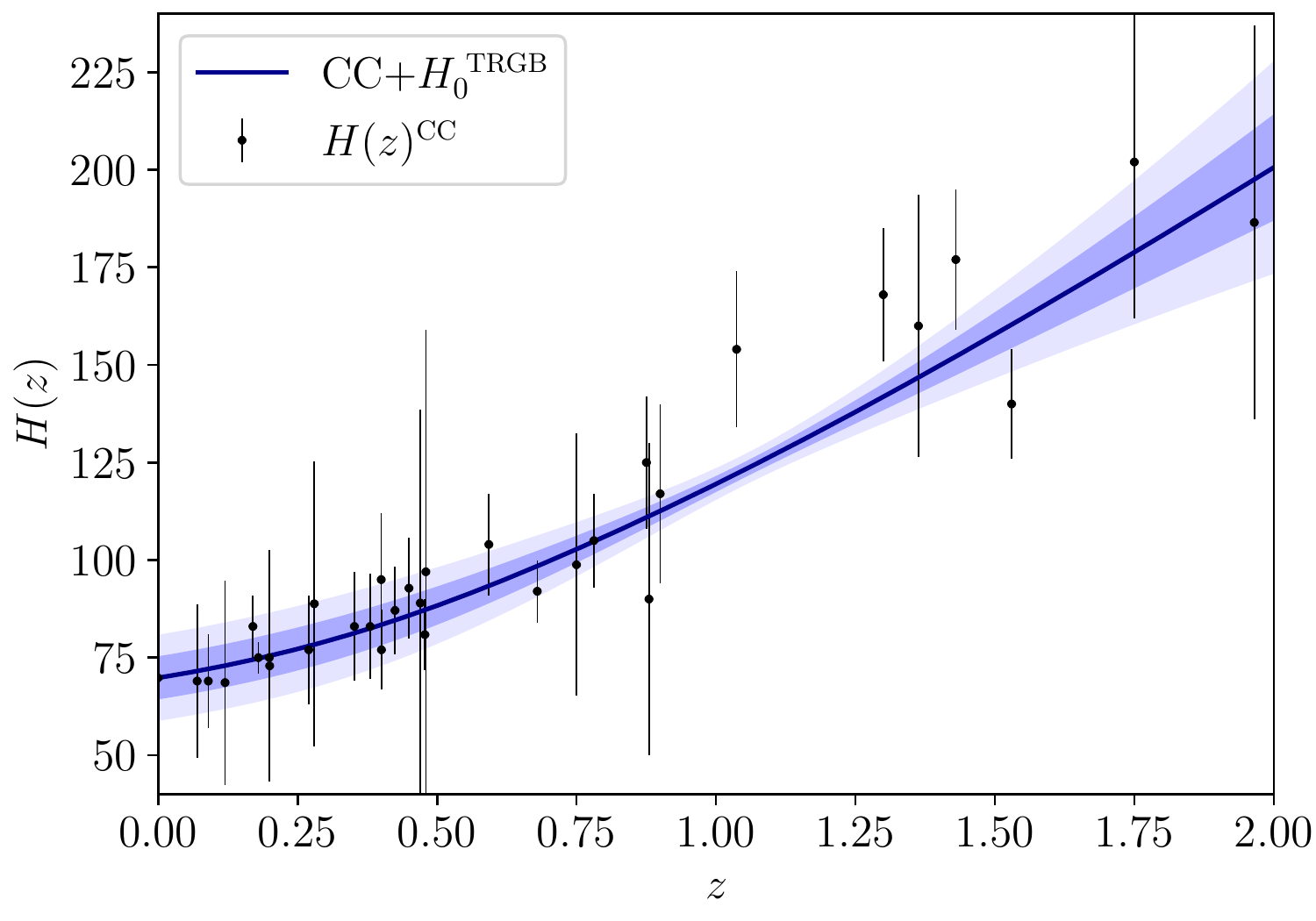} 		
		\includegraphics[angle=0, width=0.325\textwidth]{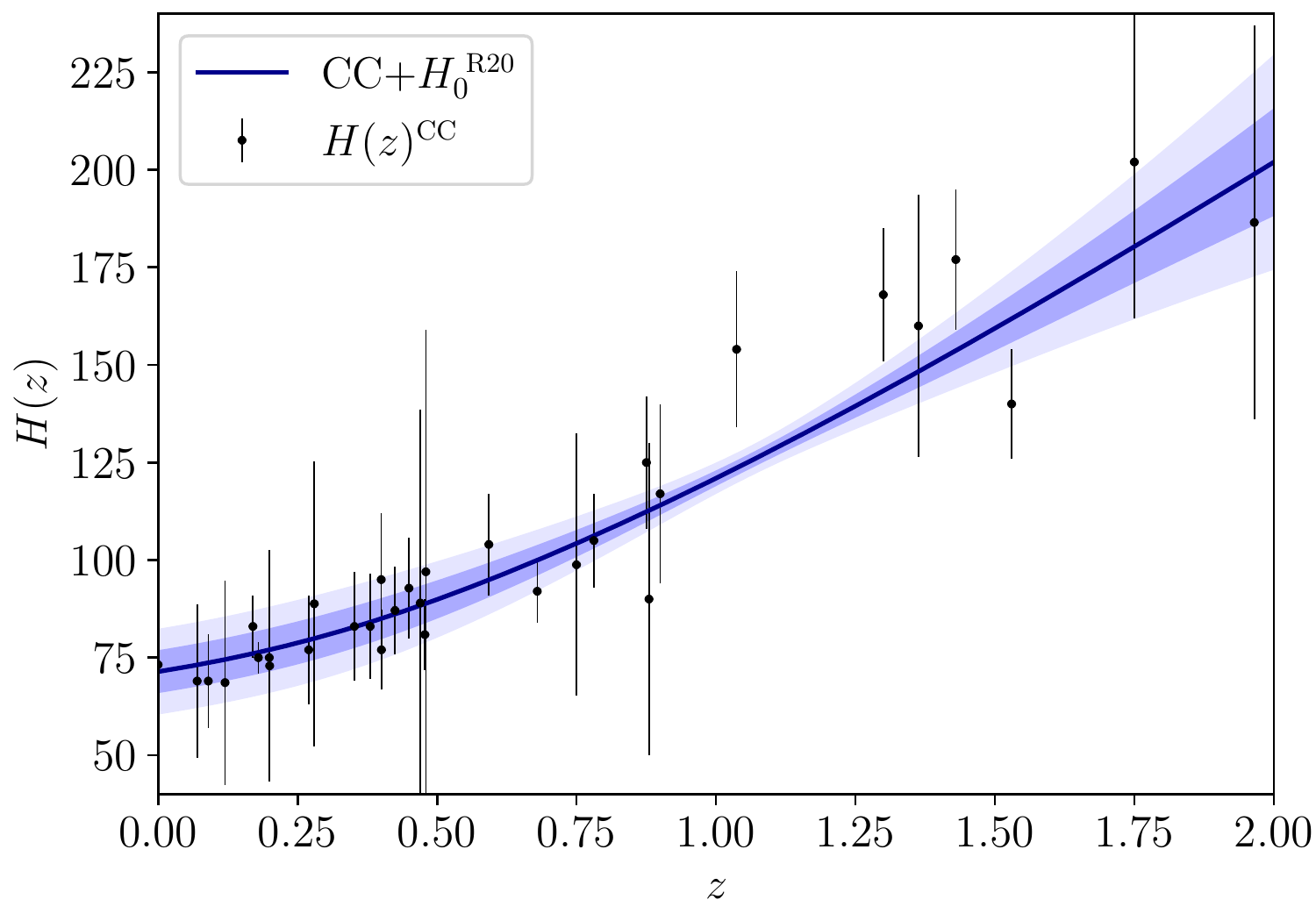} 		\\
		\includegraphics[angle=0, width=0.325\textwidth]{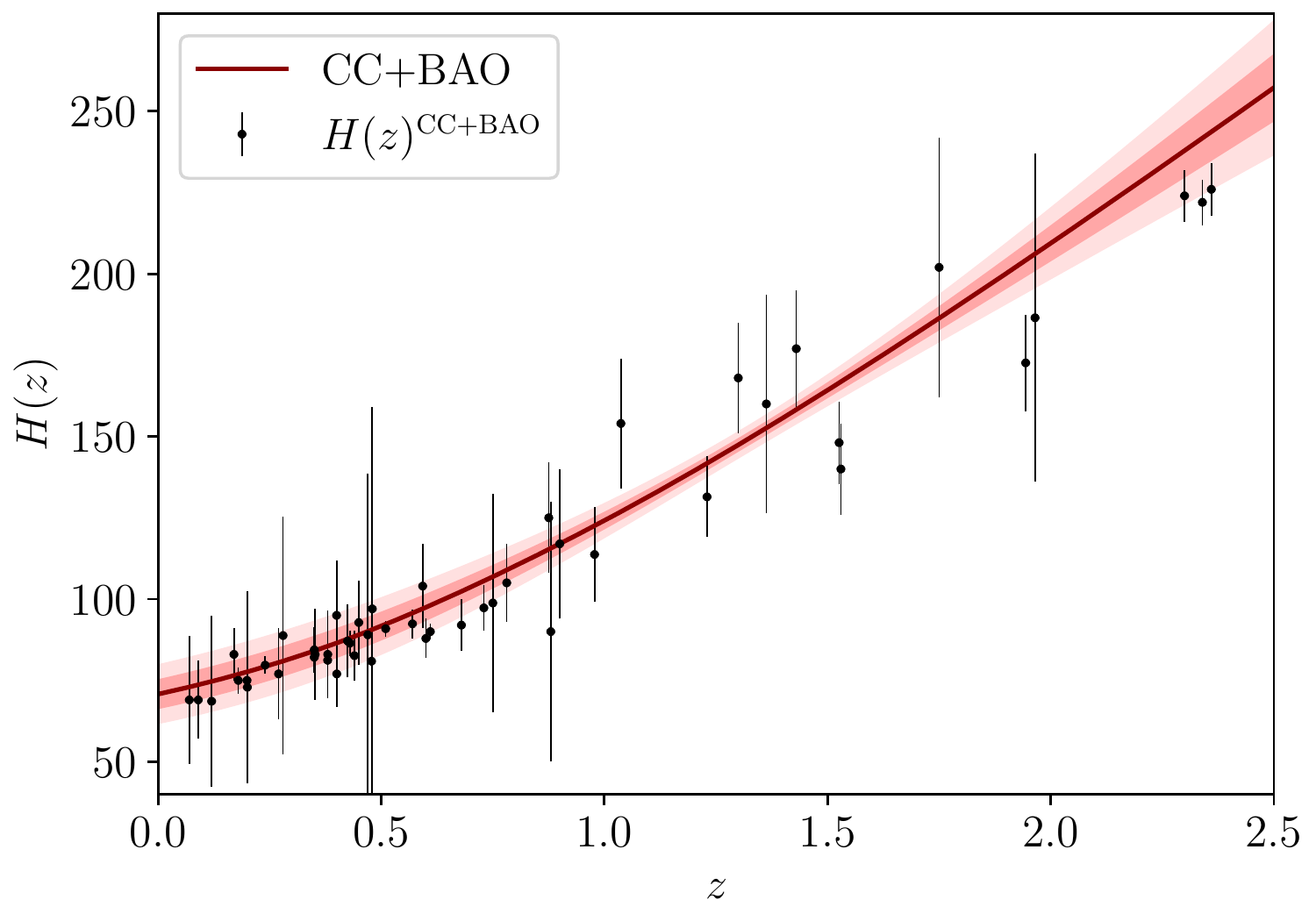}
		\includegraphics[angle=0, width=0.325\textwidth]{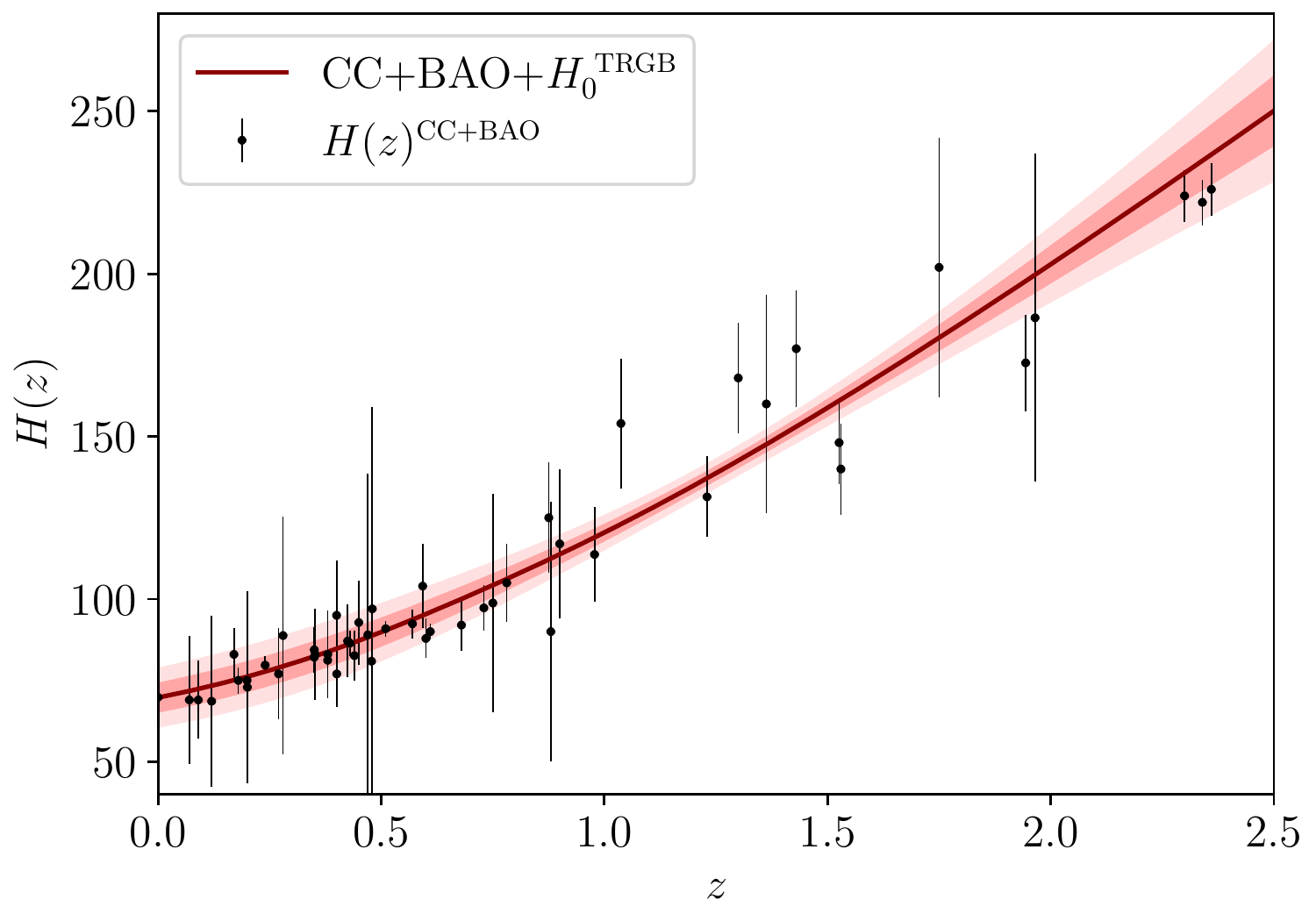} 		
		\includegraphics[angle=0, width=0.325\textwidth]{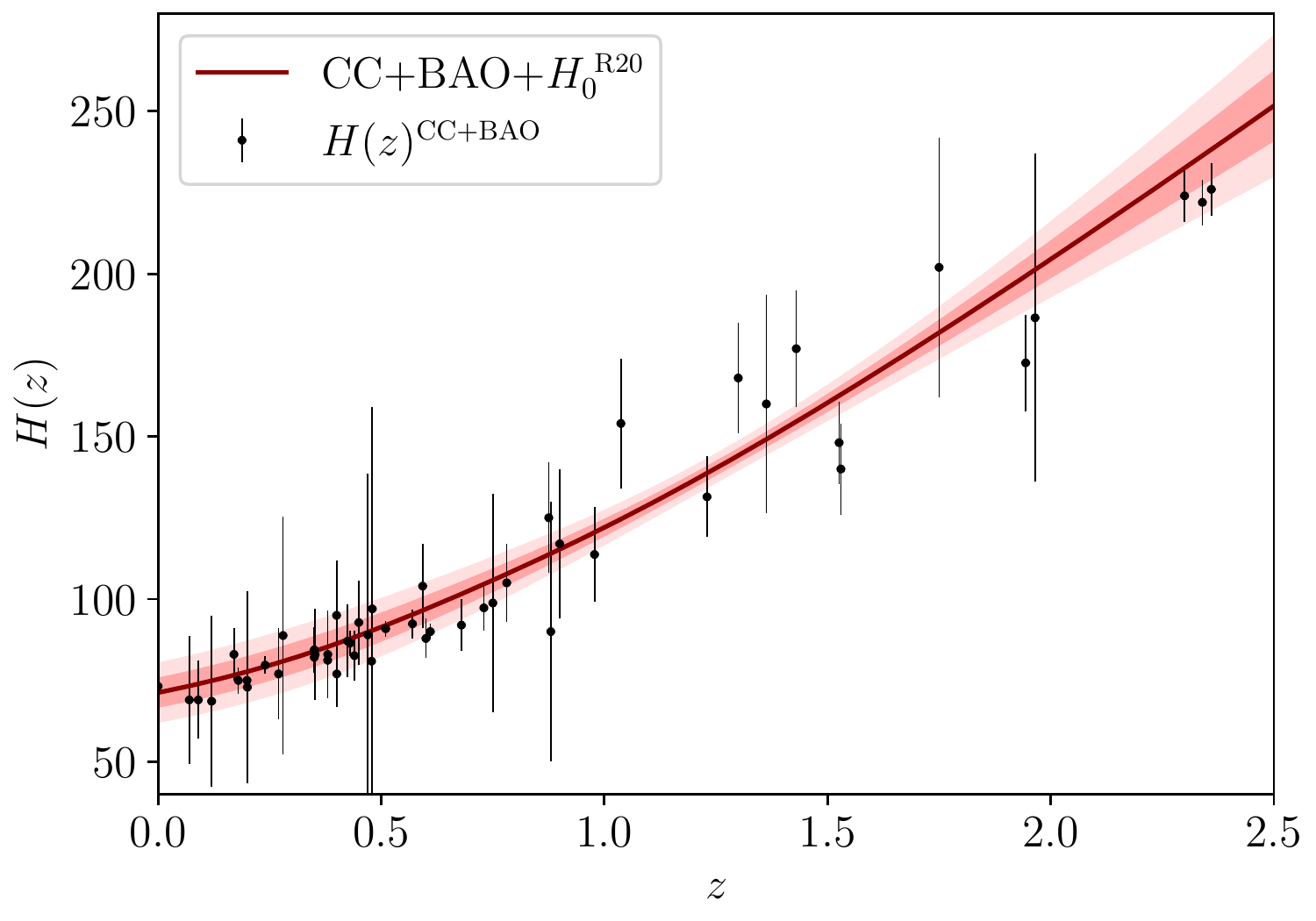} 		
	\end{center}
	\caption{{\small Plots for the reconstructed $H(z)$ using a MC estimation from 1000 neural networks. The solid line represents the mean $H(z)$ curve and the associated 1$\sigma$--2$\sigma$ confidence regions are shown in lighter shades.}}
	\label{H-mc-plot}
\end{figure}

\begin{table}[t!]
	\caption{{\small Table showing the reconstructed mean values of $H_0$ along with the associated 1$\sigma$ uncertainties using a MC estimation from 1000 neural networks.}}
	\begin{center}
		\resizebox{0.70\textwidth}{!}{\renewcommand{\arraystretch}{1.5} \setlength{\tabcolsep}{50 pt} \centering  
			\begin{tabular}{l c }
				\hline \hline
				\textbf{Datasets} &  $H_0$ [in km Mpc$^{-1}$ s$^{-1}$]  \\ 
				\hline 
				CC &  $70.88 \pm 5.34$ \\ 
				CC+BAO & $70.77 \pm 4.61$  \\ 
				\hline
				CC+$H_0^\text{TRGB} $&  $69.82 \pm 5.52$ \\ 
				CC+BAO+$H_0^\text{TRGB} $&  $69.71 \pm 4.63$ \\ 
				\hline
				CC+$H_0^\text{R20} $&  $71.41 \pm 5.49$ \\ 
				CC+BAO+$H_0^\text{R20} $&  $71.24 \pm 4.65$  \\ 
				\hline 
				\hline 
			\end{tabular} 
		}
	\end{center}
	\label{H0_MC_table}
\end{table}

For the sample of observational Hubble data, we can train a network model to learn to mimic the complex relationships between $z$, $H(z)$ and $\sigma_{H}(z)$ following the methodology outlined in Sec.~\ref{H-recon}. So, any arbitrary number of $H(z)$ samples can be reconstructed by feeding a sequence of redshifts to this network model. Here, we focus on the reconstruction of $H'(z)$ in a novel way, where this prime denotes derivative with respect to the redshift $z$. In our implementation, we take a Monte Carlo approach based on the reconstruction of the Hubble parameter for our range of data.

To begin with, we generate 1000 realizations of redshift samples from the fitted Gamma distribution given in equation~\eqref{gamma} of the real Hubble parameter measurements. With these redshift samples, 1000 simulated $H(z)$ data samples are randomly generated. Using each of these simulated $H(z)$ samples, eight network models are trained with $2^n$ number of neurons, where $7 \leq n \leq 14$ for each realization. So, a total of 8000 network models are trained for every real Hubble parameter data set, i.e., CC, CC+BAO, CC+$H_0^\text{TRGB}$,  CC+$H_0^\text{R20}$, CC+BAO+$H_0^\text{TRGB}$ and CC+BAO+$H_0^\text{R20}$. 

Using these sets of trained networks, we determine the optimal network configuration for each realization and make predictions by feeding the sequence of redshifts from the real data sets to the input layer. Consequently, we obtain 1000 realizations of the reconstructed $H(z)$ functions for the respective data sets. From these 1000 reconstructed $H(z)$ functions, we obtain the best fit values of reconstructed $H(z)$ along with the associated confidence levels using a MC routine. Plots for the reconstructed $H(z)$ using the MC routine on the 1000 reconstructed $H(z)$ realizations are shown in Fig.~\ref{H-mc-plot}. The reconstructed $H_0$ values using this MC approach for the respective data set combinations are given in Table~\ref{H0_MC_table}.

\begin{figure}[t!]
	\begin{center}
		\includegraphics[angle=0, width=0.325\textwidth]{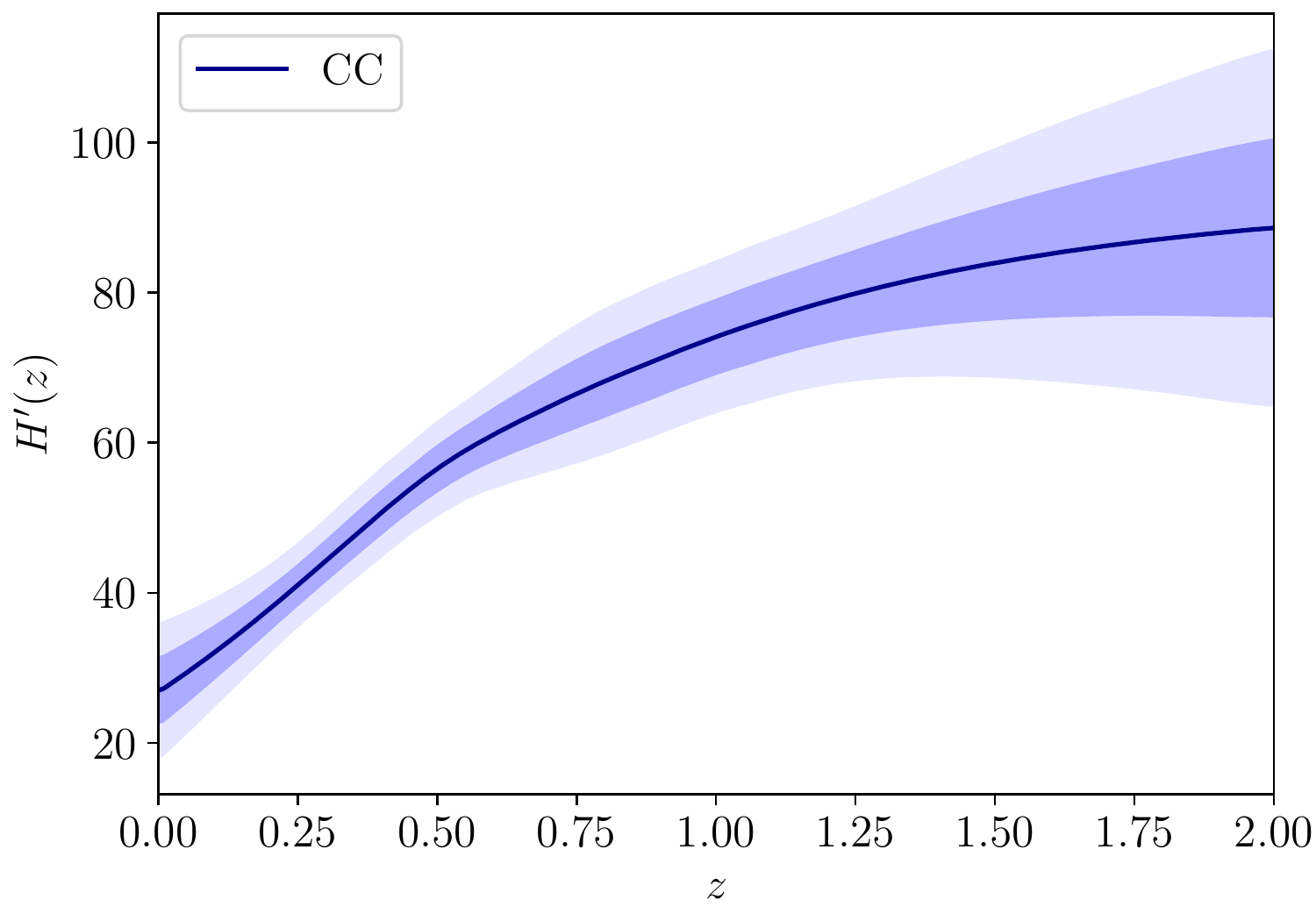}
		\includegraphics[angle=0, width=0.325\textwidth]{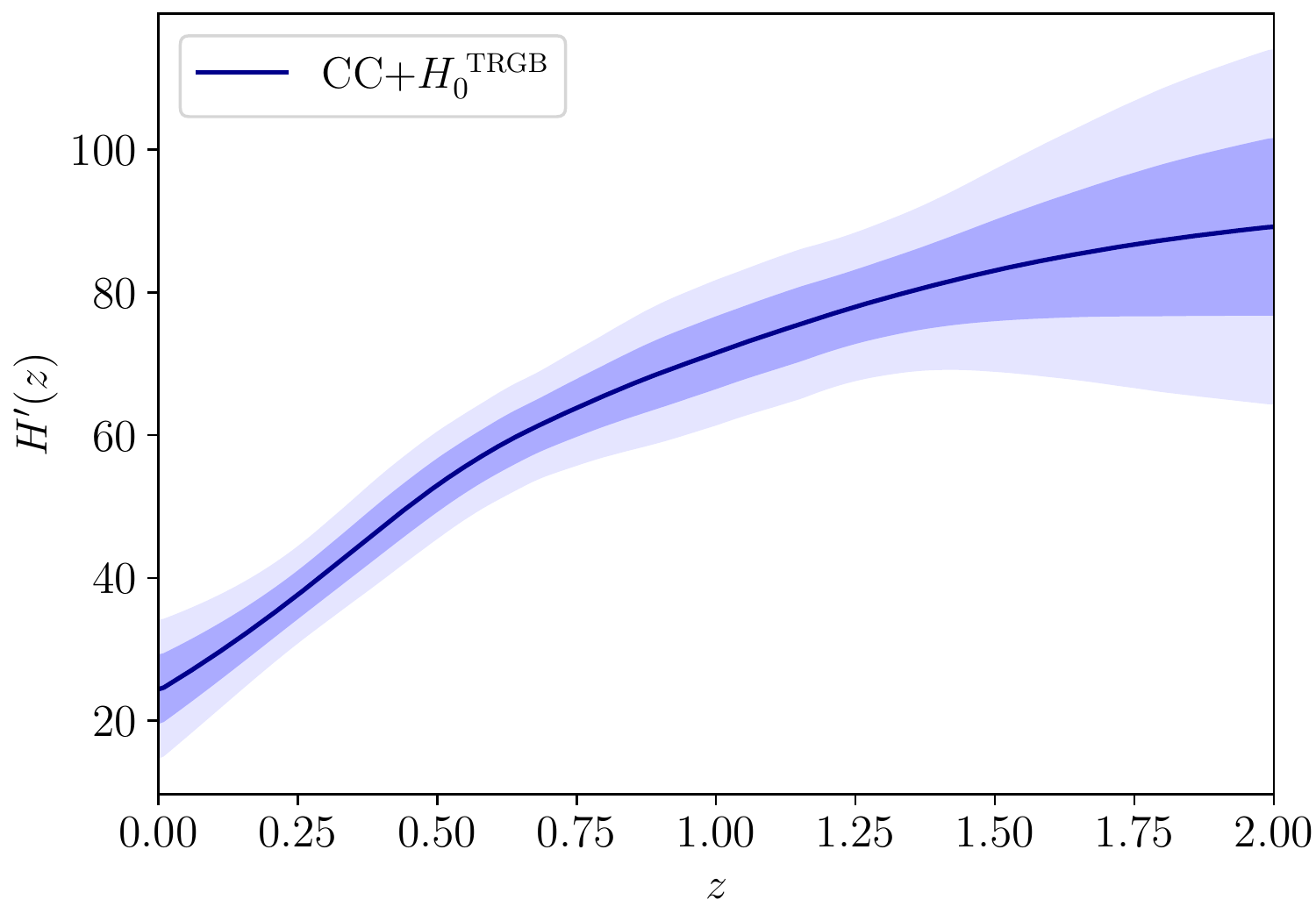} 		
		\includegraphics[angle=0, width=0.325\textwidth]{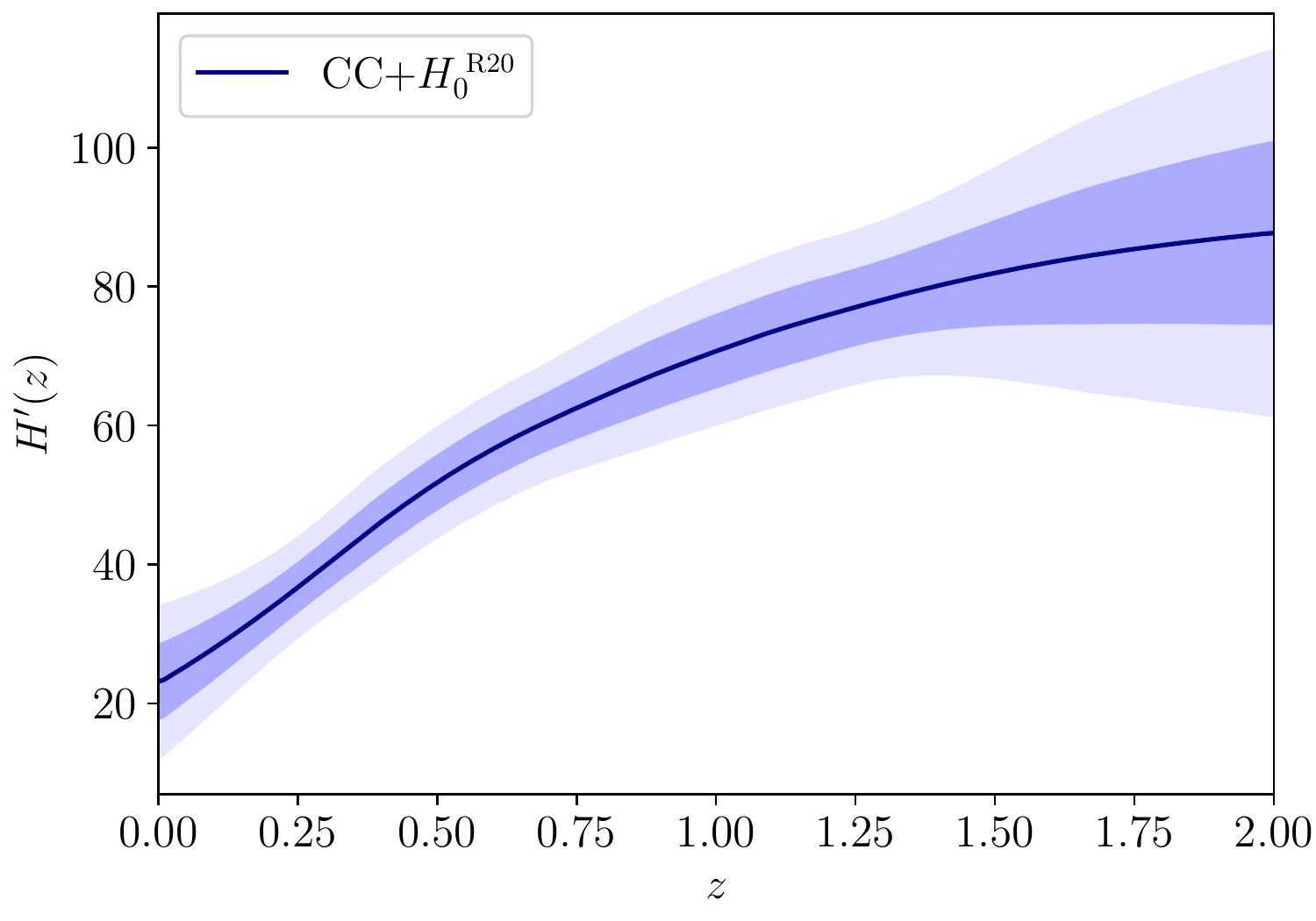} 		\\
		\includegraphics[angle=0, width=0.325\textwidth]{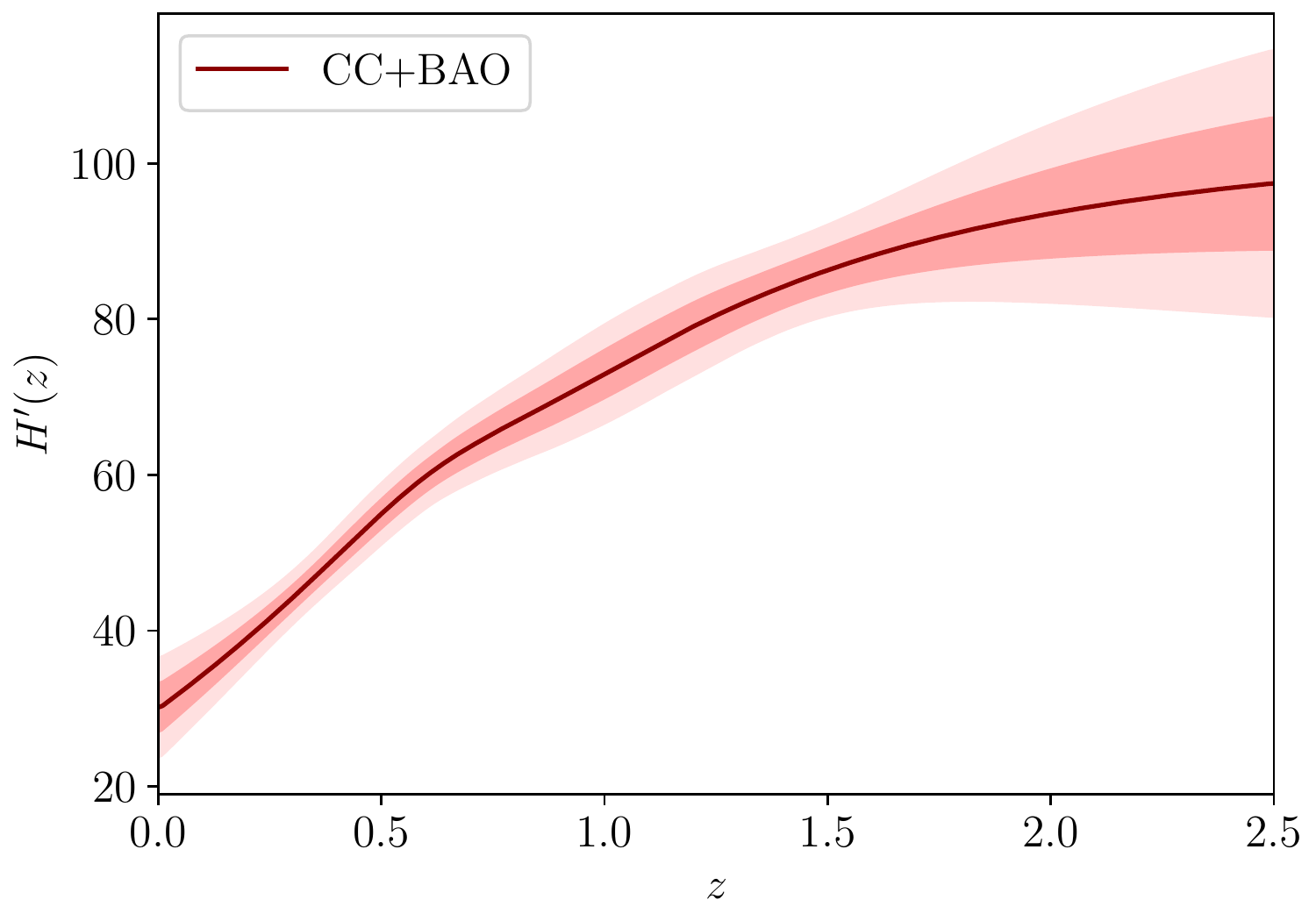}
		\includegraphics[angle=0, width=0.325\textwidth]{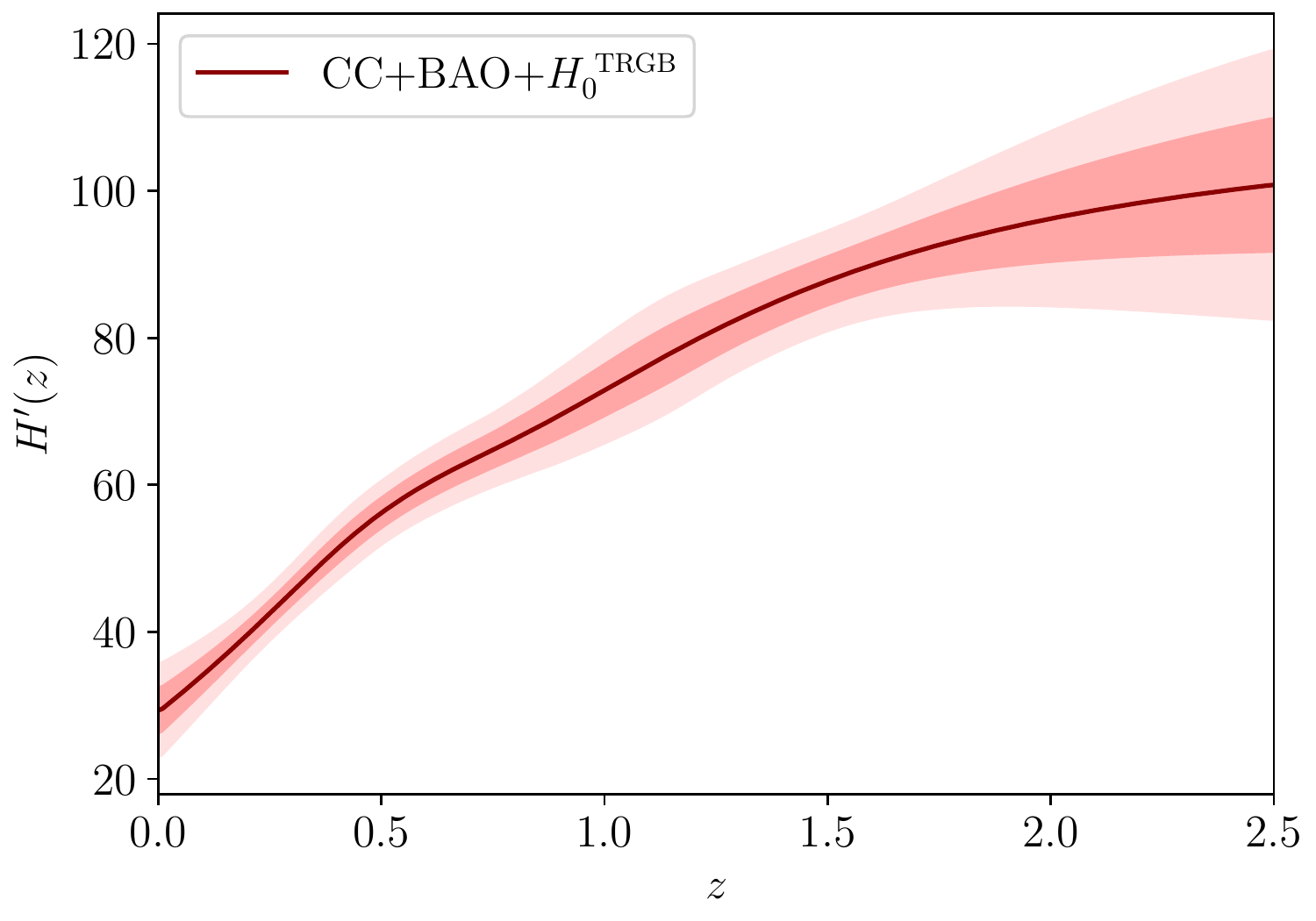} 		
		\includegraphics[angle=0, width=0.325\textwidth]{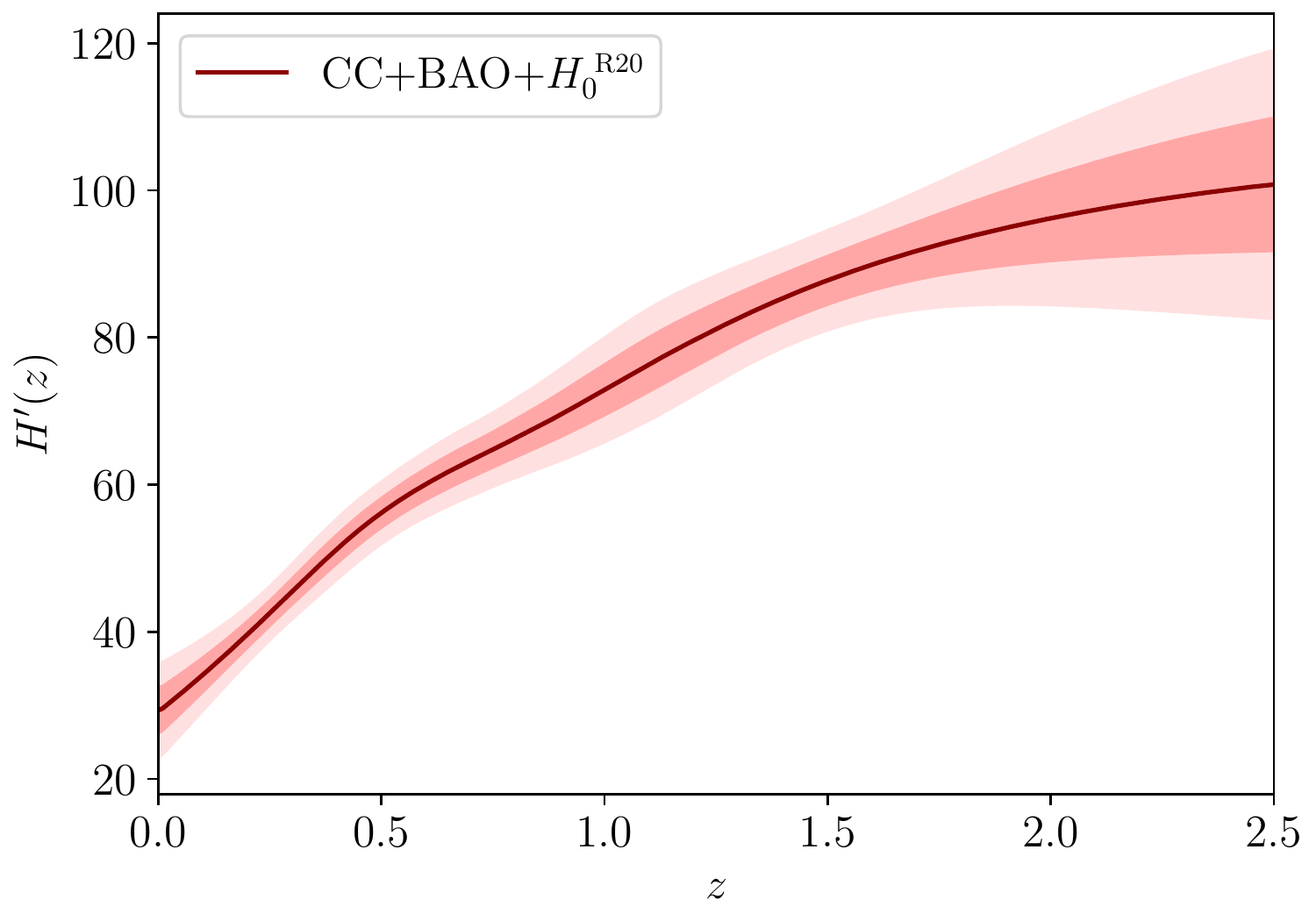} 		
	\end{center}
	\caption{{\small Plots for the reconstructed $H'(z)$ as a function of redshift. The solid line represents the mean $H'(z)$ curve and the associated 1$\sigma$-2$\sigma$ confidence regions are shown in lighter shades.}}
	\label{dH-plot}
\end{figure}

\begin{table}[t!]
	\caption{{\small Table showing the reconstructed mean values of $H'(z=0)$ along with the associated 1$\sigma$ uncertainties.}}
	\begin{center}
		\resizebox{0.70\textwidth}{!}{\renewcommand{\arraystretch}{1.5} \setlength{\tabcolsep}{50 pt} \centering  
			\begin{tabular}{l c }
				\hline \hline
				\textbf{Datasets} &  $H'(0)$ [in km Mpc$^{-1}$ s$^{-1}$]  \\ 
				\hline 
				CC &  $27.01 \pm 4.55$ \\ 
				CC+BAO & $30.13 \pm 3.29$  \\ 
				\hline
				CC+$H_0^\text{TRGB} $&  $24.41 \pm 4.86$ \\ 
				CC+BAO+$H_0^\text{TRGB} $&  $29.32 \pm 3.29$ \\ 
				\hline
				CC+$H_0^\text{R20} $&  $23.14 \pm 5.55$ \\ 
				CC+BAO+$H_0^\text{R20} $&  $29.33 \pm 3.31$  \\ 
				\hline 
				\hline 
			\end{tabular} 
		}
	\end{center}
	\label{dH0_table}
\end{table}

The immediate follow-up is the reconstruction $H'(z)$ by differentiating these 1000 reconstructed $H(z)$ realizations. We obtain these 1000 $H'(z_i)$ realizations from these reconstructed Hubble functions corresponding to each redshift $z_i$, numerically via the central differencing method as  
\begin{equation}
    H'(z_i) \simeq \frac{H(z_{i+1}) - H(z_{i-1})}{z_{i+1} - z_{i-1}}\,. \label{Hprime}
\end{equation}
This will produce smaller uncertainties $\mathcal{O}(\Delta z^2)$ rather than $\mathcal{O}(\Delta z)$ which occur for the forward and backward differencing methods, where $\Delta z = z_{i+1} - z_{i-1}$. 

From these 1000 reconstructed $H'(z)$ functions, we obtain the best fit values of reconstructed $H'(z)$ along with the associated confidence levels using another MC routine. Plots for the reconstructed $H'(z)$ using the MC routine on the 1000 reconstructed $H'(z)$ realizations are shown in Fig.~\ref{dH-plot}. The reconstructed values of $H'(z=0)$ using this MC approach for the respective data set combinations are given in Table~\ref{dH0_table}. Thus for 6 different Hubble data samples, have simultaneously obtained the reconstructed $H(z)$ and $H'(z)$ functions in a non-parametric model-independent way employing neural networks.

\subsection{Cosmological Null tests \label{null-test}}

\begin{figure}[t!]
	\begin{center}
		\includegraphics[angle=0, width=0.325\textwidth]{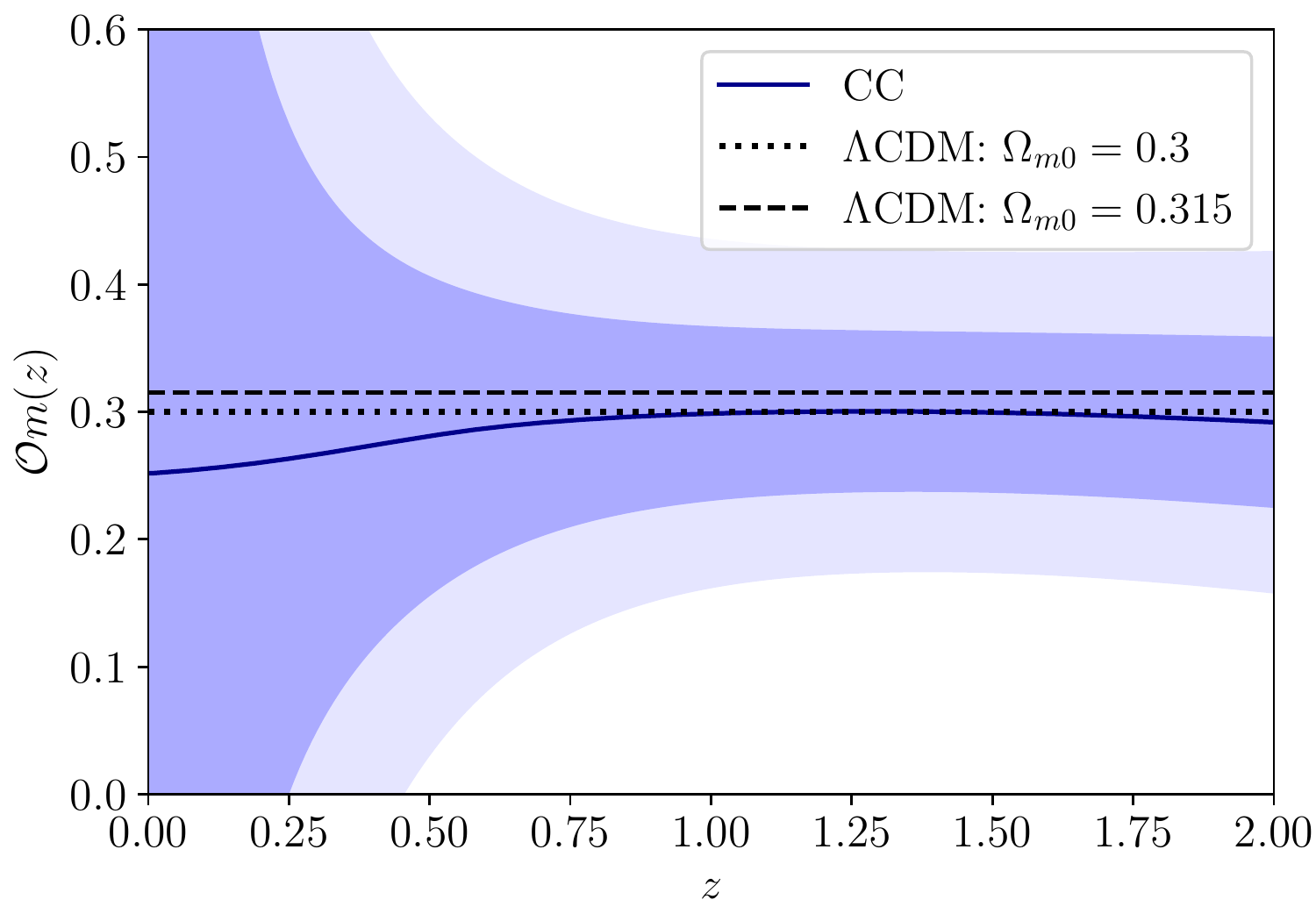}
		\includegraphics[angle=0, width=0.325\textwidth]{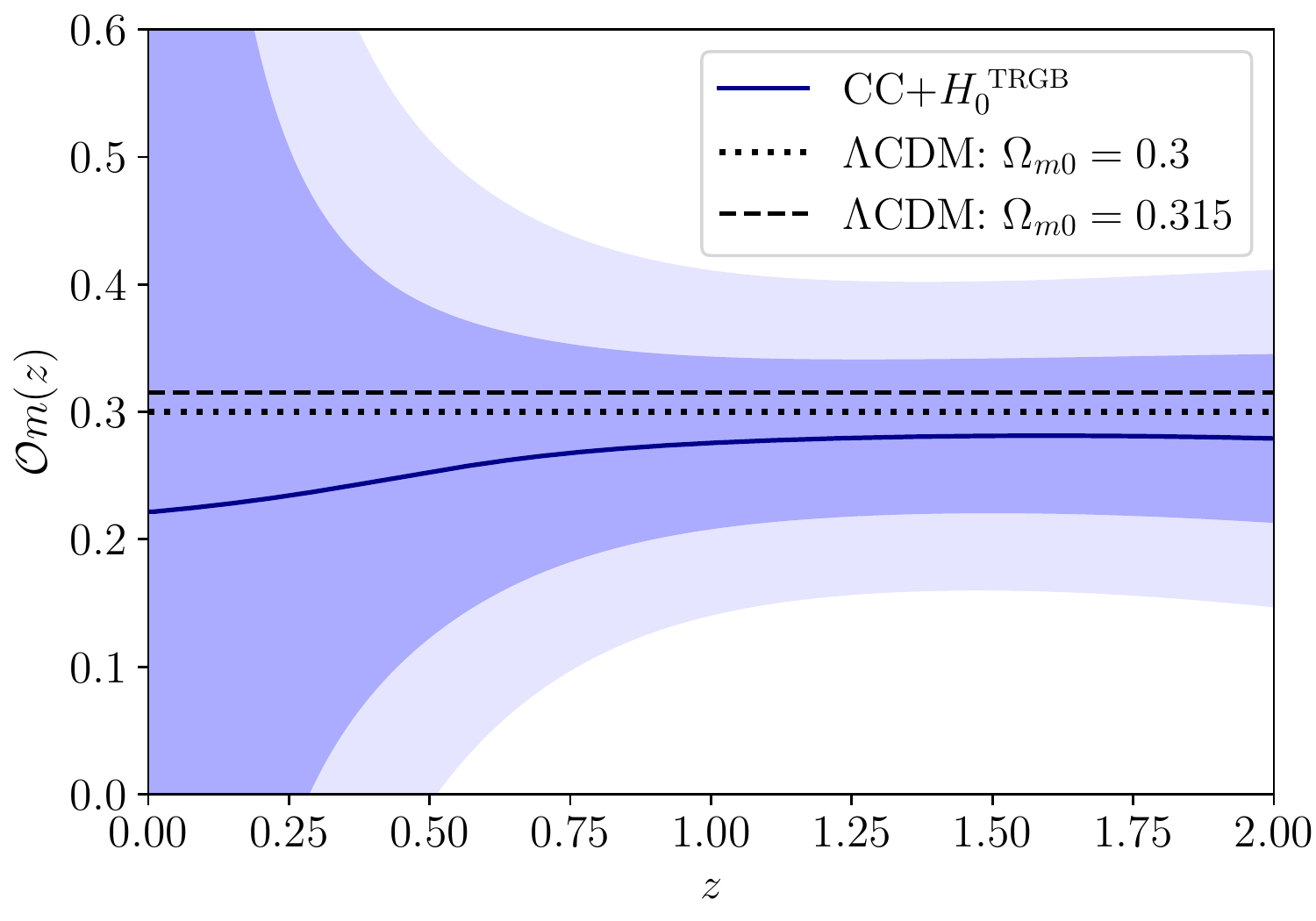} 		
		\includegraphics[angle=0, width=0.325\textwidth]{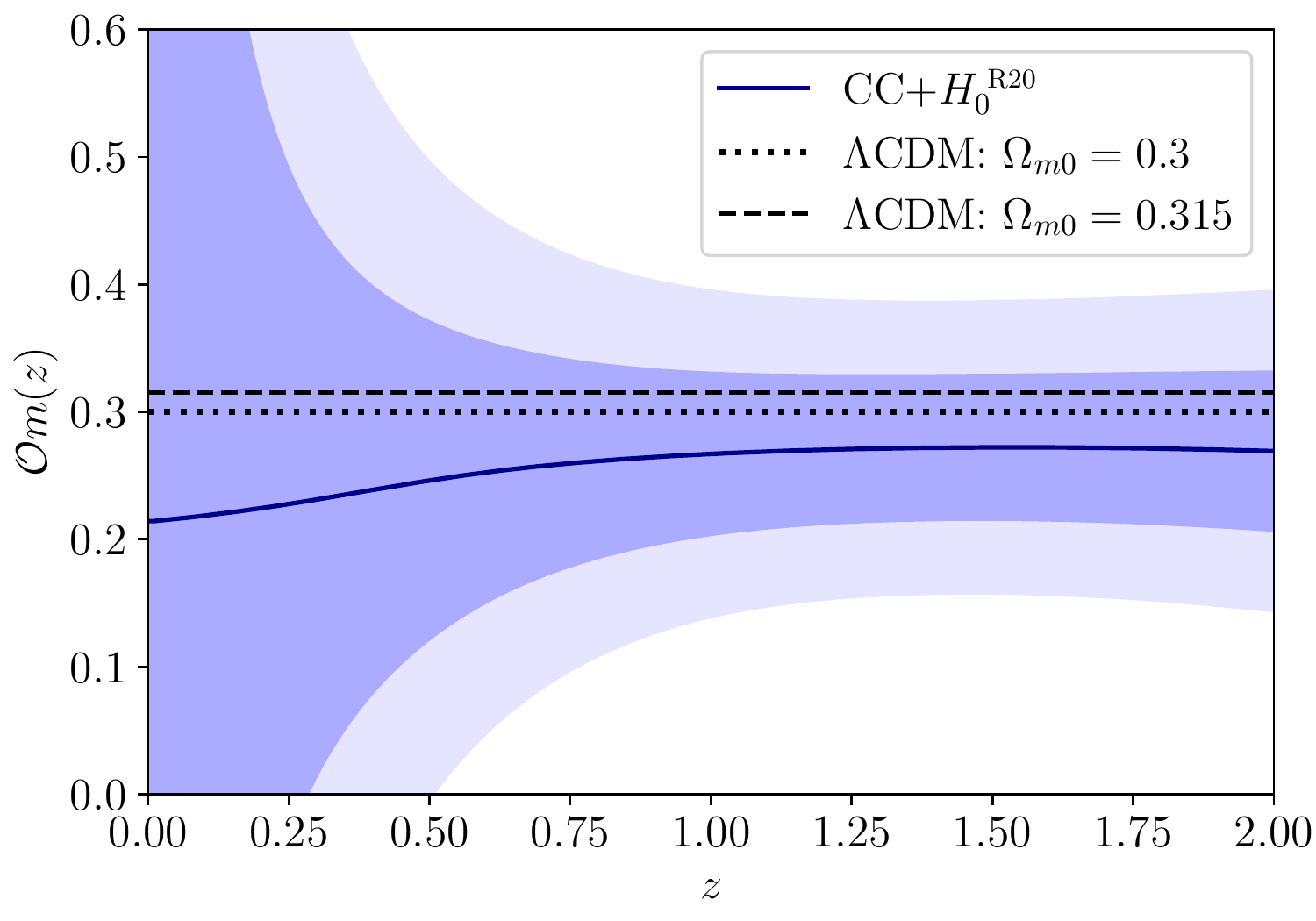} 		\\
		\includegraphics[angle=0, width=0.325\textwidth]{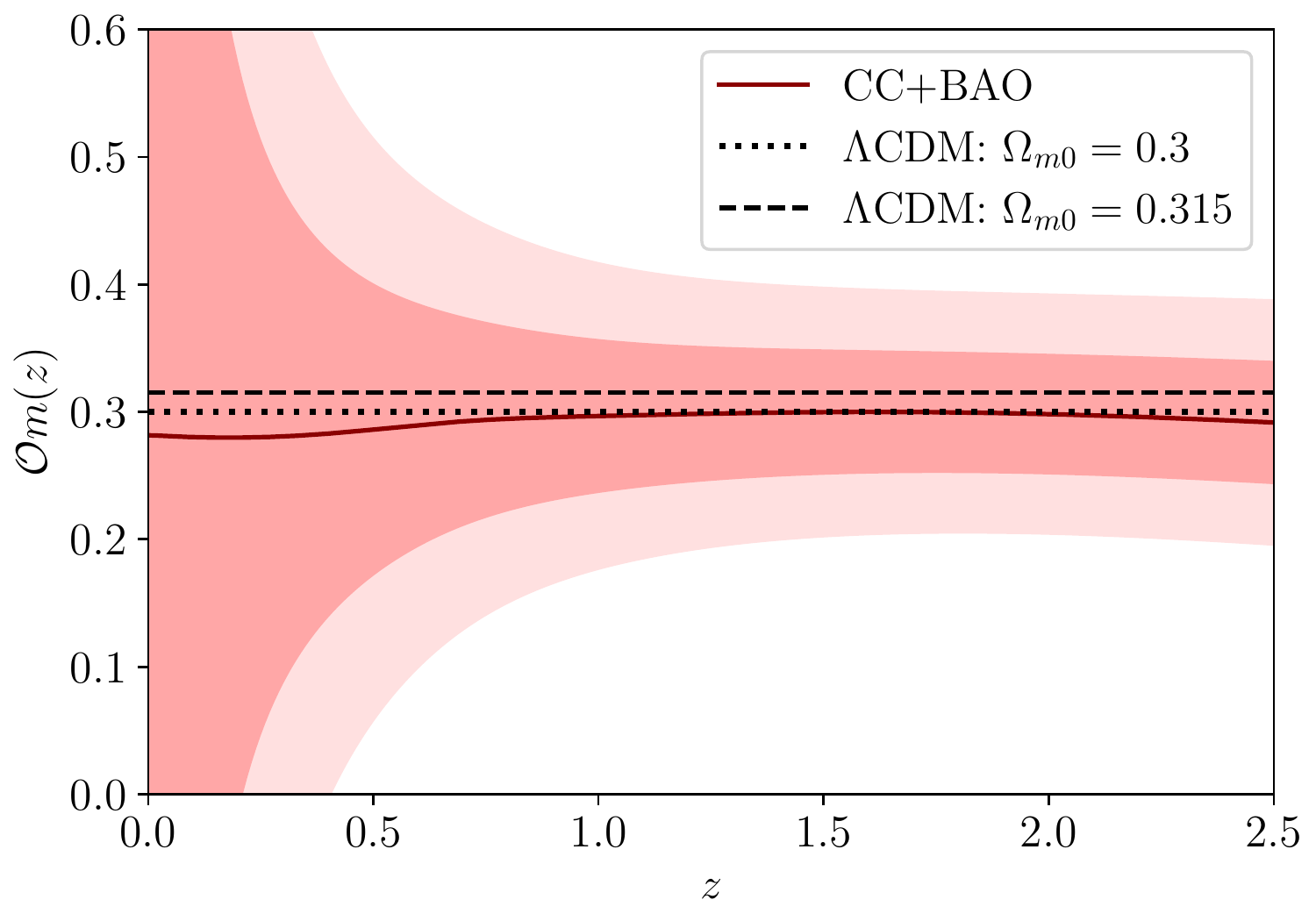}
		\includegraphics[angle=0, width=0.325\textwidth]{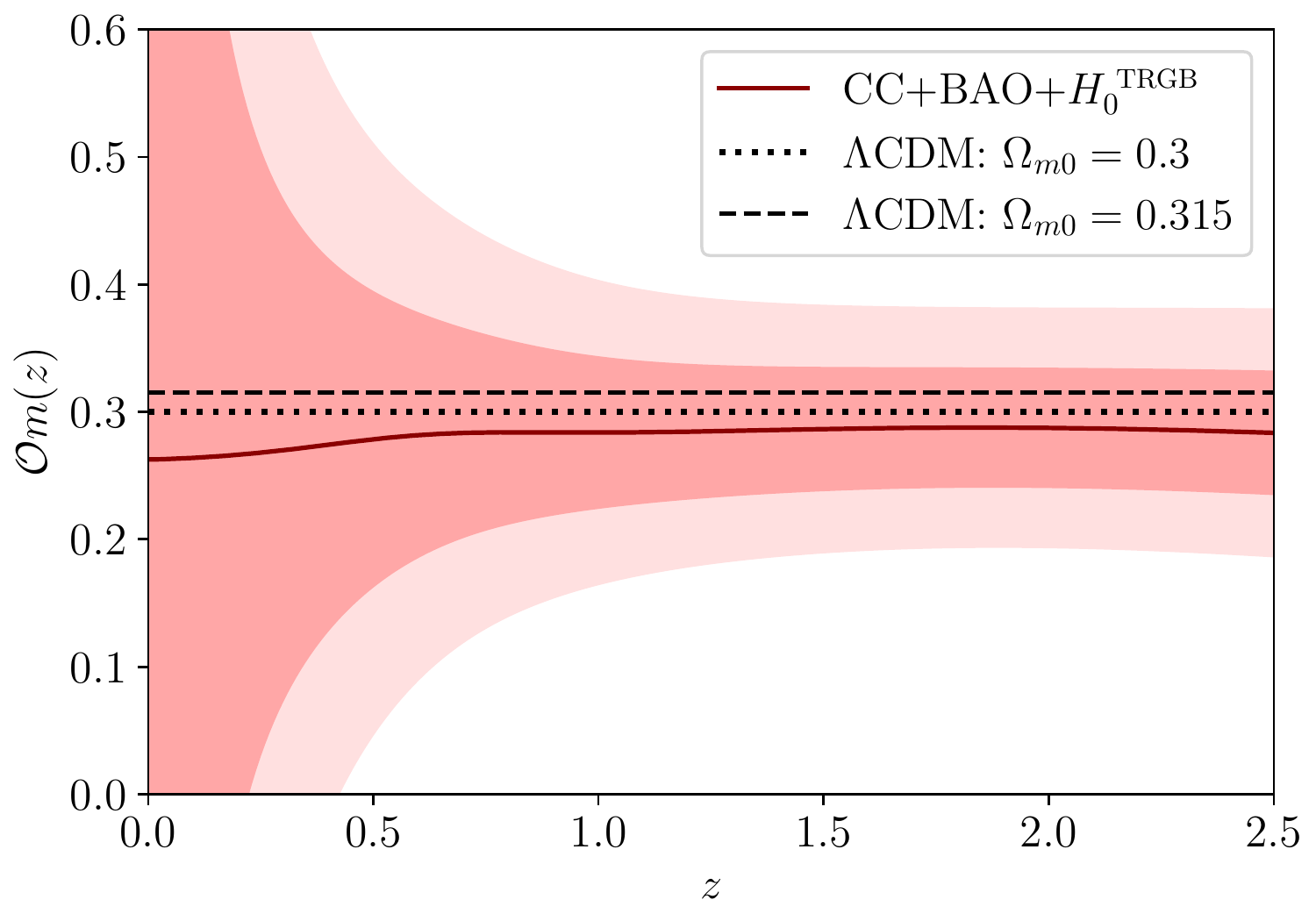} 		
		\includegraphics[angle=0, width=0.325\textwidth]{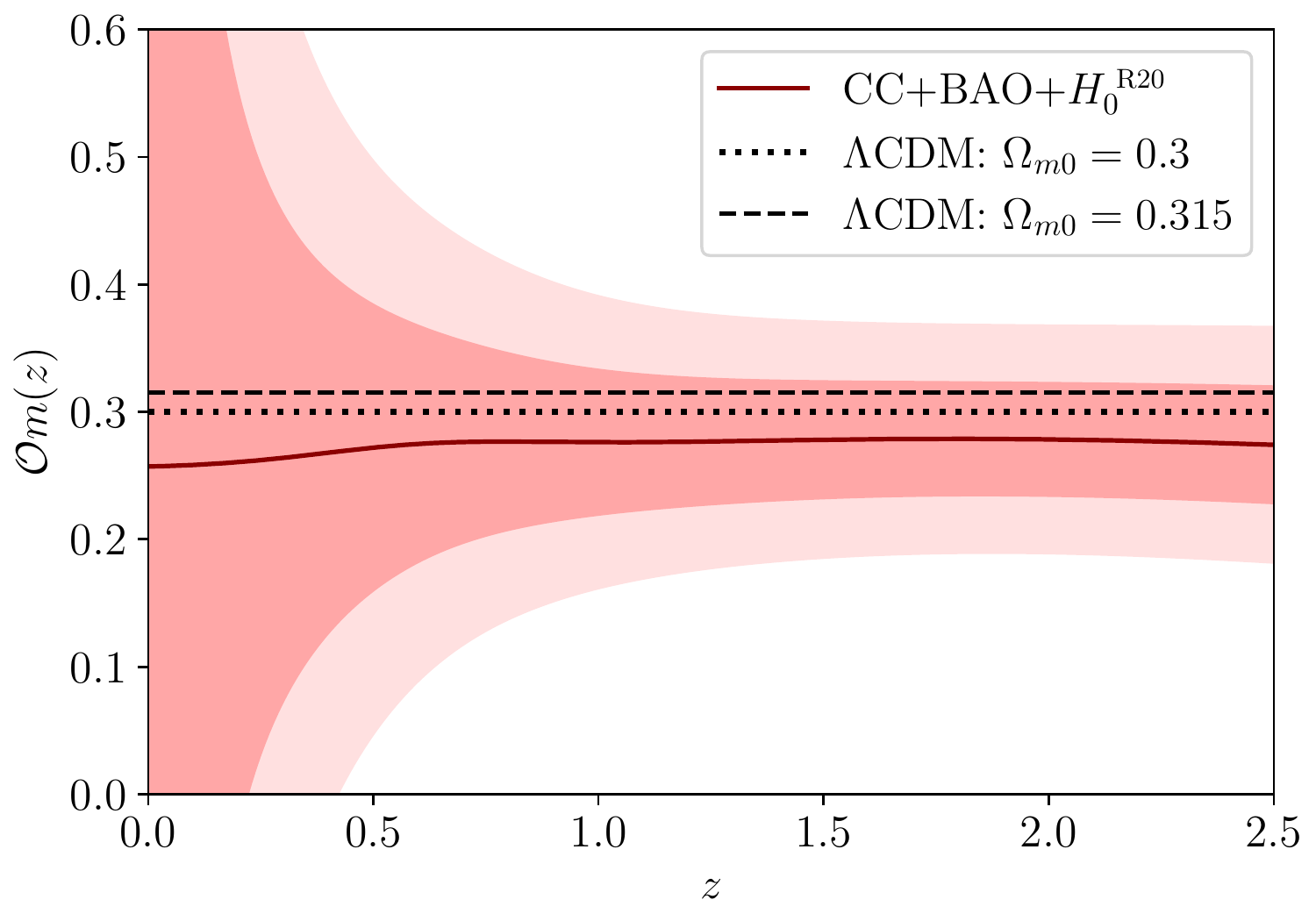} 		
	\end{center}
	\caption{{\small Plots for the reconstructed $\mathcal{O}m(z)$ as a function of redshift. The solid line represents the mean curve and the associated 1$\sigma$-2$\sigma$ confidence regions are shown in lighter shades.}} 
	\label{omz-plot}
\end{figure}

We now introduce two diagnostic functions, namely the $\mathcal{O}m$ diagnostics \cite{Sahni:2008xx, Zunckel:2008ti, Shafieloo:2009hi} followed by the $\mathcal{L}^{(1)}$ diagnostics \cite{Zunckel:2008ti, Shafieloo:2009hi}, to test the concordance model of cosmology. First, we reconstruct the $\mathcal{O}m$ diagnostics from the reconstructed $H(z)$ in Sec.~\ref{dH-recon} as a function of the redshift $z$, given by  
\begin{equation} \label{eq:null_test}
    \mathcal{O}m (z) = \frac{E^2(z) - 1}{(1+z)^3 -1}\,,
\end{equation} 
where $E (z) = {H(z)}/{H_0}$ is the reduced Hubble parameter. 

This $\mathcal{O}m$ diagnostics serves as a null test to distinguish the $\Lambda$CDM model from alternative dark energy and modified gravity models. Being a function of $H(z)$ only, which can be directly reconstructed from observational data, it is independent of the cosmic equation of state. Moreover, there is no dependence on any theory of gravity. So, this exercise serves as an alternative route towards understanding the late-time cosmic acceleration in absence of any convincing physical theory  \cite{Zunckel:2008ti,Sahni:2008xx,Shafieloo:2009hi,Clarkson:2007pz,Qi:2016wwb,Qi:2018pej,Bengaly:2020neu}.

\begin{figure}[t!]
	\begin{center}
		\includegraphics[angle=0, width=0.325\textwidth]{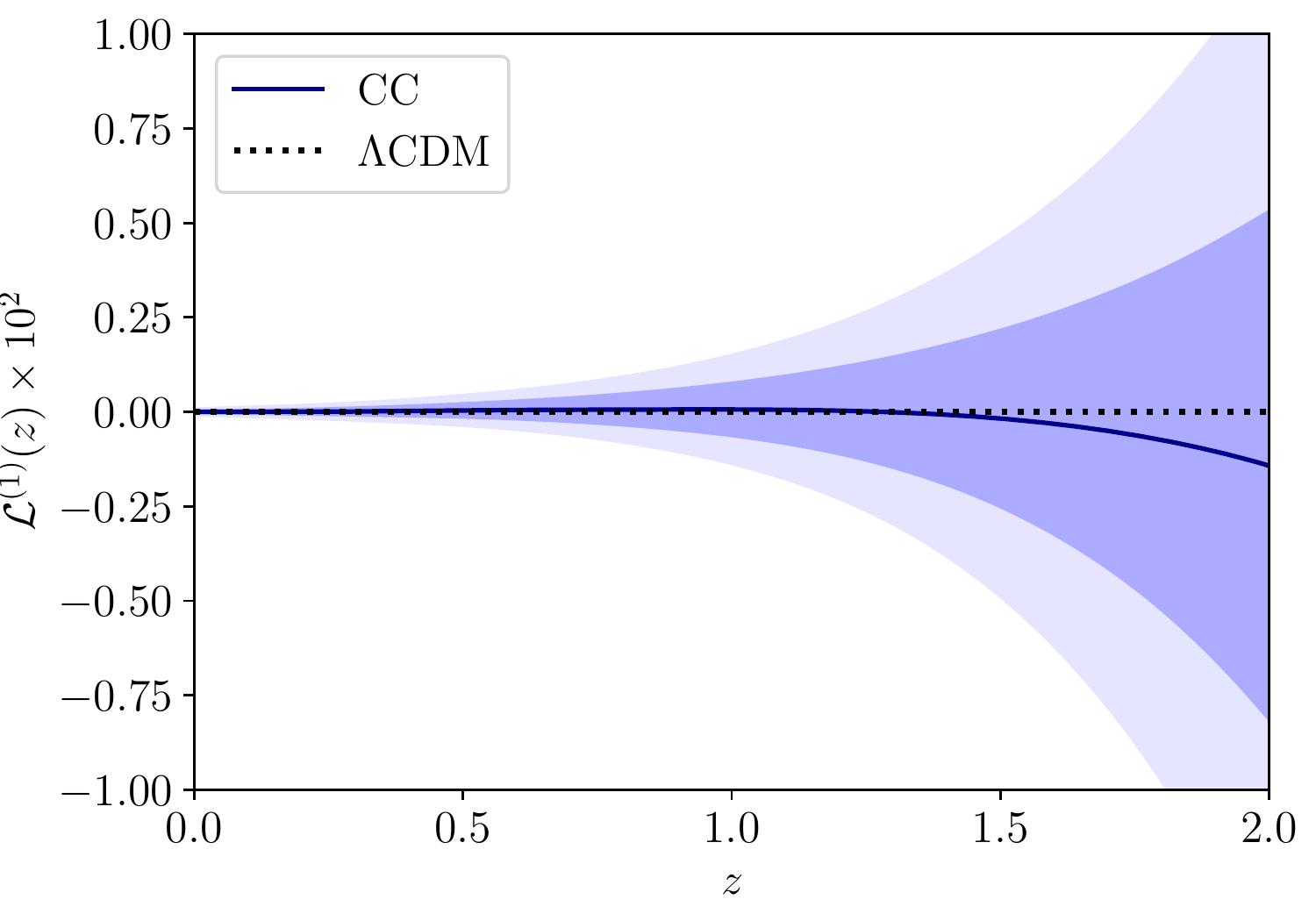}
		\includegraphics[angle=0, width=0.325\textwidth]{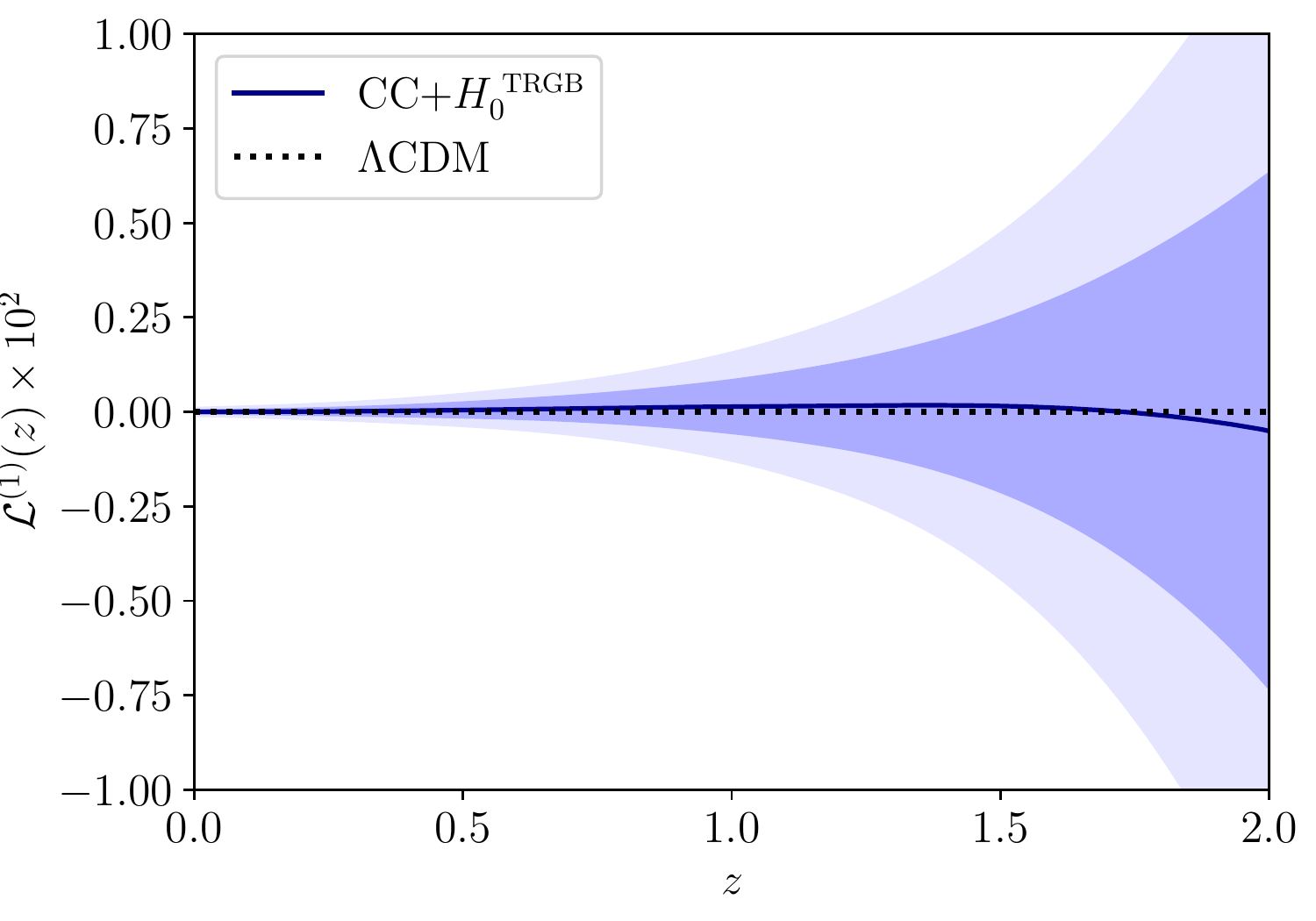} 		
		\includegraphics[angle=0, width=0.325\textwidth]{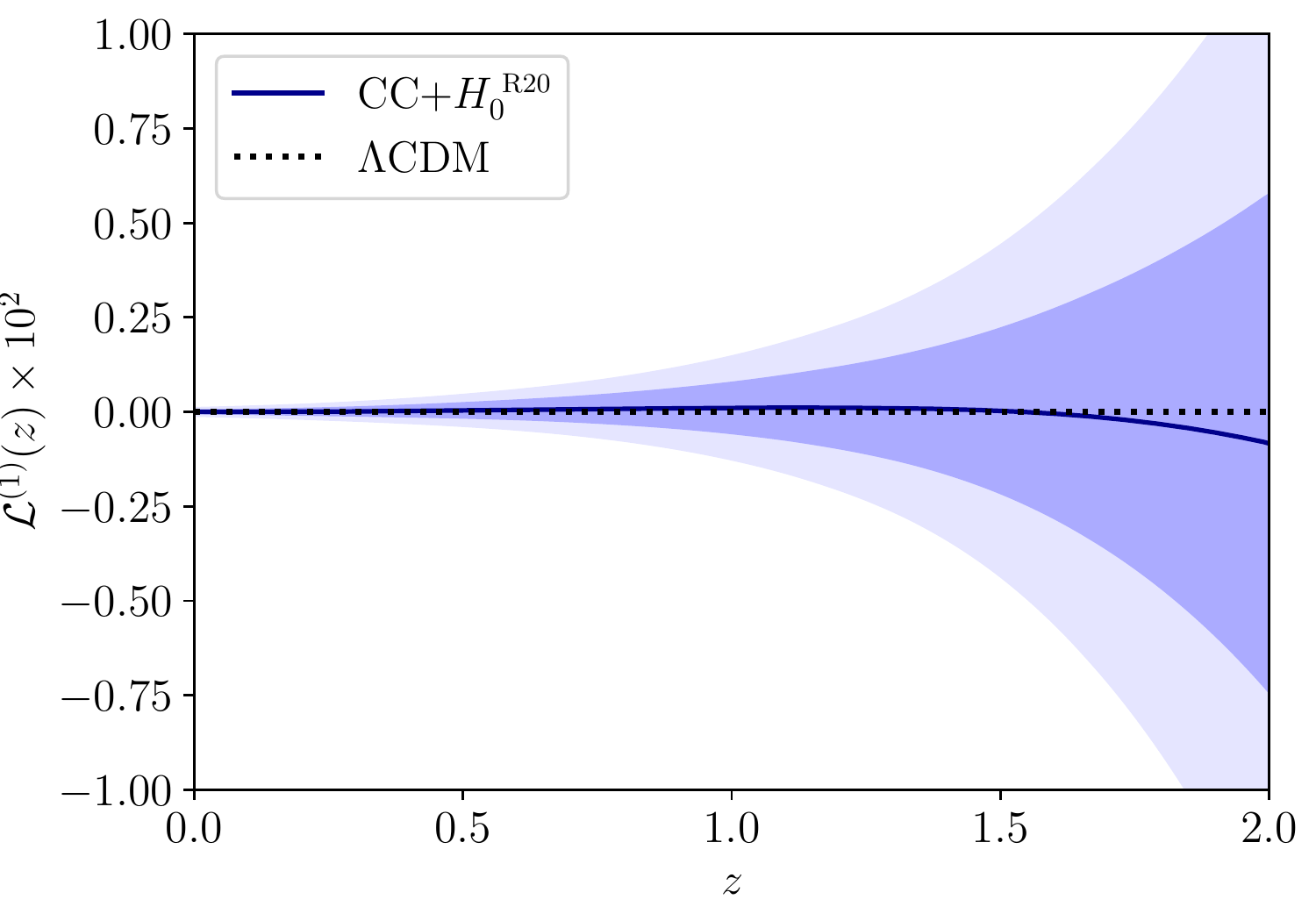} 		\\
		\includegraphics[angle=0, width=0.325\textwidth]{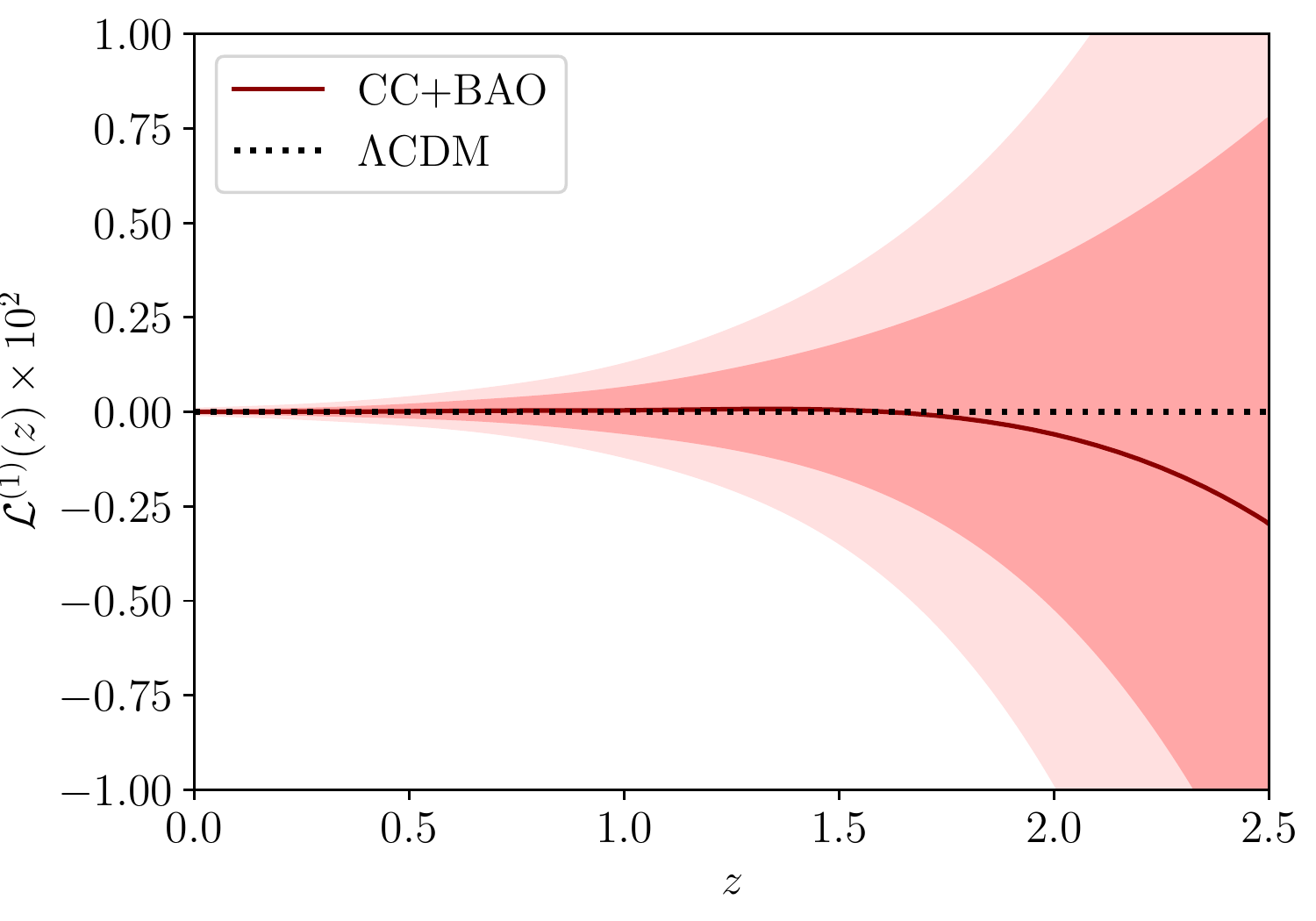}
		\includegraphics[angle=0, width=0.325\textwidth]{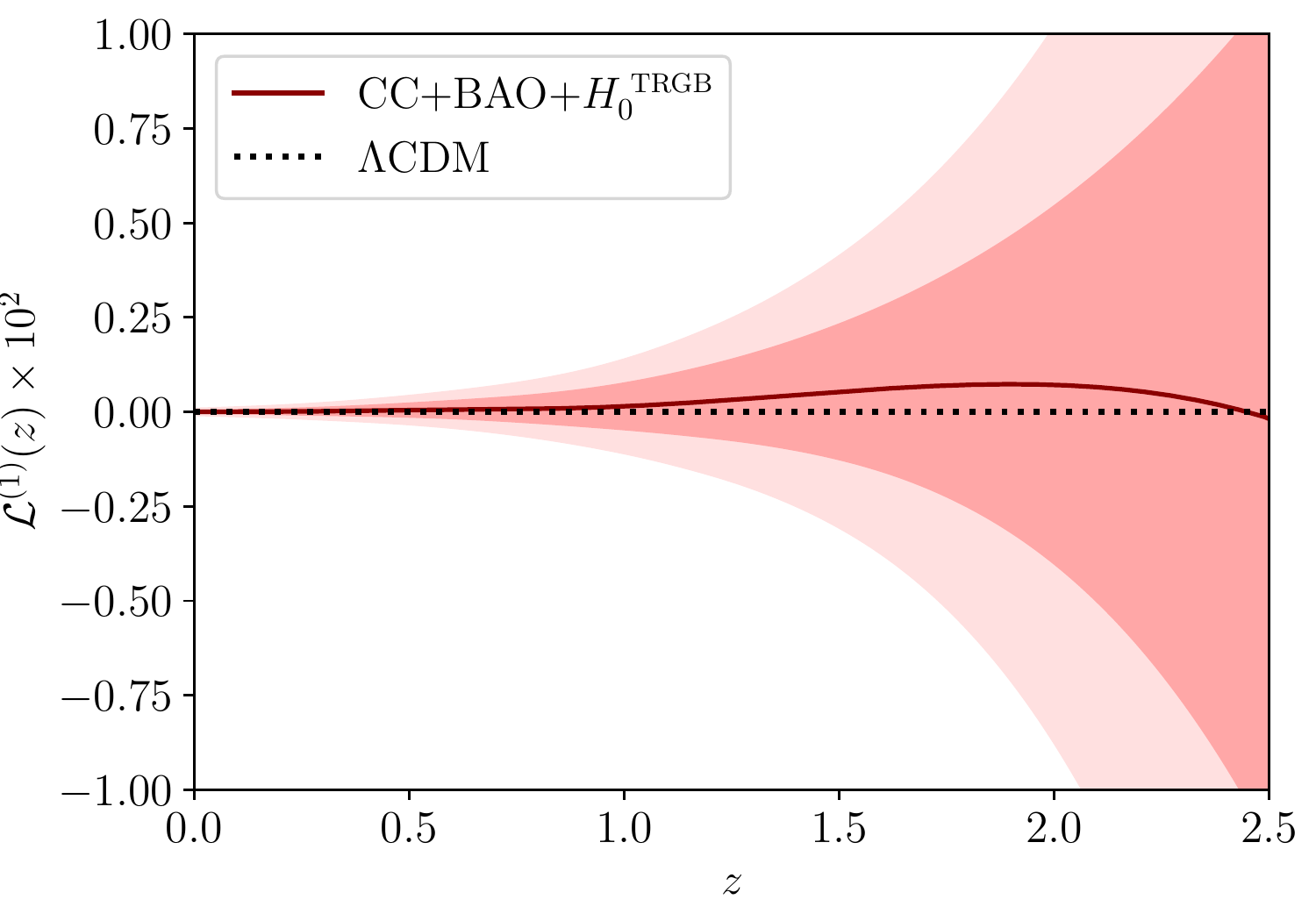} 		
		\includegraphics[angle=0, width=0.325\textwidth]{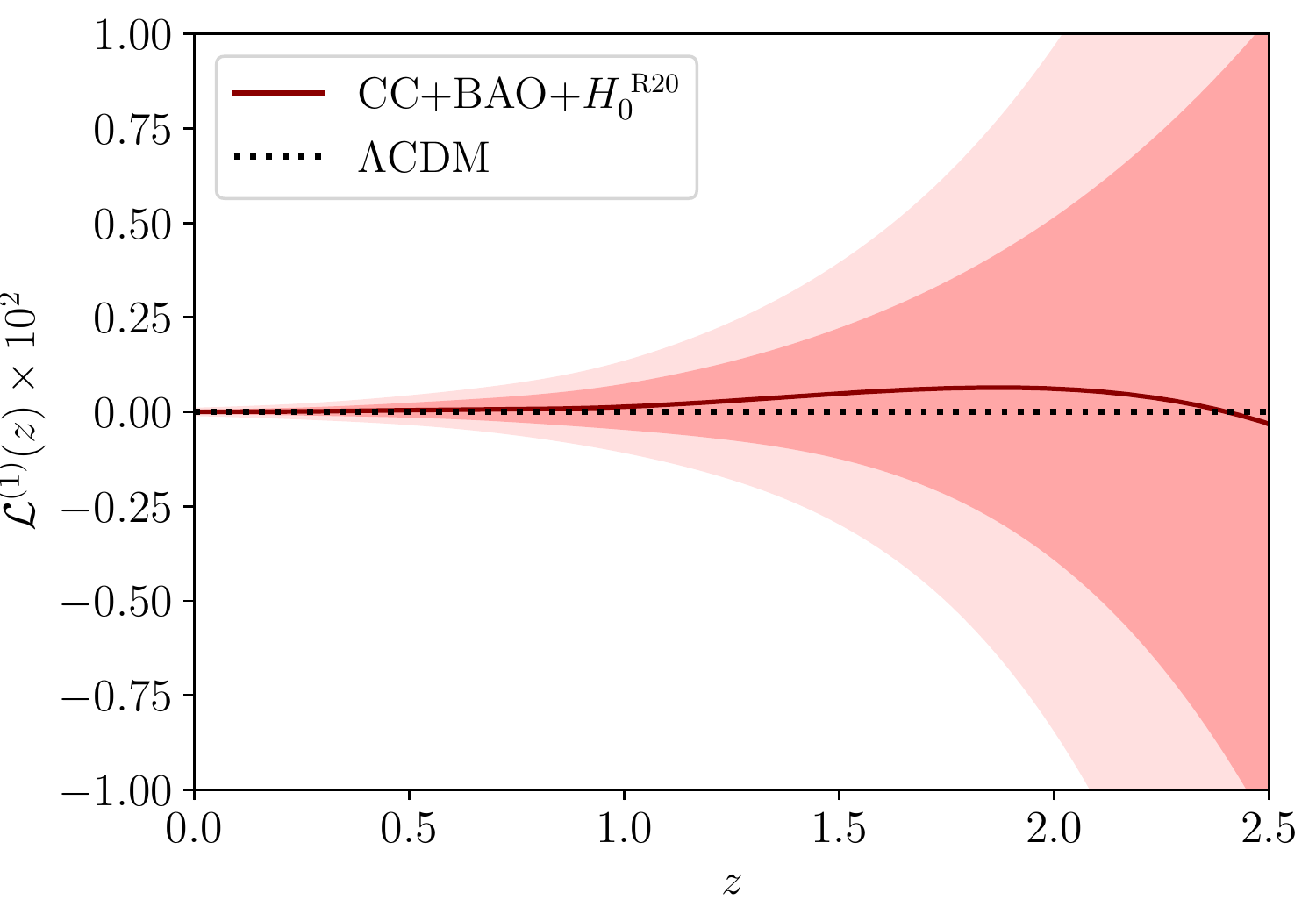} 		
	\end{center}
	\caption{{\small Plots for the reconstructed $\mathcal{L}^{(1)}(z)$ as a function of redshift. The solid line represents the mean curve and the associated 1$\sigma$-2$\sigma$ confidence regions are shown in lighter shades.}}
	\label{L1-plot}
\end{figure}

For a universe with an underlying expansion history $E(z)$, given by the $\Lambda$CDM model, $\mathcal{O}m(z)$ will essentially be a constant, exactly equal to $\Omega_{m0}$, the matter density parameter at the present epoch. The slope of $\mathcal{O}m(z)$ can differentiate between different dark energy and modified gravity models even if the $\Omega_{m0}$ is not accurately known. Therefore, any possible deviation of $\mathcal{O}m(z)$ from $\Omega_{m0}$ can be used to draw inference on the dynamics of the universe. For the phenomenological $w$CDM model, where the dark energy component is described by a constant equation of state parameter $w$, a positive slope of the $\mathcal{O}m(z)$ indicates a phantom behaviour of dark energy, whereas a negative slope points towards a quintessence dark energy model. 

Plots for the $\mathcal{O}m$ diagnostics are shown in Fig.~\ref{omz-plot}. The uncertainties associated with the reconstructed  $\mathcal{O}m$ diagnostics are obtained by an MC error propagation technique. We observe that the $\Lambda$CDM model with a constant value of $\Omega_{m0} = 0.3$ and the Planck best-fit $\Omega_{m0}= 0.315$ \cite{Aghanim:2018eyx} are consistent with the $\mathcal{O}m$ reconstruction at the 1$\sigma$ confidence level. At lower $z$, the reconstructed values are not well constrained, although at higher $z$ results show a more or less constant behaviour with respect to redshift. However, the associated uncertainties are quite large to properly distinguish between either phantom or non-phantom behaviour of dark energy \cite{Qi:2016wwb,Qi:2018pej,Bengaly:2020neu}.

The best way to measure statistical deviations from standard cosmology is by calculating whether deviations from zero appear from quantities that vanish for $\Lambda$CDM. So, an effective diagnostic is thus the vanishing of $\mathcal{O}m'(z)$, the first order derivative of $\mathcal{O}m$ diagnostics with respect to redshift. This is equivalent to $\mathcal{L}^{(1)} = 0$, where $\mathcal{L}^{(1)}$ is another null diagnostic function, defined as
\begin{equation}\label{eq:L_test}
    \mathcal{L}^{(1)}(z) = 3(1+z)^2 \left[ 1-E^2(z) \right] + 2 z (3 + 3z + z^2 )E(z) E'(z)\,,
\end{equation} 
which provides extra information regarding the possible variations in $\mathcal{O}m(z)$. Again, $\mathcal{L}^{(1)}$ utilizes the reconstructed $E(z)$ and additional input from the $E'(z) = {H'(z)}/{H_0}$ reconstruction inferred from the trained neural networks. Another crucial fact, is that this $\mathcal{L}^{(1)}$ diagnostics is independent of the matter density $\Omega_{m0}$, which makes $\mathcal{L}^{(1)}(z)$ a better diagnostic function over $\mathcal{O}m(z)$. 

For the standard $\Lambda$CDM model, $\mathcal{L}^{(1)}=0$ which serves as the null test. Hence, any deviation from this null condition represents a departure from the concordance model of cosmology.  Plots for the $\mathcal{L}^{(1)}$ diagnostics are shown in Fig.~\ref{L1-plot}. The uncertainties associated with the reconstructed $\mathcal{L}^{(1)}$ diagnostics are obtained by an MC error propagation technique. Results show that the mean reconstructed $\mathcal{L}^{(1)}$ diagnostic function shows a deviation towards negative values for higher $z$ for the CC and CC+BAO combinations. But when the $H_0$ priors are included, 
the mean values of the reconstructed $\mathcal{L}^{(1)}$ functions first shows a deviation towards positive values, which again become negative for higher redshifts. However,  these deviations are statistically not very significant as we find that the concordance model is included at the 1$\sigma$ confidence level of the reconstructed results.

\section{Reconstruction of \texorpdfstring{$f(T)$}{} Gravity \label{fT-recon}}

Model-independent reconstruction of the $f(T)$ functional form has previously been carried out first in Ref.~\cite{Cai:2019bdh} using Hubble observational data. Then in Ref.~\cite{Briffa:2020qli}, this was expanded to include more data sets and prior values for the Hubble data. Further still, Ref.~\cite{LeviSaid:2021yat} utilized different combinations of background data sets and the growth rate of structure measurements to reconstruct data-driven models of $f(T)$ gravity. Finally, Ref.~\cite{Ren:2022aeo} extended some of this work to incorporate $f(T)$ gravity as an effective field theory. All the reconstructions were undertaken via the GP approach in conjunction with a general $f(T)$ dominated universe without assuming a specific form of the arbitrary Lagrangian in Eq.~\eqref{f_T_Lagrangian}. Now, GP assumes every element of a data set is normally distributed and part of a larger stochastic process, by optimizing a covariance function between these points it can reconstruct the entire evolution of the data set for some ranges of the data. The immediate issue here is that not all cosmological data is normally distributed. Moreover, one of the most recently debated topics for non-parametric reconstruction in cosmology with GP is that this technique is exposed to several foundational issues such as overfitting and kernel consistency problems \cite{OColgain:2021pyh}. These are problems that are known to appear in GP reconstructions \cite{10.5555/1162254,Bernardo:2021mfs} but which can be quantified in a number of statistical ways such as using Automatic Relevance Determination \cite{JMLR:v8:cawley07a,10.1007/978-3-319-62416-7_14} but there are many different measures to determining overfitting.

The key element of this analysis depends on the relation between $f(T)$ gravity scalar $T$ and the Hubble parameter $H$, given by Eq.~\eqref{Tor_sca_flrw}. The cosmological dynamics of $f(T)$ gravity is given by the Friedmann equation in Eq.~\eqref{eq:Friedmann_1}. For expressing Eq.~\eqref{eq:Friedmann_1} in terms of redshift alone, we rewrite the Lagrangian derivative $f_T$ term as
\begin{equation}
    f_T = \frac{\mathrm{d}f/\mathrm{d}z}{\mathrm{d}T/\mathrm{d}z} = \frac{f'(z)}{T'(z)}\,.
\end{equation}
where $f'(z) = \mathrm{d}f/\mathrm{d}z$ and $T'(z)= 12HH'$ respectively. The immediate task to obtain $f'(z)$ for this analysis through the central differencing method, given by 
\begin{equation}
    f'(z_i) \simeq \frac{f(z_{i+1}) - f(z_{i-1})}{z_{i+1} - z_{i-1}}\,.
\end{equation}

This method produces a numerical propagation equation for $f(z)$, given by
\begin{equation}\label{prop_eq_f_T}
    f(z_{i+1}) = f(z_{i-1}) + 2\left(z_{i+1} - z_{i-1}\right) \frac{H'(z_i)}{H(z_i)} \left(3H^2(z_i) + \frac{f(z_i)}{2} - 3 H_0^2 \Omega_{m0} \left(1+z_i\right)^3\right)\,,
\end{equation}
where the propagation equation parameters $H_0$ are selected from the corresponding ANN reconstructions corresponding to the respective data sets.

\begin{figure}[t!]
	\begin{center}
		\includegraphics[angle=0, width=0.325\textwidth]{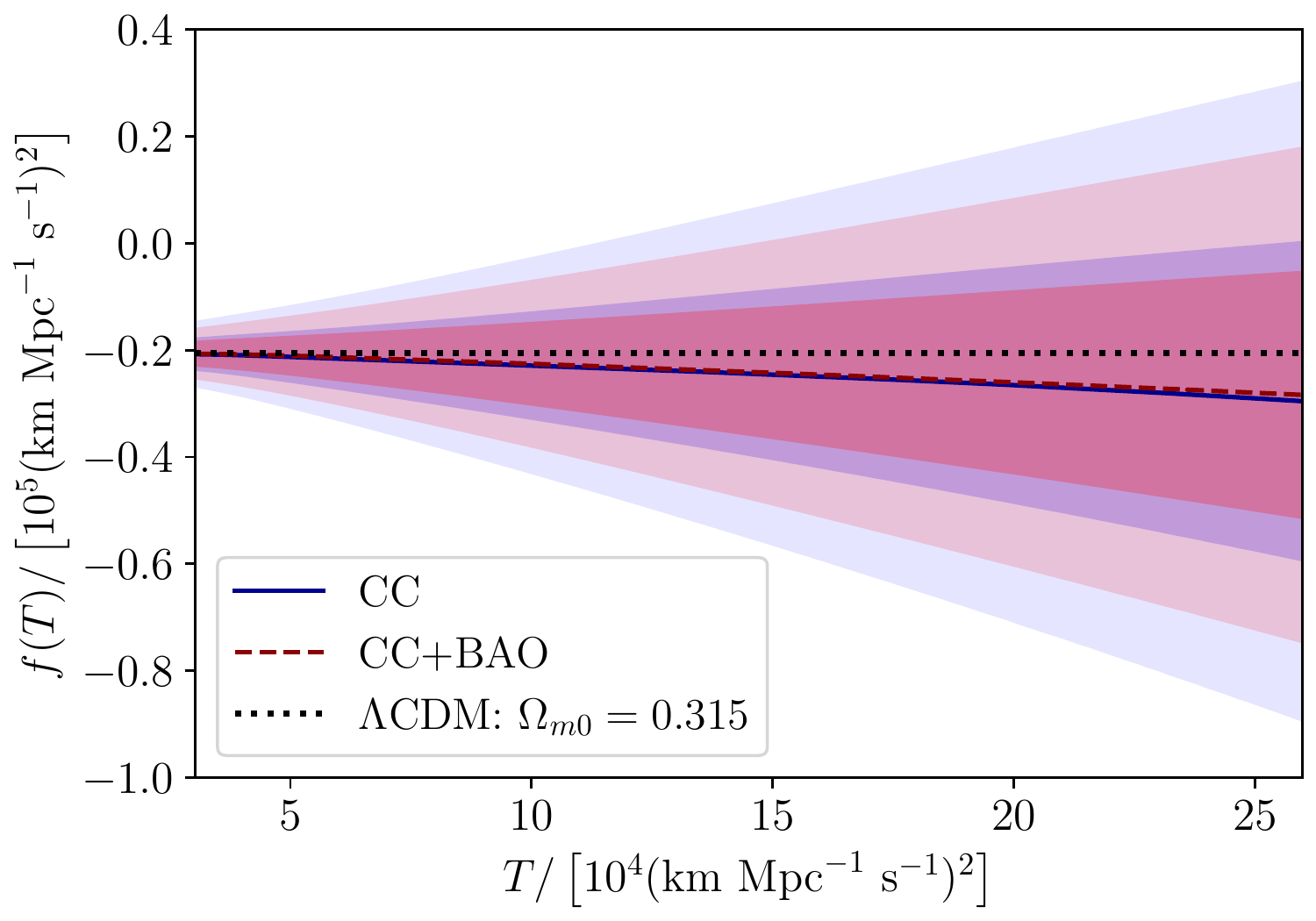}
		\includegraphics[angle=0, width=0.325\textwidth]{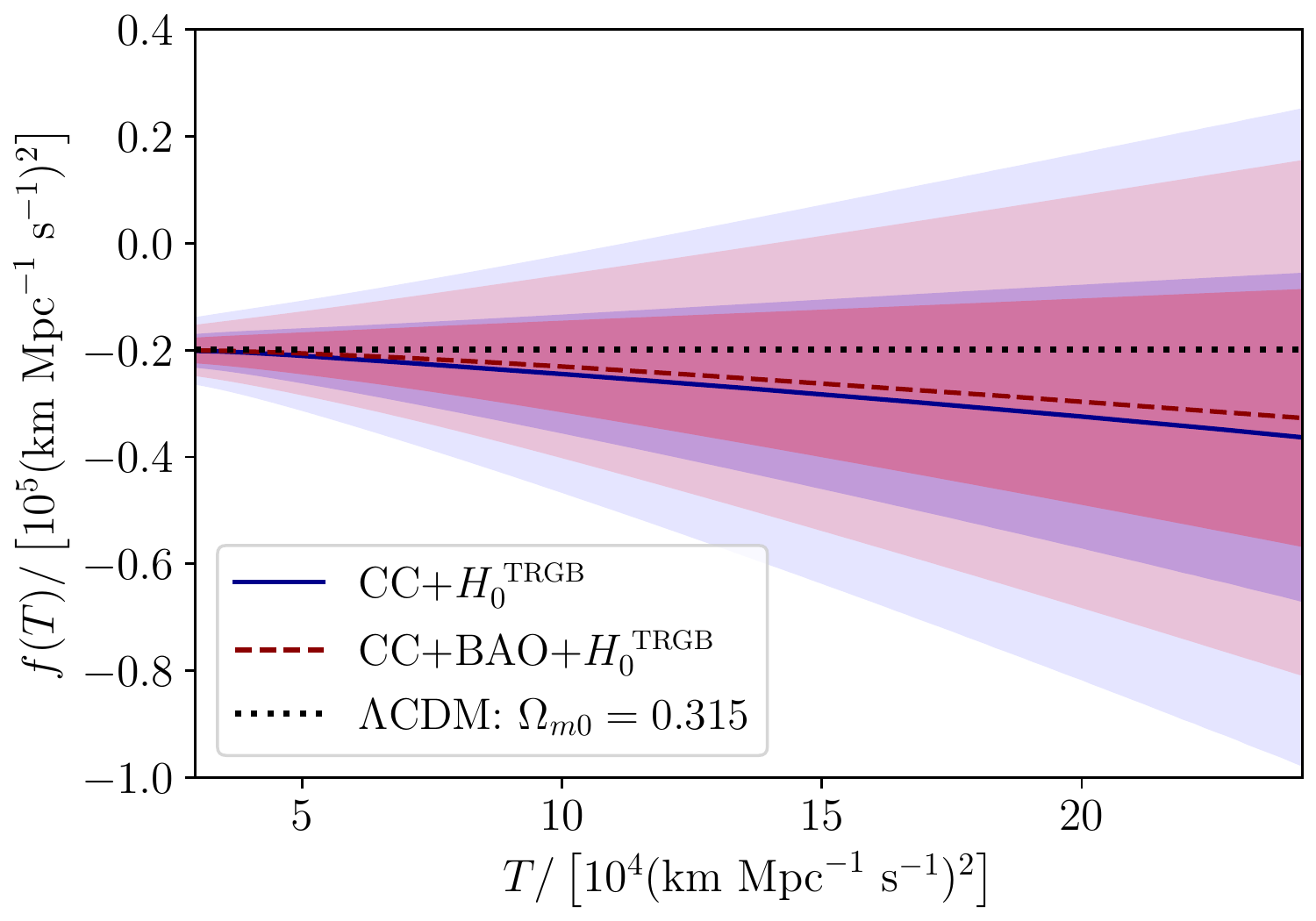} 		
		\includegraphics[angle=0, width=0.325\textwidth]{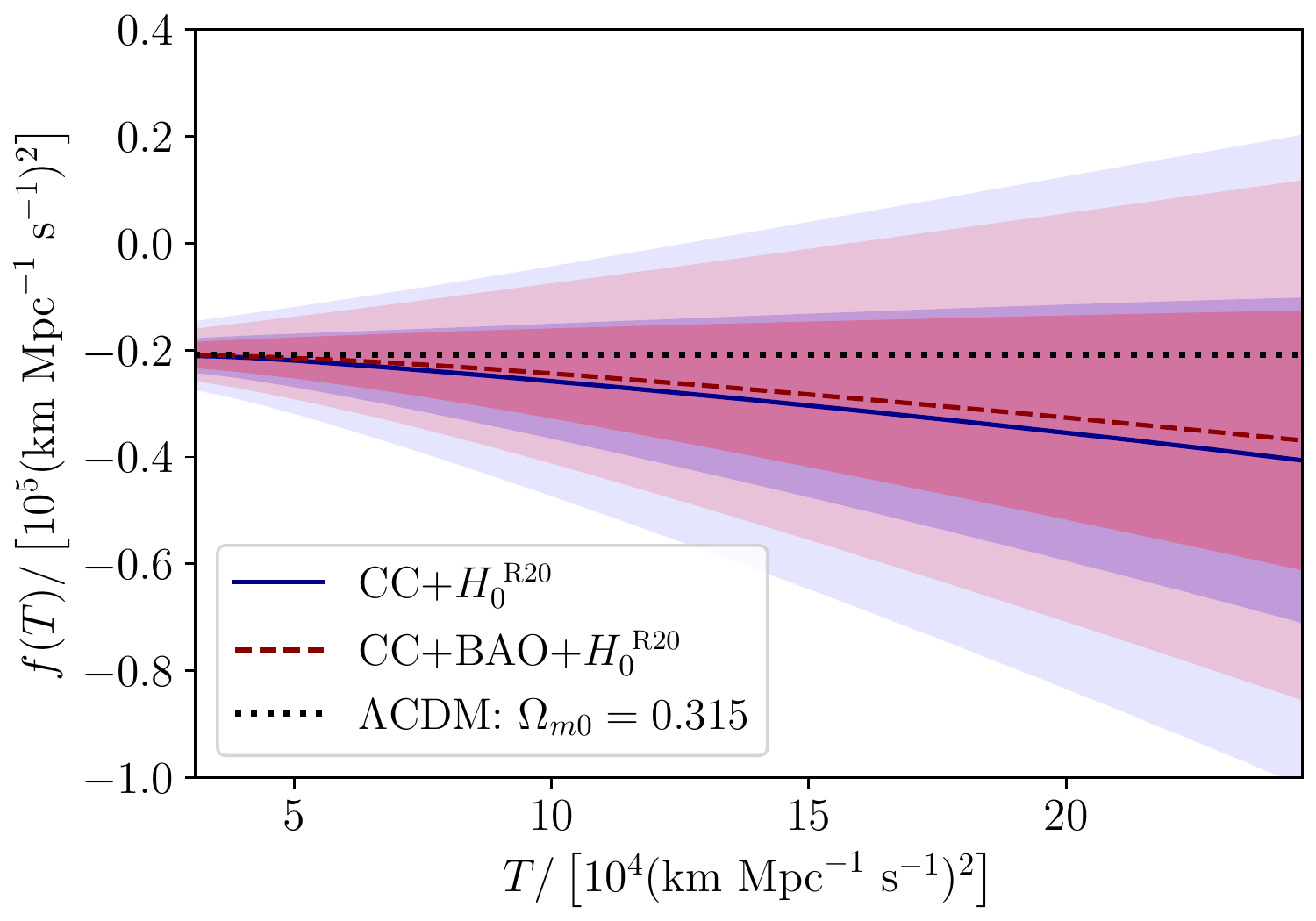} 		
	\end{center}
	\caption{{\small ANN reconstructions of $f (T )$ vs $T$ with different $H_0$ priors using $\Omega_{m0} = 0.315 \pm 0.007$\cite{Aghanim:2018eyx} for respective data set combinations.}}
	\label{f-plot2}
\end{figure}

We make use of two initial conditions to be employed for this analysis as follows
\begin{enumerate}
	\item[(i)] Evaluating the Friedmann equation~\eqref{eq:Friedmann_1} at $z=0$ gives
	\begin{equation}\label{f_T_boundary_condition}
	    f(z=0) \simeq 16\pi G \rho_{m}^0 - 6 H_0^2 = 6H_0^2\left(\Omega_{m}^0 - 1\right)\,.
	\end{equation}
	This is the Friedmann equation boundary condition, assuming that the $\Lambda$CDM model dominates at present epoch, i.e. $f_T(z=0) \simeq 0$. This further relies on the same $H_0$ values as the propagation equation itself.
	
	\item[(ii)] The second boundary condition can be obtained by using the forward differencing method through
	\begin{equation}\label{f_T_2nd_boundary_condition}
	    f'(z_i) \simeq \frac{f(z_{i+1}) - f(z_{i})}{z_{i+1} - z_{i}}\,, 
	\end{equation}
	that results in
	\begin{equation}\label{f_T_boundary_condition_2}
	    f(z_{i+1}) = f(z_{i}) + 6 \left(z_{i+1} - z_{i}\right) \frac{H'(z_{i})}{H(z_{i})} \left[H^2(z_{i}) + \frac{f(z_{i})}{6} - H_0^2 \Omega_{m}^0 \left(1+z_{i}\right)^3\right]\,,
	\end{equation}
	which straightforwardly leads to the necessary second boundary condition. 
\end{enumerate}

The uncertainties associated with the $f(T)$ function are obtained by an MC error propagation technique. Therefore, utilizing the propagation equation in Eq.~\eqref{prop_eq_f_T} along with the boundary conditions (i) and (ii), the redshift-dependent Lagrangian can be expressed as a function of $z$ in a model-independent way. Similarly, the corresponding torsion scalar can be associated with each $z$ through the Hubble parameter relation as $T(z)=6H^2(z)$. Finally, the Lagrangian function $f(T)$ and its derivative $f_T(T)$ is plot as a function of the torsion scalar $T$. We have adopted the Planck estimate for the matter density parameter, $\Omega_{m0}=0.315 \pm 0.007$\cite{Aghanim:2018eyx}, for this analysis.

\begin{figure}[t!]
	\begin{center}
		\includegraphics[angle=0, width=0.325\textwidth]{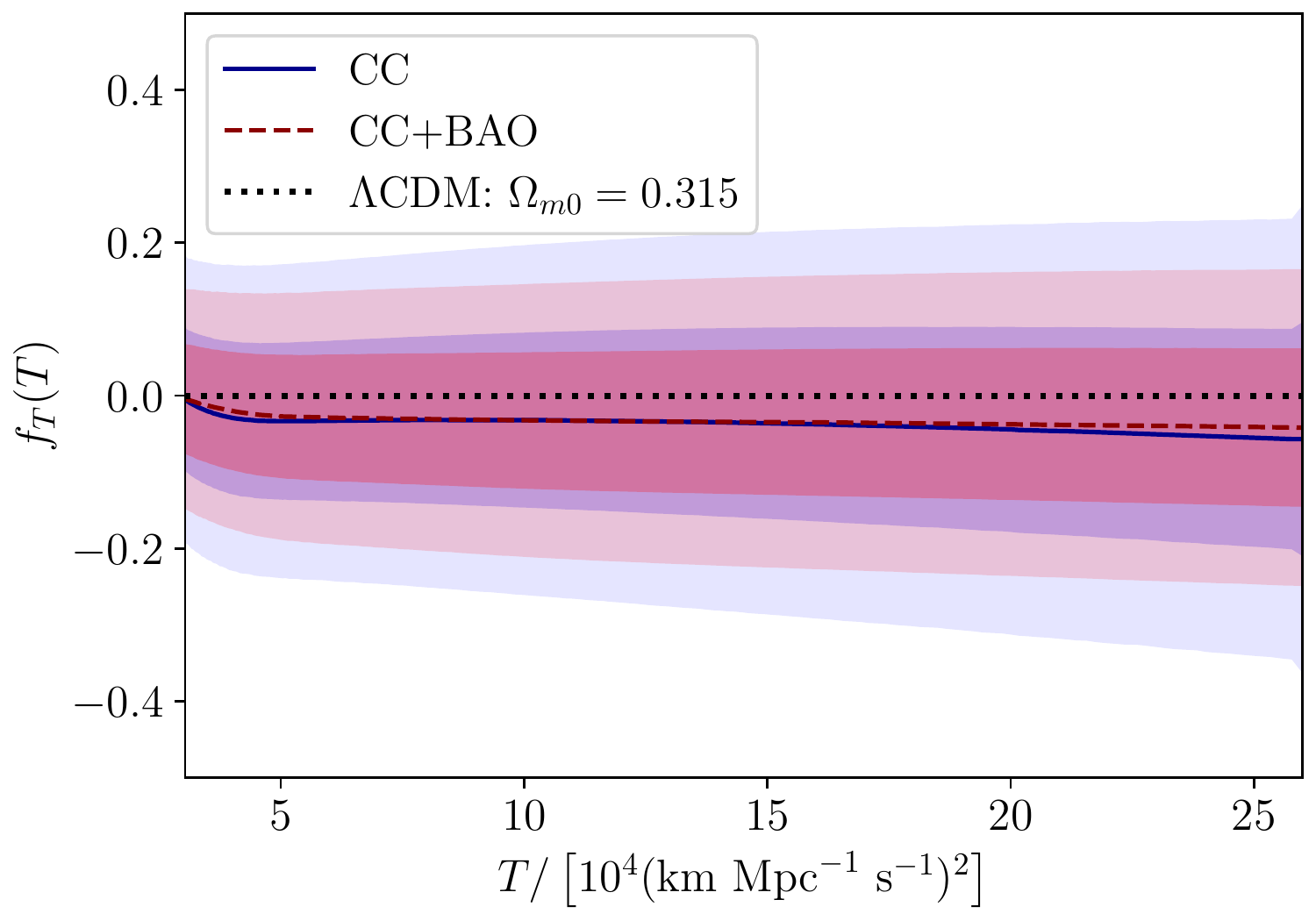}
		\includegraphics[angle=0, width=0.325\textwidth]{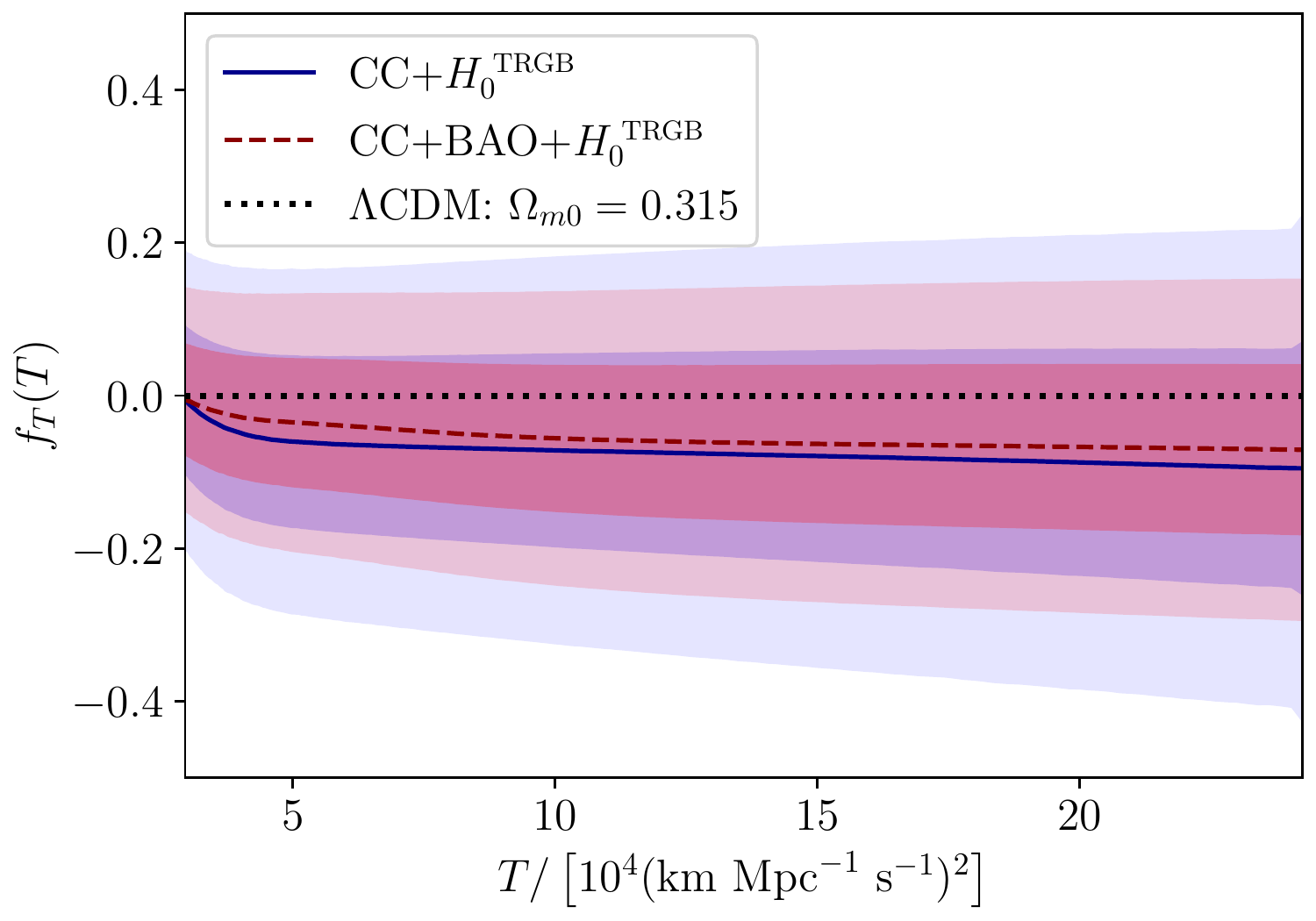} 		
		\includegraphics[angle=0, width=0.325\textwidth]{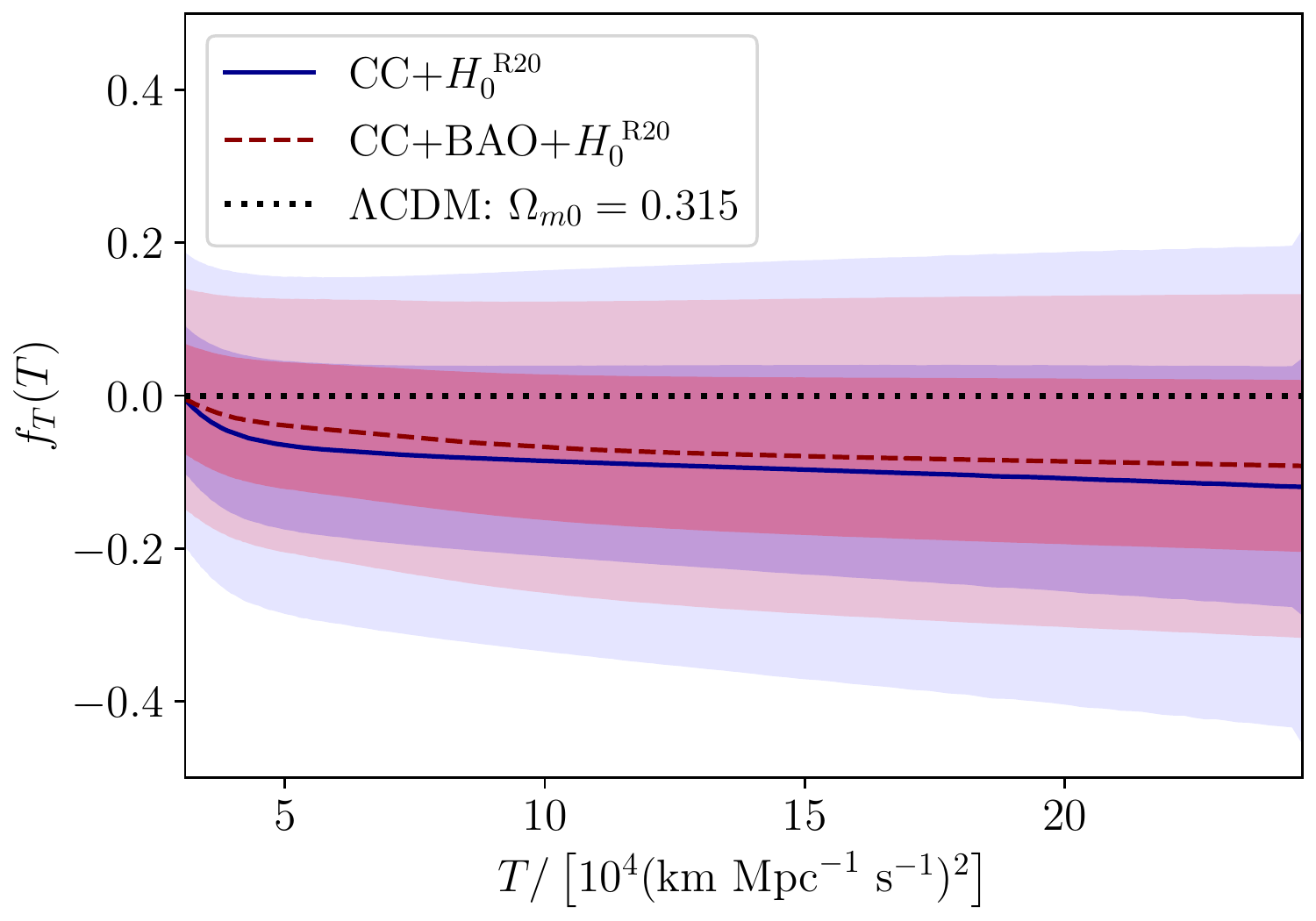} 		
	\end{center}
	\caption{{\small ANN reconstructions of $f_T(T)$ vs $T$ with different $H_0$ priors using $\Omega_{m0} = 0.315 \pm 0.007$\cite{Aghanim:2018eyx} for respective data set combinations.}}
	\label{fT-plot2}
\end{figure}

The $f(T)$ and $f_T(T)$ reconstructions against the torsion scalar $T$ corresponding to 6 sets of Hubble data combinations, namely $-$ CC, CC+$H_0^\text{TRGB}$,  CC+$H_0^\text{R20}$, CC+BAO, CC+BAO+$H_0^\text{TRGB}$ and CC+BAO+$H_0^\text{R20}$ respectively, are illustrated in Figs.~\ref{f-plot2} and \ref{fT-plot2}. It deserves mention that in the $\Lambda$CDM paradigm, $f(T) \rightarrow 6H_0^2(\Omega_m^0-1)$ and $f_T(T) \rightarrow 0$, denoted by the respective dotted horizontal lines in Fig.~\ref{f-plot2} and \ref{fT-plot2}. The mean reconstructed $f(T)$ curves are slightly decreasing functions of $T$, and the reconstructed $f_T(T)$ curves have slightly negative values. We observe that the reconstruction with the CC and CC+BAO Hubble data set only, i.e., the cases where no prior is set on the value of $H_0$, have the least deviation from $\Lambda$CDM. When introducing the R20 priors, i.e., the CC+$H_0^{\text{R20}}$ and CC+BAO+$H_0^{\text{R20}}$ combinations, this deviation from $\Lambda$CDM is highest or maximum. Moreover, we can clearly see that the joint CC+BAO data set led to tighter constraints with respect to the CC data, which are further improved by introducing the $H_0$ priors for the analysis. Nevertheless, we find that the $\Lambda$CDM scenario lies well included within the 1$\sigma$ confidence level for all the reconstructions.

\section{Discussion \label{conclusion}}

The use of learning techniques in tandem with recent observational data to reconstruct dark energy and its potential theoretical foundations has been a growing theme of research in the last few years. The topic has also led to new null tests of $\Lambda$CDM and other tenants of standard cosmology. To a large extent, these approaches have relied on GP to reconstruct various elements of arbitrary elements of these new theories. However, GP suffers from various issues such as overfitting at low redshifts and the kernel selection issue.

In this work, we have shown the reconstruction of the Hubble parameter derivative $H'(z)$ can be constructed using a combination of ANNs and the MC approach. This allows us to propose a new approach by which to perform the reconstructions of dark energy replacing GP with ANNs. This gives a better way to build observationally-driven models of gravity that can compete with the concordance models in the cosmological context. By this, we mean that this approach could conceivably be implemented for other general classes of models where the arbitrary functional $f(T)$ could be exchanged with scalar-tensor models, or other functional forms such as $f(\lc{R})$ \cite{Sotiriou:2008rp,Faraoni:2008mf,Capozziello:2011et} or $f(Q)$ \cite{BeltranJimenez:2017tkd,Harko:2018gxr,Gakis:2019rdd,Soudi:2018dhv}. As explained in detail in Sec.~\ref{sec:ANN_intro}, ANNs offer a natural way to build a system that learns how the data is behaving and to mimic that data for intermediary redshift points. this gives a powerful base on which to perform calculations using Hubble data. Here, we also describe the data that is used throughout the work and the priors on the Hubble constant that we take from the literature.

The reconstruction of the Hubble diagram using the various combinations of Hubble data and priors is explained in Sec.~\ref{H-recon}. Here, we explain how the learning process helps inform the best structure of the ANN architecture by optimizing the number of neurons and layers. In our reconstructed Hubble diagrams, the mean Hubble parameter is in agreement with other approach to reconstruction such as GP but also those others mentioned in Sec.~\ref{sec:intro}. On the other hand, the associated uncertainties are larger than these other reconstructions. This is an indication that the overfitting problem that mainly arises in GP, but also other reconstruction methods, is vastly diminished here. As an alternative that combines the power of the Monte Carlo approach together with the model-independence of ANN architectures, we show how these error bars can be reduced, in some redshift ranges, without adding further statistical assumptions such as the kernel in GP \cite{10.5555/1162254}. To achieve this, we apply the MC routine with 1000 realizations from which we determine the uncertainties at every reconstructed redshift. This combined approach is then applied to the problem of reconstructing the $H'(z)$ parameter. In this way, we not only obtain mean values for this derivative term, but also realistic values for the associated uncertainties at each of redshift points.

It is not enough to build the reconstructions of the Hubble diagram and its derivatives, we also perform diagnostic tests on the results to assess their behaviour against the concordance model. This is done in Sec.~\ref{null-test} where we principally build on the test outlined about Eq.~\eqref{eq:null_test} which is related to the matter density parameter for the $\Lambda$CDM model. We find that the mean diagnostic curve is largely consistent with $\Lambda$CDM fro low redshifts but then starts to veer away at the higher redshift range of the reconstructed data interval. It is important to highlight that the uncertainties of this diagnostic also increase in this regime making it difficult to make robust conclusions from this result. However, these potential divergences from standard cosmology show an interesting preference in the evolution of the Hubble diagram.

One of the aims of proposing this new approach to reconstructing the Hubble derivative $H'(z)$ is to be more competitive with the applications of GP. One main application is the reconstruction of dark energy within modified theories of gravity. In this work, we show using $f(T)$ cosmology how this can be done. We review the theory in Sec.~\ref{sec:f_T_intro}, where the foundations of TG and its connection to $f(T)$ gravity are briefly explained. We then use this base to explain our reconstruction approach in Sec.~\ref{fT-recon}. Here, we use a central differencing method in Eq.~\eqref{prop_eq_f_T} to propagate the arbitrary functional assuming only that $\Lambda$CDM dominates in the late Universe. This is possible since $f(T)$ gravity is a second order gravitational theory. It may be possible to extend this approach to higher derivative theories, but this may be limited due to possible drastic increases in the associated uncertainties of the reconstructed Hubble parameter. In the reconstructions of the arbitrary $f(T)$ Lagrangian functional, we find a similar behaviour with low redshift regions pointing to a cosmology similar to $\Lambda$CDM which then starts to diverge at higher redshifts. This is an initial result since the associated uncertainties are quite large in these intervals. However, it is interesting to understand how this first reconstruction of a modified gravity model performs through an ANN architecture pipeline. On the other hand, for the reconstruction of $f_{T}(T)$, the uncertainties remain largely flat across the redshift interval. However, the mean values do show an immediate preference to an evolution that deviates from $\Lambda$CDM to a non-negligible extent.

The approach which ANNs rely on is altogether different from that of GPs where the over-fitting and kernel selection issue has been largely replaced by the large size of the neuron system which learns how to mimic the observational data in a more natural way. In this setting, the much larger number of hyperparameters helps optimize better how the system approaches the data being used to learn. The resulting trained ANN system can be competitive with GP in terms of the breadth of applications. In this work, we show how ANNs can be used to directly build modified gravity models built on observational data. It would be interesting to apply this approach to other models of dark energy and modified cosmology models

\appendix
\section{Appendix}

\label{app:mock_data}

\begin{figure*}[htb!]
		\begin{center}
			\includegraphics[angle=0, width=0.49\textwidth]{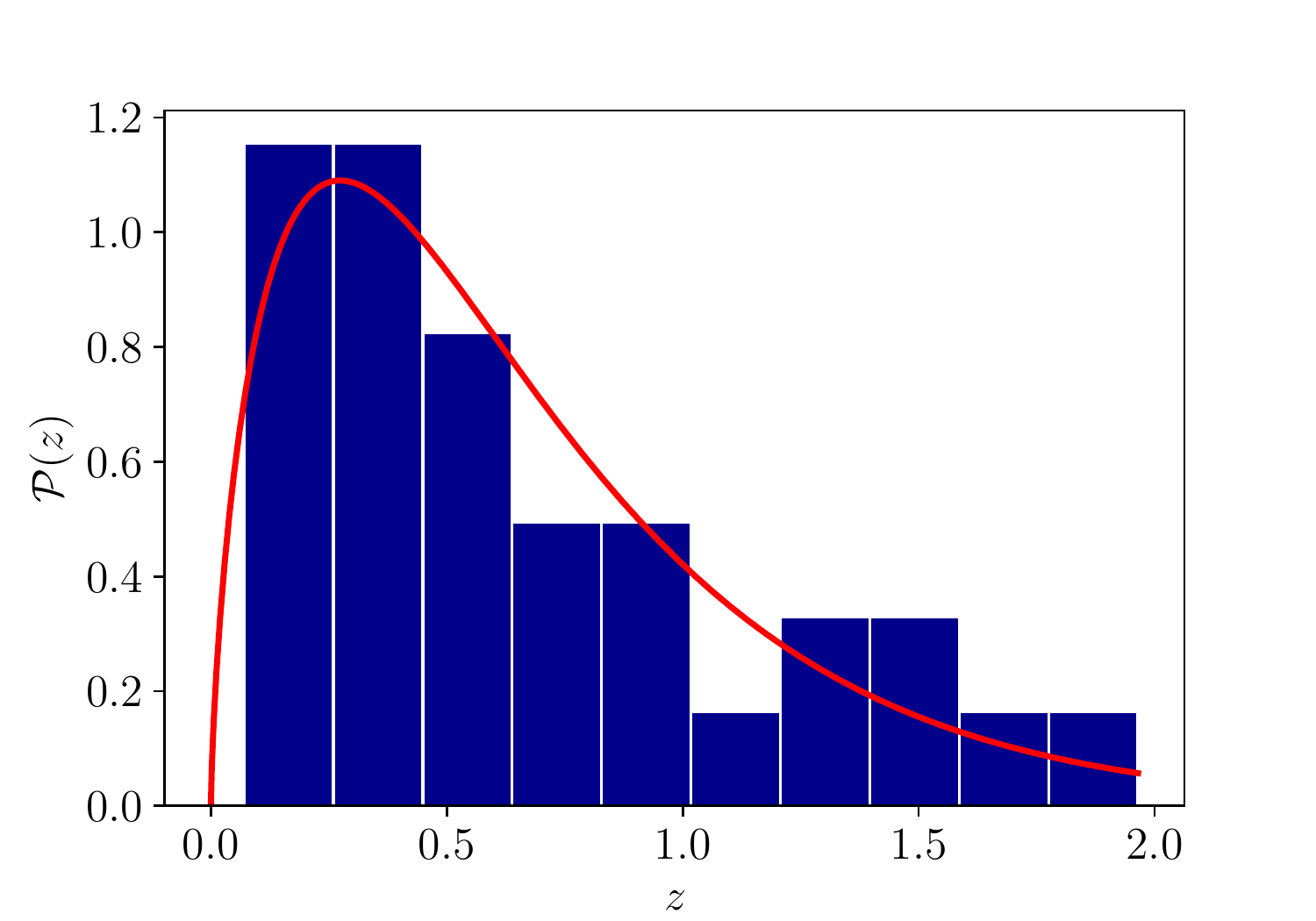}
			\includegraphics[angle=0, width=0.49\textwidth]{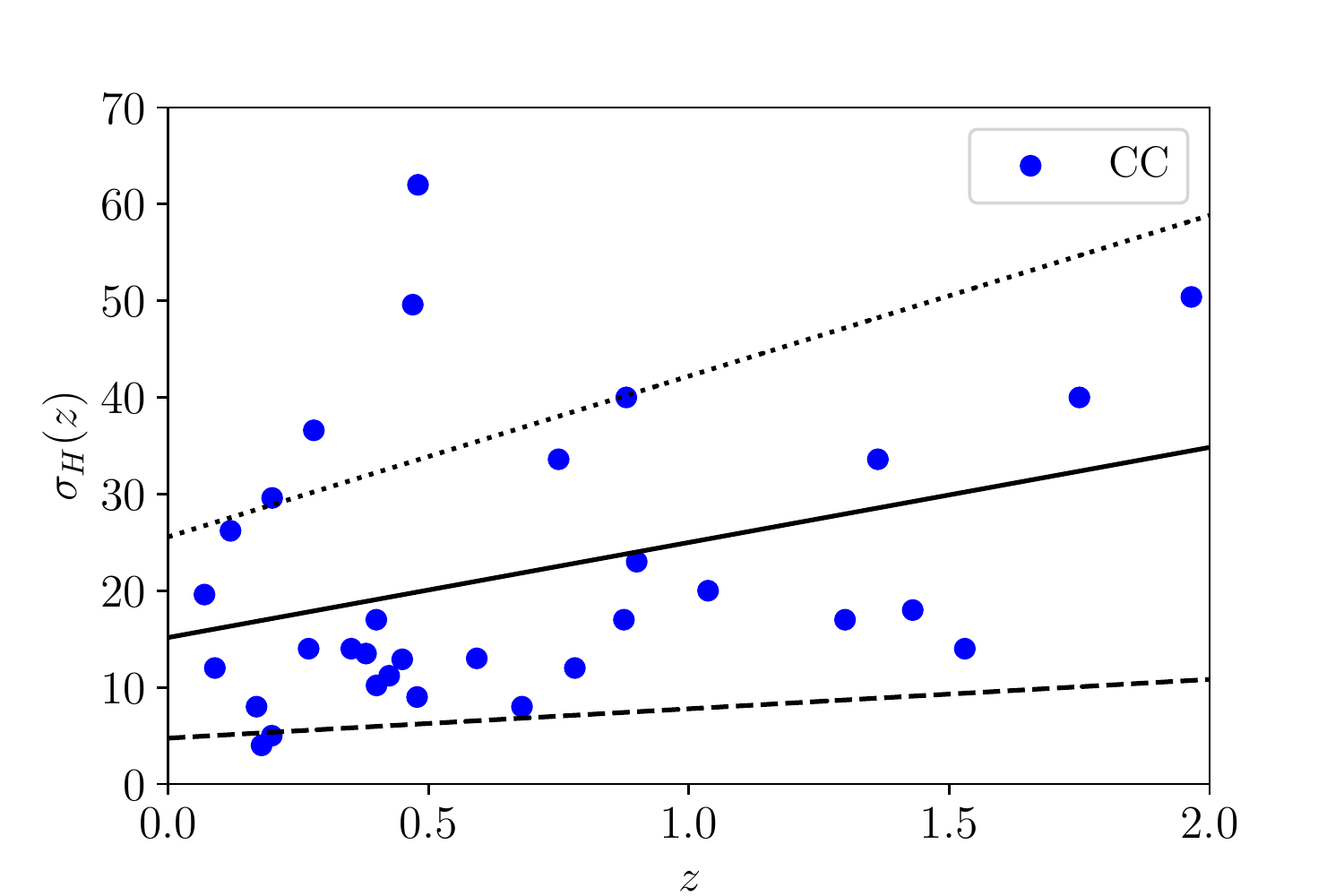} 		
		\end{center}
		\caption{{\small Plot showing the redshift distribution of the observational CC $H(z)$ dataset (left) and the errors associated to the observational CC $H(z)$ measurements (right).}}
		\label{z-plot}
	\end{figure*}

\begin{figure*}[htb!]
	\begin{center}
		\includegraphics[angle=0, width=0.49\textwidth]{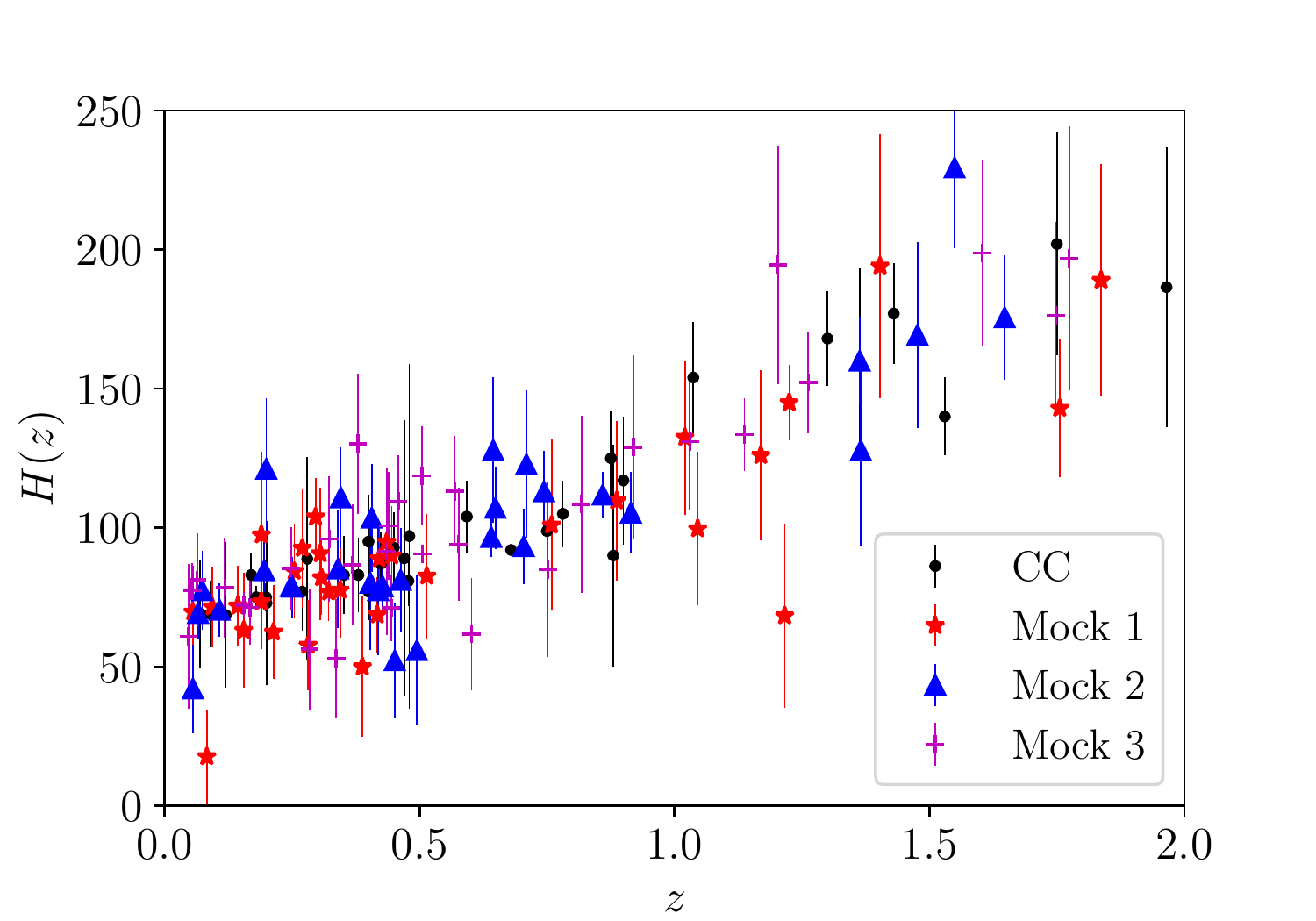}
		\includegraphics[angle=0, width=0.49\textwidth]{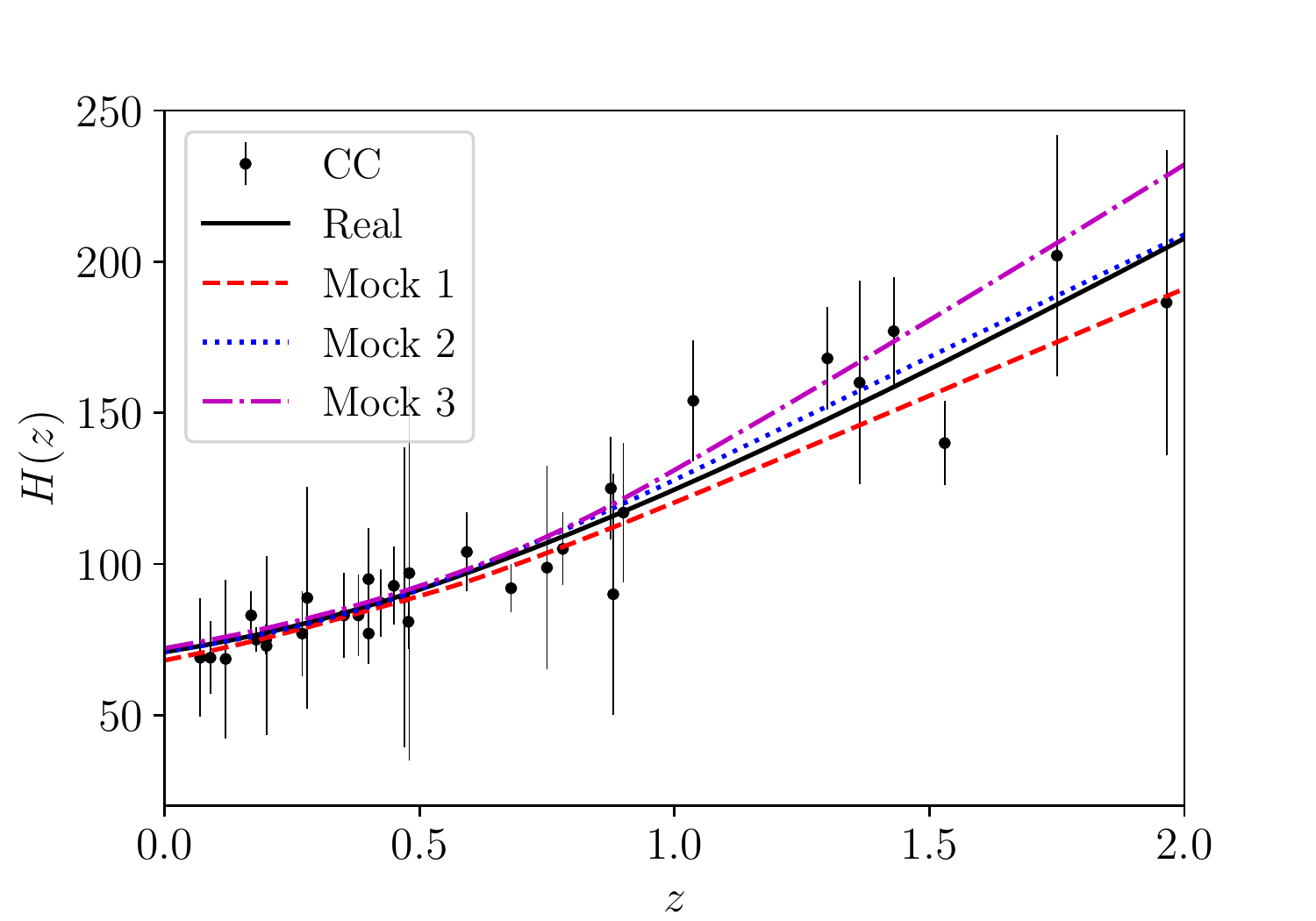}
	\end{center}
	\caption{{\small Plot showing the three sets of mock data-like $H(z)$ samples in comparison to the CC $H(z)$ data set (left). Plot for the trained ANNs using the corresponding mock samples of $H(z)$ in comparison to a trained ANN on the real data.}}
	\label{mock-plot}
\end{figure*}

Here, we show the distribution of the data point themselves together with the associated uncertainties in Fig.~\ref{z-plot}. We also show the mock data that is based on the real data in Fig.~\ref{mock-plot}. Again, we emphasize that the final ANN is only structured through this mock data and not actually trained on it. Thus, this is a vehicle to construct the ANN, that is, to select the optimal number of neurons and layers, and nothing more.

	\begin{figure*}[htb!]
	\begin{center}
		\includegraphics[angle=0, width=0.49\textwidth]{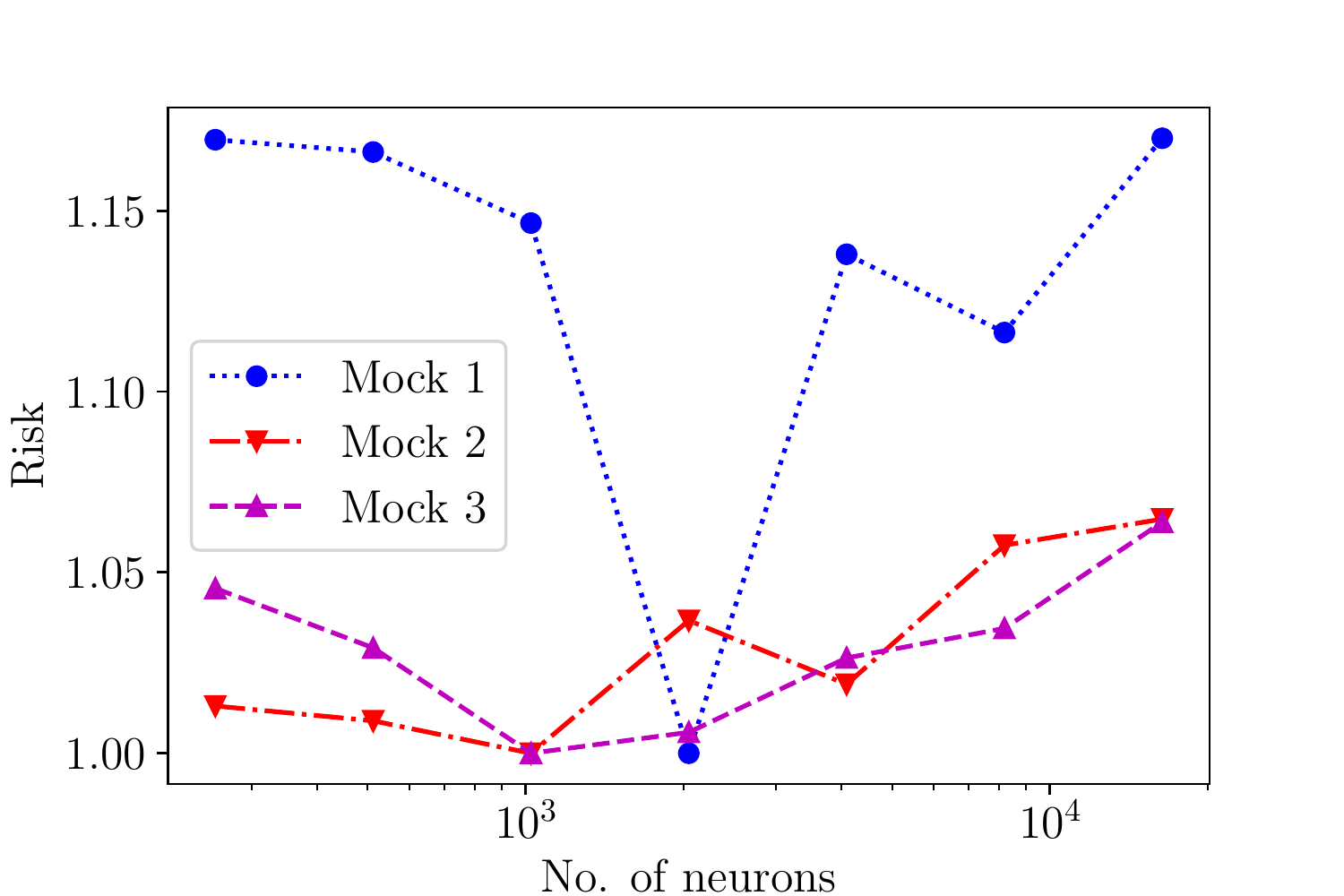}
	\end{center}
	\caption{{\small Plot showing the normalised risk for three mock $H(z)$ ANN models that have one hidden layer and a corresponding number of neurons in their hidden layer.}}
	\label{risk-plot}
\end{figure*}

\acknowledgments

JLS and JM would like to acknowledge support from the Malta Digital Innovation Authority through the IntelliVerse grant. This research has been partly carried out using computational facilities procured through the European Regional Development Fund, Project No. ERDF-080 ``A supercomputing laboratory for the University of Malta''. JLS and JM would also like to acknowledge funding from ``The Malta Council for Science and Technology'' in project IPAS-2020-007. PM would like to acknowledge the use of ``Dirac Supercomputing Facility'' of IISER Kolkata. This article is based upon work from COST Action CA21136 Addressing observational tensions in cosmology with systematics and fundamental physics (CosmoVerse) supported by COST (European Cooperation in Science and Technology).

\bibliographystyle{JHEP}
\bibliography{references}

\providecommand{\href}[2]{#2}\begingroup\raggedright\begin{thebibliography}{100}

\bibitem{Peebles:2002gy}
P.J.E.~Peebles and B.~Ratra, \emph{{The Cosmological Constant and Dark
  Energy}}, \href{https://doi.org/10.1103/RevModPhys.75.559}{\emph{Rev. Mod.
  Phys.} {\bfseries 75} (2003) 559}
  [\href{https://arxiv.org/abs/astro-ph/0207347}{{\ttfamily
  astro-ph/0207347}}].

\bibitem{Copeland:2006wr}
E.J.~Copeland, M.~Sami and S.~Tsujikawa, \emph{{Dynamics of dark energy}},
  \href{https://doi.org/10.1142/S021827180600942X}{\emph{Int. J. Mod. Phys. D}
  {\bfseries 15} (2006) 1753}
  [\href{https://arxiv.org/abs/hep-th/0603057}{{\ttfamily hep-th/0603057}}].

\bibitem{Baudis:2016qwx}
L.~Baudis, \emph{{Dark matter detection}},
  \href{https://doi.org/10.1088/0954-3899/43/4/044001}{\emph{J. Phys. G}
  {\bfseries 43} (2016) 044001}.

\bibitem{Bertone:2004pz}
G.~Bertone, D.~Hooper and J.~Silk, \emph{{Particle dark matter: Evidence,
  candidates and constraints}},
  \href{https://doi.org/10.1016/j.physrep.2004.08.031}{\emph{Phys. Rept.}
  {\bfseries 405} (2005) 279}
  [\href{https://arxiv.org/abs/hep-ph/0404175}{{\ttfamily hep-ph/0404175}}].

\bibitem{Riess:1998cb}
{\scshape Supernova Search Team} collaboration, \emph{{Observational evidence
  from supernovae for an accelerating universe and a cosmological constant}},
  \href{https://doi.org/10.1086/300499}{\emph{Astron. J.} {\bfseries 116}
  (1998) 1009} [\href{https://arxiv.org/abs/astro-ph/9805201}{{\ttfamily
  astro-ph/9805201}}].

\bibitem{Perlmutter:1998np}
{\scshape Supernova Cosmology Project} collaboration, \emph{{Measurements of
  $\Omega$ and $\Lambda$ from 42 high redshift supernovae}},
  \href{https://doi.org/10.1086/307221}{\emph{Astrophys. J.} {\bfseries 517}
  (1999) 565} [\href{https://arxiv.org/abs/astro-ph/9812133}{{\ttfamily
  astro-ph/9812133}}].

\bibitem{Guth:1980zm}
A.H.~Guth, \emph{{The Inflationary Universe: A Possible Solution to the Horizon
  and Flatness Problems}},
  \href{https://doi.org/10.1103/PhysRevD.23.347}{\emph{Phys. Rev. D} {\bfseries
  23} (1981) 347}.

\bibitem{Linde:1981mu}
A.D.~Linde, \emph{{A New Inflationary Universe Scenario: A Possible Solution of
  the Horizon, Flatness, Homogeneity, Isotropy and Primordial Monopole
  Problems}}, \href{https://doi.org/10.1016/0370-2693(82)91219-9}{\emph{Phys.
  Lett. B} {\bfseries 108} (1982) 389}.

\bibitem{Weinberg:1988cp}
S.~Weinberg, \emph{{The Cosmological Constant Problem}},
  \href{https://doi.org/10.1103/RevModPhys.61.1}{\emph{Rev. Mod. Phys.}
  {\bfseries 61} (1989) 1}.

\bibitem{Addazi:2021xuf}
A.~Addazi et~al., \emph{{Quantum gravity phenomenology at the dawn of the
  multi-messenger era\textemdash{}A review}},
  \href{https://doi.org/10.1016/j.ppnp.2022.103948}{\emph{Prog. Part. Nucl.
  Phys.} {\bfseries 125} (2022) 103948}
  [\href{https://arxiv.org/abs/2111.05659}{{\ttfamily 2111.05659}}].

\bibitem{CANTATA:2021ktz}
{\scshape CANTATA} collaboration, \emph{{Modified Gravity and Cosmology: An
  Update by the CANTATA Network}},
  \href{https://arxiv.org/abs/2105.12582}{{\ttfamily 2105.12582}}.

\bibitem{LUX:2016ggv}
{\scshape LUX} collaboration, \emph{{Results from a search for dark matter in
  the complete LUX exposure}},
  \href{https://doi.org/10.1103/PhysRevLett.118.021303}{\emph{Phys. Rev. Lett.}
  {\bfseries 118} (2017) 021303}
  [\href{https://arxiv.org/abs/1608.07648}{{\ttfamily 1608.07648}}].

\bibitem{Gaitskell:2004gd}
R.J.~Gaitskell, \emph{{Direct detection of dark matter}},
  \href{https://doi.org/10.1146/annurev.nucl.54.070103.181244}{\emph{Ann. Rev.
  Nucl. Part. Sci.} {\bfseries 54} (2004) 315}.

\bibitem{DiValentino:2020vhf}
E.~Di~Valentino et~al., \emph{{Snowmass2021 - Letter of interest cosmology
  intertwined I: Perspectives for the next decade}},
  \href{https://doi.org/10.1016/j.astropartphys.2021.102606}{\emph{Astropart.
  Phys.} {\bfseries 131} (2021) 102606}
  [\href{https://arxiv.org/abs/2008.11283}{{\ttfamily 2008.11283}}].

\bibitem{DiValentino:2020zio}
E.~Di~Valentino et~al., \emph{{Snowmass2021 - Letter of interest cosmology
  intertwined II: The hubble constant tension}},
  \href{https://doi.org/10.1016/j.astropartphys.2021.102605}{\emph{Astropart.
  Phys.} {\bfseries 131} (2021) 102605}
  [\href{https://arxiv.org/abs/2008.11284}{{\ttfamily 2008.11284}}].

\bibitem{DiValentino:2020vvd}
E.~Di~Valentino et~al., \emph{{Cosmology intertwined III: $f\sigma_8$ and
  $S_8$}},
  \href{https://doi.org/10.1016/j.astropartphys.2021.102604}{\emph{Astropart.
  Phys.} {\bfseries 131} (2021) 102604}
  [\href{https://arxiv.org/abs/2008.11285}{{\ttfamily 2008.11285}}].

\bibitem{Staicova:2021ajb}
D.~Staicova, \emph{{Hints of the $H_0-r_d$ tension in uncorrelated Baryon
  Acoustic Oscillations dataset}},  in \emph{{16th Marcel Grossmann Meeting
  on~Recent Developments in Theoretical and Experimental General Relativity,
  Astrophysics and Relativistic Field Theories}}, 11, 2021
  [\href{https://arxiv.org/abs/2111.07907}{{\ttfamily 2111.07907}}].

\bibitem{DiValentino:2021izs}
E.~Di~Valentino, O.~Mena, S.~Pan, L.~Visinelli, W.~Yang, A.~Melchiorri et~al.,
  \emph{{In the realm of the Hubble tension\textemdash{}a review of
  solutions}}, \href{https://doi.org/10.1088/1361-6382/ac086d}{\emph{Class.
  Quant. Grav.} {\bfseries 38} (2021) 153001}
  [\href{https://arxiv.org/abs/2103.01183}{{\ttfamily 2103.01183}}].

\bibitem{Perivolaropoulos:2021jda}
L.~Perivolaropoulos and F.~Skara, \emph{{Challenges for
  \ensuremath{\Lambda}CDM: An update}},
  \href{https://doi.org/10.1016/j.newar.2022.101659}{\emph{New Astron. Rev.}
  {\bfseries 95} (2022) 101659}
  [\href{https://arxiv.org/abs/2105.05208}{{\ttfamily 2105.05208}}].

\bibitem{DiValentino:2022oon}
E.~Di~Valentino, W.~Giar\`e, A.~Melchiorri and J.~Silk, \emph{{Health checkup
  test of the standard cosmological model in view of recent cosmic microwave
  background anisotropies experiments}},
  \href{https://doi.org/10.1103/PhysRevD.106.103506}{\emph{Phys. Rev. D}
  {\bfseries 106} (2022) 103506}
  [\href{https://arxiv.org/abs/2209.12872}{{\ttfamily 2209.12872}}].

\bibitem{Aghanim:2018eyx}
{\scshape Planck} collaboration, \emph{{Planck 2018 results. VI. Cosmological
  parameters}},
  \href{https://doi.org/10.1051/0004-6361/201833910}{\emph{Astron. Astrophys.}
  {\bfseries 641} (2020) A6}
  [\href{https://arxiv.org/abs/1807.06209}{{\ttfamily 1807.06209}}].

\bibitem{ACT:2020gnv}
{\scshape ACT} collaboration, \emph{{The Atacama Cosmology Telescope: DR4 Maps
  and Cosmological Parameters}},
  \href{https://doi.org/10.1088/1475-7516/2020/12/047}{\emph{JCAP} {\bfseries
  12} (2020) 047} [\href{https://arxiv.org/abs/2007.07288}{{\ttfamily
  2007.07288}}].

\bibitem{Riess:2020fzl}
A.G.~Riess, S.~Casertano, W.~Yuan, J.B.~Bowers, L.~Macri, J.C.~Zinn et~al.,
  \emph{{Cosmic Distances Calibrated to 1\% Precision with Gaia EDR3 Parallaxes
  and Hubble Space Telescope Photometry of 75 Milky Way Cepheids Confirm
  Tension with $\Lambda$CDM}},
  \href{https://doi.org/10.3847/2041-8213/abdbaf}{\emph{Astrophys. J. Lett.}
  {\bfseries 908} (2021) L6}
  [\href{https://arxiv.org/abs/2012.08534}{{\ttfamily 2012.08534}}].

\bibitem{Wong:2019kwg}
K.C.~Wong et~al., \emph{{H0LiCOW \textendash{} XIII. A 2.4 per cent measurement
  of H0 from lensed quasars: 5.3\ensuremath{\sigma} tension between early- and
  late-Universe probes}},
  \href{https://doi.org/10.1093/mnras/stz3094}{\emph{Mon. Not. Roy. Astron.
  Soc.} {\bfseries 498} (2020) 1420}
  [\href{https://arxiv.org/abs/1907.04869}{{\ttfamily 1907.04869}}].

\bibitem{Freedman:2020dne}
W.L.~Freedman, B.F.~Madore, T.~Hoyt, I.S.~Jang, R.~Beaton, M.G.~Lee et~al.,
  \emph{{Calibration of the Tip of the Red Giant Branch (TRGB)}},
  \href{https://arxiv.org/abs/2002.01550}{{\ttfamily 2002.01550}}.

\bibitem{Abbott:2017xzu}
{\scshape LIGO Scientific, Virgo, 1M2H, Dark Energy Camera GW-E, DES, DLT40,
  Las Cumbres Observatory, VINROUGE, MASTER} collaboration, \emph{{A
  gravitational-wave standard siren measurement of the Hubble constant}},
  \href{https://doi.org/10.1038/nature24471}{\emph{Nature} {\bfseries 551}
  (2017) 85} [\href{https://arxiv.org/abs/1710.05835}{{\ttfamily 1710.05835}}].

\bibitem{Bargiacchi:2021fow}
G.~Bargiacchi, G.~Risaliti, M.~Benetti, S.~Capozziello, E.~Lusso, A.~Saccardi
  et~al., \emph{{Cosmography by orthogonalized logarithmic polynomials}},
  \href{https://doi.org/10.1051/0004-6361/202140386}{\emph{Astron. Astrophys.}
  {\bfseries 649} (2021) A65}
  [\href{https://arxiv.org/abs/2101.08278}{{\ttfamily 2101.08278}}].

\bibitem{Capozziello:2019cav}
S.~Capozziello, R.~D'Agostino and O.~Luongo, \emph{{Extended Gravity
  Cosmography}}, \href{https://doi.org/10.1142/S0218271819300167}{\emph{Int. J.
  Mod. Phys. D} {\bfseries 28} (2019) 1930016}
  [\href{https://arxiv.org/abs/1904.01427}{{\ttfamily 1904.01427}}].

\bibitem{Bamba:2012cp}
K.~Bamba, S.~Capozziello, S.~Nojiri and S.D.~Odintsov, \emph{{Dark energy
  cosmology: the equivalent description via different theoretical models and
  cosmography tests}},
  \href{https://doi.org/10.1007/s10509-012-1181-8}{\emph{Astrophys. Space Sci.}
  {\bfseries 342} (2012) 155}
  [\href{https://arxiv.org/abs/1205.3421}{{\ttfamily 1205.3421}}].

\bibitem{Cai:2019bdh}
Y.-F.~Cai, M.~Khurshudyan and E.N.~Saridakis, \emph{{Model-independent
  reconstruction of $f(T)$ gravity from Gaussian Processes}},
  \href{https://doi.org/10.3847/1538-4357/ab5a7f}{\emph{Astrophys. J.}
  {\bfseries 888} (2020) 62}
  [\href{https://arxiv.org/abs/1907.10813}{{\ttfamily 1907.10813}}].

\bibitem{Ren:2022aeo}
X.~Ren, S.-F.~Yan, Y.~Zhao, Y.-F.~Cai and E.N.~Saridakis, \emph{{Gaussian
  processes and effective field theory of $f(T)$ gravity under the $H_0$
  tension}}, \href{https://doi.org/10.3847/1538-4357/ac6ba5}{\emph{Astrophys.
  J.} {\bfseries 932} (2022) 131}
  [\href{https://arxiv.org/abs/2203.01926}{{\ttfamily 2203.01926}}].

\bibitem{Bernardo:2021qhu}
R.C.~Bernardo and J.~Levi~Said, \emph{{A data-driven reconstruction of
  Horndeski gravity via the Gaussian processes}},
  \href{https://doi.org/10.1088/1475-7516/2021/09/014}{\emph{JCAP} {\bfseries
  09} (2021) 014} [\href{https://arxiv.org/abs/2105.12970}{{\ttfamily
  2105.12970}}].

\bibitem{Briffa:2020qli}
R.~Briffa, S.~Capozziello, J.~Levi~Said, J.~Mifsud and E.N.~Saridakis,
  \emph{{Constraining teleparallel gravity through Gaussian processes}},
  \href{https://doi.org/10.1088/1361-6382/abd4f5}{\emph{Class. Quant. Grav.}
  {\bfseries 38} (2020) 055007}
  [\href{https://arxiv.org/abs/2009.14582}{{\ttfamily 2009.14582}}].

\bibitem{LeviSaid:2021yat}
J.~Levi~Said, J.~Mifsud, J.~Sultana and K.Z.~Adami, \emph{{Reconstructing
  teleparallel gravity with cosmic structure growth and expansion rate data}},
  \href{https://doi.org/10.1088/1475-7516/2021/06/015}{\emph{JCAP} {\bfseries
  06} (2021) 015} [\href{https://arxiv.org/abs/2103.05021}{{\ttfamily
  2103.05021}}].

\bibitem{Aldrovandi:2013wha}
R.~Aldrovandi and J.G.~Pereira, \emph{{Teleparallel Gravity}: {An
  Introduction}}, Springer (2013),
  \href{https://doi.org/10.1007/978-94-007-5143-9}{10.1007/978-94-007-5143-9}.

\bibitem{Bahamonde:2021gfp}
S.~Bahamonde, K.F.~Dialektopoulos, C.~Escamilla-Rivera, G.~Farrugia, V.~Gakis,
  M.~Hendry et~al., \emph{{Teleparallel Gravity: From Theory to Cosmology}},
  \href{https://arxiv.org/abs/2106.13793}{{\ttfamily 2106.13793}}.

\bibitem{Krssak:2018ywd}
M.~Krssak, R.J.~van~den Hoogen, J.G.~Pereira, C.G.~B\"ohmer and A.A.~Coley,
  \emph{{Teleparallel theories of gravity: illuminating a fully invariant
  approach}}, \href{https://doi.org/10.1088/1361-6382/ab2e1f}{\emph{Class.
  Quant. Grav.} {\bfseries 36} (2019) 183001}
  [\href{https://arxiv.org/abs/1810.12932}{{\ttfamily 1810.12932}}].

\bibitem{Cai:2015emx}
Y.-F.~Cai, S.~Capozziello, M.~De~Laurentis and E.N.~Saridakis, \emph{{f(T)
  teleparallel gravity and cosmology}},
  \href{https://doi.org/10.1088/0034-4885/79/10/106901}{\emph{Rept. Prog.
  Phys.} {\bfseries 79} (2016) 106901}
  [\href{https://arxiv.org/abs/1511.07586}{{\ttfamily 1511.07586}}].

\bibitem{Sotiriou:2008rp}
T.P.~Sotiriou and V.~Faraoni, \emph{{$f(R)$ Theories Of Gravity}},
  \href{https://doi.org/10.1103/RevModPhys.82.451}{\emph{Rev. Mod. Phys.}
  {\bfseries 82} (2010) 451} [\href{https://arxiv.org/abs/0805.1726}{{\ttfamily
  0805.1726}}].

\bibitem{Faraoni:2008mf}
V.~Faraoni, \emph{{$f(R)$ gravity: Successes and challenges}},  in \emph{{18th
  SIGRAV Conference}}, 10, 2008
  [\href{https://arxiv.org/abs/0810.2602}{{\ttfamily 0810.2602}}].

\bibitem{Capozziello:2011et}
S.~Capozziello and M.~De~Laurentis, \emph{{Extended Theories of Gravity}},
  \href{https://doi.org/10.1016/j.physrep.2011.09.003}{\emph{Phys. Rept.}
  {\bfseries 509} (2011) 167}
  [\href{https://arxiv.org/abs/1108.6266}{{\ttfamily 1108.6266}}].

\bibitem{Ferraro:2006jd}
R.~Ferraro and F.~Fiorini, \emph{{Modified teleparallel gravity: Inflation
  without inflaton}},
  \href{https://doi.org/10.1103/PhysRevD.75.084031}{\emph{Phys. Rev. D}
  {\bfseries 75} (2007) 084031}
  [\href{https://arxiv.org/abs/gr-qc/0610067}{{\ttfamily gr-qc/0610067}}].

\bibitem{Ferraro:2008ey}
R.~Ferraro and F.~Fiorini, \emph{{On Born-Infeld Gravity in Weitzenbock
  spacetime}}, \href{https://doi.org/10.1103/PhysRevD.78.124019}{\emph{Phys.
  Rev. D} {\bfseries 78} (2008) 124019}
  [\href{https://arxiv.org/abs/0812.1981}{{\ttfamily 0812.1981}}].

\bibitem{Bengochea:2008gz}
G.R.~Bengochea and R.~Ferraro, \emph{{Dark torsion as the cosmic speed-up}},
  \href{https://doi.org/10.1103/PhysRevD.79.124019}{\emph{Phys. Rev. D}
  {\bfseries 79} (2009) 124019}
  [\href{https://arxiv.org/abs/0812.1205}{{\ttfamily 0812.1205}}].

\bibitem{Linder:2010py}
E.V.~Linder, \emph{{Einstein's Other Gravity and the Acceleration of the
  Universe}}, \href{https://doi.org/10.1103/PhysRevD.81.127301}{\emph{Phys.
  Rev. D} {\bfseries 81} (2010) 127301}
  [\href{https://arxiv.org/abs/1005.3039}{{\ttfamily 1005.3039}}].

\bibitem{Chen:2010va}
S.-H.~Chen, J.B.~Dent, S.~Dutta and E.N.~Saridakis, \emph{{Cosmological
  perturbations in $f(T)$ gravity}},
  \href{https://doi.org/10.1103/PhysRevD.83.023508}{\emph{Phys. Rev. D}
  {\bfseries 83} (2011) 023508}
  [\href{https://arxiv.org/abs/1008.1250}{{\ttfamily 1008.1250}}].

\bibitem{Bahamonde:2019zea}
S.~Bahamonde, K.~Flathmann and C.~Pfeifer, \emph{{Photon sphere and perihelion
  shift in weak $f(T)$ gravity}},
  \href{https://doi.org/10.1103/PhysRevD.100.084064}{\emph{Phys. Rev. D}
  {\bfseries 100} (2019) 084064}
  [\href{https://arxiv.org/abs/1907.10858}{{\ttfamily 1907.10858}}].

\bibitem{10.5555/1162254}
C.E.~Rasmussen and C.K.I.~Williams, \emph{Gaussian Processes for Machine
  Learning (Adaptive Computation and Machine Learning)}, The MIT Press (2005).

\bibitem{Busti:2014aoa}
V.C.~Busti, C.~Clarkson and M.~Seikel, \emph{{The Value of $H_0$ from Gaussian
  Processes}}, \href{https://doi.org/10.1017/S1743921314013751}{\emph{IAU
  Symp.} {\bfseries 306} (2014) 25}
  [\href{https://arxiv.org/abs/1407.5227}{{\ttfamily 1407.5227}}].

\bibitem{Busti:2014dua}
V.C.~Busti, C.~Clarkson and M.~Seikel, \emph{{Evidence for a Lower Value for
  $H_0$ from Cosmic Chronometers Data?}},
  \href{https://doi.org/10.1093/mnrasl/slu035}{\emph{Mon. Not. Roy. Astron.
  Soc.} {\bfseries 441} (2014) 11}
  [\href{https://arxiv.org/abs/1402.5429}{{\ttfamily 1402.5429}}].

\bibitem{Seikel:2013fda}
M.~Seikel and C.~Clarkson, \emph{{Optimising Gaussian processes for
  reconstructing dark energy dynamics from supernovae}},
  \href{https://arxiv.org/abs/1311.6678}{{\ttfamily 1311.6678}}.

\bibitem{Bernardo:2021mfs}
R.C.~Bernardo and J.~Levi~Said, \emph{{Towards a model-independent
  reconstruction approach for late-time Hubble data}},
  \href{https://doi.org/10.1088/1475-7516/2021/08/027}{\emph{JCAP} {\bfseries
  08} (2021) 027} [\href{https://arxiv.org/abs/2106.08688}{{\ttfamily
  2106.08688}}].

\bibitem{Yahya:2013xma}
S.~Yahya, M.~Seikel, C.~Clarkson, R.~Maartens and M.~Smith, \emph{{Null tests
  of the cosmological constant using supernovae}},
  \href{https://doi.org/10.1103/PhysRevD.89.023503}{\emph{Phys. Rev. D}
  {\bfseries 89} (2014) 023503}
  [\href{https://arxiv.org/abs/1308.4099}{{\ttfamily 1308.4099}}].

\bibitem{2012JCAP...06..036S}
M.~{Seikel}, C.~{Clarkson} and M.~{Smith}, \emph{{Reconstruction of dark energy
  and expansion dynamics using Gaussian processes}},
  \href{https://doi.org/10.1088/1475-7516/2012/06/036}{\emph{JCAP} {\bfseries
  2012} (2012) 036} [\href{https://arxiv.org/abs/1204.2832}{{\ttfamily
  1204.2832}}].

\bibitem{Shafieloo:2012ht}
A.~Shafieloo, A.G.~Kim and E.V.~Linder, \emph{{Gaussian Process Cosmography}},
  \href{https://doi.org/10.1103/PhysRevD.85.123530}{\emph{Phys. Rev. D}
  {\bfseries 85} (2012) 123530}
  [\href{https://arxiv.org/abs/1204.2272}{{\ttfamily 1204.2272}}].

\bibitem{Benisty:2020kdt}
D.~Benisty, \emph{{Quantifying the $S_8$ tension with the Redshift Space
  Distortion data set}},
  \href{https://doi.org/10.1016/j.dark.2020.100766}{\emph{Phys. Dark Univ.}
  {\bfseries 31} (2021) 100766}
  [\href{https://arxiv.org/abs/2005.03751}{{\ttfamily 2005.03751}}].

\bibitem{Mukherjee:2021epjc}
P.~Mukherjee and N.~Banerjee, \emph{{Non-parametric reconstruction of the
  cosmological \textit{jerk} parameter}},
  \href{https://doi.org/10.1140/epjc/s10052-021-08830-5}{\emph{Eur. Phys. J. C}
  {\bfseries 81} (2021) 36} [\href{https://arxiv.org/abs/2007.10124}{{\ttfamily
  2007.10124}}].

\bibitem{Mukherjee:2022pdu}
P.~Mukherjee and N.~Banerjee, \emph{{Revisiting a non-parametric reconstruction
  of the deceleration parameter from combined background and the growth rate
  data}}, \href{https://doi.org/10.1016/j.dark.2022.100998}{\emph{Phys. Dark
  Univ.} {\bfseries 36} (2022) 100998}
  [\href{https://arxiv.org/abs/2007.15941}{{\ttfamily 2007.15941}}].

\bibitem{Bernardo:2021cxi}
R.C.~Bernardo, D.~Grand\'on, J.~Levi~Said and V.H.~C\'ardenas,
  \emph{{Parametric and nonparametric methods hint dark energy evolution}},
  \href{https://doi.org/10.1016/j.dark.2022.101017}{\emph{Phys. Dark Univ.}
  {\bfseries 36} (2022) 101017}
  [\href{https://arxiv.org/abs/2111.08289}{{\ttfamily 2111.08289}}].

\bibitem{Montiel:2014fpa}
A.~Montiel, R.~Lazkoz, I.~Sendra, C.~Escamilla-Rivera and V.~Salzano,
  \emph{{Nonparametric reconstruction of the cosmic expansion with local
  regression smoothing and simulation extrapolation}},
  \href{https://doi.org/10.1103/PhysRevD.89.043007}{\emph{Phys. Rev. D}
  {\bfseries 89} (2014) 043007}
  [\href{https://arxiv.org/abs/1401.4188}{{\ttfamily 1401.4188}}].

\bibitem{2011A&A...527A..49I}
E.E.O.~{Ishida} and R.S.~{de Souza}, \emph{{Hubble parameter reconstruction
  from a principal component analysis: minimizing the bias}},
  \href{https://doi.org/10.1051/0004-6361/201015281}{\emph{Astronomy \&
  Astrophysics} {\bfseries 527} (2011) A49}
  [\href{https://arxiv.org/abs/1012.5335}{{\ttfamily 1012.5335}}].

\bibitem{Shafieloo:2005nd}
A.~Shafieloo, U.~Alam, V.~Sahni and A.A.~Starobinsky, \emph{{Smoothing
  Supernova Data to Reconstruct the Expansion History of the Universe and its
  Age}}, \href{https://doi.org/10.1111/j.1365-2966.2005.09911.x}{\emph{Mon.
  Not. Roy. Astron. Soc.} {\bfseries 366} (2006) 1081}
  [\href{https://arxiv.org/abs/astro-ph/0505329}{{\ttfamily
  astro-ph/0505329}}].

\bibitem{Porqueres:2016kfv}
N.~Porqueres, T.A.~En\ss{}lin, M.~Greiner, V.~B\"ohm, S.~Dorn, P.~Ruiz-Lapuente
  et~al., \emph{{Cosmic expansion history from SNe Ia data via information
  field theory -- the charm code}},
  \href{https://doi.org/10.1051/0004-6361/201629527}{\emph{Astron. Astrophys.}
  {\bfseries 599} (2017) A92}
  [\href{https://arxiv.org/abs/1608.04007}{{\ttfamily 1608.04007}}].

\bibitem{Escamilla-Rivera:2021rbe}
C.~Escamilla-Rivera, J.~Levi~Said and J.~Mifsud, \emph{{Performance of
  non-parametric reconstruction techniques in the late-time universe}},
  \href{https://doi.org/10.1088/1475-7516/2021/10/016}{\emph{JCAP} {\bfseries
  10} (2021) 016} [\href{https://arxiv.org/abs/2105.14332}{{\ttfamily
  2105.14332}}].

\bibitem{10.2307/j.ctt4cgbdj}
Željko Ivezić, A.J.~Connolly, J.T.~VanderPlas and A.~Gray, \emph{Statistics,
  Data Mining, and Machine Learning in Astronomy: A Practical Python Guide for
  the Analysis of Survey Data}, Princeton University Press, stu - student
  edition~ed. (2014).

\bibitem{aggarwal2018neural}
C.~Aggarwal, \emph{Neural Networks and Deep Learning: A Textbook}, Springer
  International Publishing (2018).

\bibitem{Wang:2020sxl}
Y.-C.~Wang, Y.-B.~Xie, T.-J.~Zhang, H.-C.~Huang, T.~Zhang and K.~Liu,
  \emph{{Likelihood-free Cosmological Constraints with Artificial Neural
  Networks: An Application on Hubble Parameters and SNe Ia}},
  \href{https://doi.org/10.3847/1538-4365/abf8aa}{\emph{Astrophys. J. Supp.}
  {\bfseries 254} (2021) 43}
  [\href{https://arxiv.org/abs/2005.10628}{{\ttfamily 2005.10628}}].

\bibitem{Gomez-Vargas:2021zyl}
I.~G\'omez-Vargas, J.A.~V\'azquez, R.M.~Esquivel and R.~Garc\'\i{}a-Salcedo,
  \emph{{Cosmological Reconstructions with Artificial Neural Networks}},
  \href{https://arxiv.org/abs/2104.00595}{{\ttfamily 2104.00595}}.

\bibitem{Dialektopoulos:2021wde}
K.~Dialektopoulos, J.L.~Said, J.~Mifsud, J.~Sultana and K.Z.~Adami,
  \emph{{Neural network reconstruction of late-time cosmology and null tests}},
  \href{https://doi.org/10.1088/1475-7516/2022/02/023}{\emph{JCAP} {\bfseries
  02} (2022) 023} [\href{https://arxiv.org/abs/2111.11462}{{\ttfamily
  2111.11462}}].

\bibitem{Auld:2007qz}
T.~Auld, M.~Bridges and M.P.~Hobson, \emph{{CosmoNet: Fast cosmological
  parameter estimation in non-flat models using neural networks}},
  \href{https://doi.org/10.1111/j.1365-2966.2008.13279.x}{\emph{Mon. Not. Roy.
  Astron. Soc.} {\bfseries 387} (2008) 1575}
  [\href{https://arxiv.org/abs/astro-ph/0703445}{{\ttfamily
  astro-ph/0703445}}].

\bibitem{Auld:2006pm}
T.~Auld, M.~Bridges, M.P.~Hobson and S.F.~Gull, \emph{{Fast cosmological
  parameter estimation using neural networks}},
  \href{https://doi.org/10.1111/j.1745-3933.2006.00276.x}{\emph{Mon. Not. Roy.
  Astron. Soc.} {\bfseries 376} (2007) L11}
  [\href{https://arxiv.org/abs/astro-ph/0608174}{{\ttfamily
  astro-ph/0608174}}].

\bibitem{2012MNRAS.421..169G}
P.~{Graff}, F.~{Feroz}, M.P.~{Hobson} and A.~{Lasenby}, \emph{{BAMBI: blind
  accelerated multimodal Bayesian inference}},
  \href{https://doi.org/10.1111/j.1365-2966.2011.20288.x}{\emph{Monthly Notices
  of the Royal Astronomical Society} {\bfseries 421} (2012) 169}
  [\href{https://arxiv.org/abs/1110.2997}{{\ttfamily 1110.2997}}].

\bibitem{Escamilla-Rivera:2019hqt}
C.~Escamilla-Rivera, M.A.C.~Quintero and S.~Capozziello, \emph{{A deep learning
  approach to cosmological dark energy models}},
  \href{https://doi.org/10.1088/1475-7516/2020/03/008}{\emph{JCAP} {\bfseries
  03} (2020) 008} [\href{https://arxiv.org/abs/1910.02788}{{\ttfamily
  1910.02788}}].

\bibitem{Aragon-Calvo:2018kfw}
M.A.~Aragon-Calvo, \emph{{Classifying the Large Scale Structure of the Universe
  with Deep Neural Networks}},
  \href{https://arxiv.org/abs/1804.00816}{{\ttfamily 1804.00816}}.

\bibitem{Ntampaka:2019ole}
M.~Ntampaka, D.J.~Eisenstein, S.~Yuan and L.H.~Garrison, \emph{{A Hybrid Deep
  Learning Approach to Cosmological Constraints From Galaxy Redshift Surveys}},
   \href{https://arxiv.org/abs/1909.10527}{{\ttfamily 1909.10527}}.

\bibitem{Ribli:2019wtw}
D.~Ribli, B.A.~Pataki, J.M.~Zorrilla~Matilla, D.~Hsu, Z.~Haiman and I.~Csabai,
  \emph{{Weak lensing cosmology with convolutional neural networks on noisy
  data}}, \href{https://doi.org/10.1093/mnras/stz2610}{\emph{Mon. Not. Roy.
  Astron. Soc.} {\bfseries 490} (2019) 1843}
  [\href{https://arxiv.org/abs/1902.03663}{{\ttfamily 1902.03663}}].

\bibitem{Fluri:2019qtp}
J.~Fluri, T.~Kacprzak, A.~Lucchi, A.~Refregier, A.~Amara, T.~Hofmann et~al.,
  \emph{{Cosmological constraints with deep learning from KiDS-450 weak lensing
  maps}}, \href{https://doi.org/10.1103/PhysRevD.100.063514}{\emph{Phys. Rev.
  D} {\bfseries 100} (2019) 063514}
  [\href{https://arxiv.org/abs/1906.03156}{{\ttfamily 1906.03156}}].

\bibitem{Fluri:2018hoy}
J.~Fluri, T.~Kacprzak, A.~Refregier, A.~Amara, A.~Lucchi and T.~Hofmann,
  \emph{{Cosmological constraints from noisy convergence maps through deep
  learning}}, \href{https://doi.org/10.1103/PhysRevD.98.123518}{\emph{Phys.
  Rev. D} {\bfseries 98} (2018) 123518}
  [\href{https://arxiv.org/abs/1807.08732}{{\ttfamily 1807.08732}}].

\bibitem{Hayashi:1979qx}
K.~Hayashi and T.~Shirafuji, \emph{{New General Relativity}},
  \href{https://doi.org/10.1103/PhysRevD.19.3524}{\emph{Phys. Rev. D}
  {\bfseries 19} (1979) 3524}.

\bibitem{nakahara2003geometry}
M.~Nakahara, \emph{Geometry, Topology and Physics, Second Edition}, Graduate
  student series in physics, Taylor \& Francis (2003).

\bibitem{ortin2004gravity}
T.~Ort{\'\i}n, \emph{Gravity and Strings}, Cambridge Monographs on Mathematical
  Physics, Cambridge University Press (2004),
  \href{https://doi.org/10.1017/CBO9780511616563}{10.1017/CBO9780511616563}.

\bibitem{Krssak:2015oua}
M.~Kr\v{s}\v{s}\'ak and E.N.~Saridakis, \emph{{The covariant formulation of
  $f(T)$ gravity}},
  \href{https://doi.org/10.1088/0264-9381/33/11/115009}{\emph{Class. Quant.
  Grav.} {\bfseries 33} (2016) 115009}
  [\href{https://arxiv.org/abs/1510.08432}{{\ttfamily 1510.08432}}].

\bibitem{chandrasekhar1998mathematical}
S.~Chandrasekhar and S.~Chandrasekhar, \emph{The Mathematical Theory of Black
  Holes}, International series of monographs on physics, Clarendon Press
  (1998).

\bibitem{Misner:1973prb}
C.W.~Misner, K.S.~Thorne and J.A.~Wheeler, \emph{{Gravitation}}, W. H. Freeman,
  San Francisco (1973).

\bibitem{DeFelice:2010aj}
A.~De~Felice and S.~Tsujikawa, \emph{{f(R) theories}},
  \href{https://doi.org/10.12942/lrr-2010-3}{\emph{Living Rev. Rel.} {\bfseries
  13} (2010) 3} [\href{https://arxiv.org/abs/1002.4928}{{\ttfamily
  1002.4928}}].

\bibitem{Bamba:2013ooa}
K.~Bamba, S.~Capozziello, M.~De~Laurentis, S.~Nojiri and D.~S\'aez-G\'omez,
  \emph{{No further gravitational wave modes in $F(T)$ gravity}},
  \href{https://doi.org/10.1016/j.physletb.2013.10.022}{\emph{Phys. Lett. B}
  {\bfseries 727} (2013) 194}
  [\href{https://arxiv.org/abs/1309.2698}{{\ttfamily 1309.2698}}].

\bibitem{Farrugia:2018gyz}
G.~Farrugia, J.~Levi~Said, V.~Gakis and E.N.~Saridakis, \emph{{Gravitational
  Waves in Modified Teleparallel Theories}},
  \href{https://doi.org/10.1103/PhysRevD.97.124064}{\emph{Phys. Rev. D}
  {\bfseries 97} (2018) 124064}
  [\href{https://arxiv.org/abs/1804.07365}{{\ttfamily 1804.07365}}].

\bibitem{Cai:2018rzd}
Y.-F.~Cai, C.~Li, E.N.~Saridakis and L.~Xue, \emph{{$f(T)$ gravity after
  GW170817 and GRB170817A}},
  \href{https://doi.org/10.1103/PhysRevD.97.103513}{\emph{Phys. Rev. D}
  {\bfseries 97} (2018) 103513}
  [\href{https://arxiv.org/abs/1801.05827}{{\ttfamily 1801.05827}}].

\bibitem{Abedi:2017jqx}
H.~Abedi and S.~Capozziello, \emph{{Gravitational waves in modified
  teleparallel theories of gravity}},
  \href{https://doi.org/10.1140/epjc/s10052-018-5967-x}{\emph{Eur. Phys. J. C}
  {\bfseries 78} (2018) 474}
  [\href{https://arxiv.org/abs/1712.05933}{{\ttfamily 1712.05933}}].

\bibitem{Chen:2019ftv}
Z.~Chen, W.~Luo, Y.-F.~Cai and E.N.~Saridakis, \emph{{New test on general
  relativity and $f(T)$ torsional gravity from galaxy-galaxy weak lensing
  surveys}}, \href{https://doi.org/10.1103/PhysRevD.102.104044}{\emph{Phys.
  Rev. D} {\bfseries 102} (2020) 104044}
  [\href{https://arxiv.org/abs/1907.12225}{{\ttfamily 1907.12225}}].

\bibitem{Tamanini:2012hg}
N.~Tamanini and C.G.~Boehmer, \emph{{Good and bad tetrads in $f(T)$ gravity}},
  \href{https://doi.org/10.1103/PhysRevD.86.044009}{\emph{Phys. Rev. D}
  {\bfseries 86} (2012) 044009}
  [\href{https://arxiv.org/abs/1204.4593}{{\ttfamily 1204.4593}}].

\bibitem{Bahamonde:2016cul}
S.~Bahamonde, M.~Zubair and G.~Abbas, \emph{{Thermodynamics and cosmological
  reconstruction in $f(T,B)$ gravity}},
  \href{https://doi.org/10.1016/j.dark.2017.12.005}{\emph{Phys. Dark Univ.}
  {\bfseries 19} (2018) 78} [\href{https://arxiv.org/abs/1609.08373}{{\ttfamily
  1609.08373}}].

\bibitem{Escamilla-Rivera:2019ulu}
C.~Escamilla-Rivera and J.~Levi~Said, \emph{{Cosmological viable models in
  $f(T,B)$ theory as solutions to the $H_0$ tension}},
  \href{https://doi.org/10.1088/1361-6382/ab939c}{\emph{Class. Quant. Grav.}
  {\bfseries 37} (2020) 165002}
  [\href{https://arxiv.org/abs/1909.10328}{{\ttfamily 1909.10328}}].

\bibitem{2015arXiv151107289C}
D.-A.~{Clevert}, T.~{Unterthiner} and S.~{Hochreiter}, \emph{{Fast and Accurate
  Deep Network Learning by Exponential Linear Units (ELUs)}}, {\emph{arXiv
  e-prints} (2015) arXiv:1511.07289}
  [\href{https://arxiv.org/abs/1511.07289}{{\ttfamily 1511.07289}}].

\bibitem{Wang:2019vxv}
G.-J.~Wang, X.-J.~Ma, S.-Y.~Li and J.-Q.~Xia, \emph{{Reconstructing Functions
  and Estimating Parameters with Artificial Neural Networks: A Test with a
  Hubble Parameter and SNe Ia}},
  \href{https://doi.org/10.3847/1538-4365/ab620b}{\emph{Astrophys. J. Suppl.}
  {\bfseries 246} (2020) 13}
  [\href{https://arxiv.org/abs/1910.03636}{{\ttfamily 1910.03636}}].

\bibitem{2014arXiv1412.6980K}
D.P.~{Kingma} and J.~{Ba}, \emph{{Adam: A Method for Stochastic Optimization}},
  {\emph{arXiv e-prints} (2014) arXiv:1412.6980}
  [\href{https://arxiv.org/abs/1412.6980}{{\ttfamily 1412.6980}}].

\bibitem{Wang:2020hmn}
G.-J.~Wang, S.-Y.~Li and J.-Q.~Xia, \emph{{ECoPANN: A Framework for Estimating
  Cosmological Parameters using Artificial Neural Networks}},
  \href{https://doi.org/10.3847/1538-4365/aba190}{\emph{Astrophys. J. Suppl.}
  {\bfseries 249} (2020) 25}
  [\href{https://arxiv.org/abs/2005.07089}{{\ttfamily 2005.07089}}].

\bibitem{Jimenez:2003iv}
R.~Jimenez, L.~Verde, T.~Treu and D.~Stern, \emph{{Constraints on the equation
  of state of dark energy and the Hubble constant from stellar ages and the
  CMB}}, \href{https://doi.org/10.1086/376595}{\emph{Astrophys. J.} {\bfseries
  593} (2003) 622} [\href{https://arxiv.org/abs/astro-ph/0302560}{{\ttfamily
  astro-ph/0302560}}].

\bibitem{Simon:2004tf}
J.~Simon, L.~Verde and R.~Jimenez, \emph{{Constraints on the redshift
  dependence of the dark energy potential}},
  \href{https://doi.org/10.1103/PhysRevD.71.123001}{\emph{Phys. Rev. D}
  {\bfseries 71} (2005) 123001}
  [\href{https://arxiv.org/abs/astro-ph/0412269}{{\ttfamily
  astro-ph/0412269}}].

\bibitem{Stern:2009ep}
D.~Stern, R.~Jimenez, L.~Verde, M.~Kamionkowski and S.A.~Stanford,
  \emph{{Cosmic Chronometers: Constraining the Equation of State of Dark
  Energy. I: $H(z)$ Measurements}},
  \href{https://doi.org/10.1088/1475-7516/2010/02/008}{\emph{JCAP} {\bfseries
  02} (2010) 008} [\href{https://arxiv.org/abs/0907.3149}{{\ttfamily
  0907.3149}}].

\bibitem{Moresco:2012jh}
M.~Moresco et~al., \emph{{Improved constraints on the expansion rate of the
  Universe up to $z$\textasciitilde{}1.1 from the spectroscopic evolution of
  cosmic chronometers}},
  \href{https://doi.org/10.1088/1475-7516/2012/08/006}{\emph{JCAP} {\bfseries
  08} (2012) 006} [\href{https://arxiv.org/abs/1201.3609}{{\ttfamily
  1201.3609}}].

\bibitem{Zhang:2012mp}
C.~Zhang, H.~Zhang, S.~Yuan, T.-J.~Zhang and Y.-C.~Sun, \emph{{Four new
  observational $H(z)$ data from luminous red galaxies in the Sloan Digital Sky
  Survey data release seven}},
  \href{https://doi.org/10.1088/1674-4527/14/10/002}{\emph{Res. Astron.
  Astrophys.} {\bfseries 14} (2014) 1221}
  [\href{https://arxiv.org/abs/1207.4541}{{\ttfamily 1207.4541}}].

\bibitem{Moresco:2015cya}
M.~Moresco, \emph{{Raising the bar: new constraints on the Hubble parameter
  with cosmic chronometers at $z$ \ensuremath{\sim} 2}},
  \href{https://doi.org/10.1093/mnrasl/slv037}{\emph{Mon. Not. Roy. Astron.
  Soc.} {\bfseries 450} (2015) L16}
  [\href{https://arxiv.org/abs/1503.01116}{{\ttfamily 1503.01116}}].

\bibitem{Moresco:2016mzx}
M.~Moresco, L.~Pozzetti, A.~Cimatti, R.~Jimenez, C.~Maraston, L.~Verde et~al.,
  \emph{{A 6\% measurement of the Hubble parameter at $z\sim0.45$: direct
  evidence of the epoch of cosmic re-acceleration}},
  \href{https://doi.org/10.1088/1475-7516/2016/05/014}{\emph{JCAP} {\bfseries
  05} (2016) 014} [\href{https://arxiv.org/abs/1601.01701}{{\ttfamily
  1601.01701}}].

\bibitem{Ratsimbazafy:2017vga}
A.L.~Ratsimbazafy, S.I.~Loubser, S.M.~Crawford, C.M.~Cress, B.A.~Bassett,
  R.C.~Nichol et~al., \emph{{Age-dating Luminous Red Galaxies observed with the
  Southern African Large Telescope}},
  \href{https://doi.org/10.1093/mnras/stx301}{\emph{Mon. Not. Roy. Astron.
  Soc.} {\bfseries 467} (2017) 3239}
  [\href{https://arxiv.org/abs/1702.00418}{{\ttfamily 1702.00418}}].

\bibitem{Borghi:2022apj}
N.~Borghi, M.~Moresco and A.~Cimatti, \emph{{Toward a Better Understanding of
  Cosmic Chronometers: A New Measurement of $H(z)$ at $z$ \ensuremath{\sim}
  0.7}}, \href{https://doi.org/10.3847/2041-8213/ac3fb2}{\emph{Astrophys. J.
  Lett.} {\bfseries 928} (2022) L4}
  [\href{https://arxiv.org/abs/2110.04304}{{\ttfamily 2110.04304}}].

\bibitem{Jimenez:2001gg}
R.~Jimenez and A.~Loeb, \emph{{Constraining cosmological parameters based on
  relative galaxy ages}},
  \href{https://doi.org/10.1086/340549}{\emph{Astrophys. J.} {\bfseries 573}
  (2002) 37} [\href{https://arxiv.org/abs/astro-ph/0106145}{{\ttfamily
  astro-ph/0106145}}].

\bibitem{Zhao:2018gvb}
G.-B.~Zhao et~al., \emph{{The clustering of the SDSS-IV extended Baryon
  Oscillation Spectroscopic Survey DR14 quasar sample: a tomographic
  measurement of cosmic structure growth and expansion rate based on optimal
  redshift weights}}, \href{https://doi.org/10.1093/mnras/sty2845}{\emph{Mon.
  Not. Roy. Astron. Soc.} {\bfseries 482} (2019) 3497}
  [\href{https://arxiv.org/abs/1801.03043}{{\ttfamily 1801.03043}}].

\bibitem{Gaztanaga:2008xz}
E.~Gaztanaga, A.~Cabre and L.~Hui, \emph{{Clustering of Luminous Red Galaxies
  IV: Baryon Acoustic Peak in the Line-of-Sight Direction and a Direct
  Measurement of $H(z)$}},
  \href{https://doi.org/10.1111/j.1365-2966.2009.15405.x}{\emph{Mon. Not. Roy.
  Astron. Soc.} {\bfseries 399} (2009) 1663}
  [\href{https://arxiv.org/abs/0807.3551}{{\ttfamily 0807.3551}}].

\bibitem{Blake:2012pj}
C.~Blake et~al., \emph{{The WiggleZ Dark Energy Survey: Joint measurements of
  the expansion and growth history at $z$ \ensuremath{<} 1}},
  \href{https://doi.org/10.1111/j.1365-2966.2012.21473.x}{\emph{Mon. Not. Roy.
  Astron. Soc.} {\bfseries 425} (2012) 405}
  [\href{https://arxiv.org/abs/1204.3674}{{\ttfamily 1204.3674}}].

\bibitem{Samushia:2012iq}
L.~Samushia et~al., \emph{{The Clustering of Galaxies in the SDSS-III DR9
  Baryon Oscillation Spectroscopic Survey: Testing Deviations from $\Lambda$
  and General Relativity using anisotropic clustering of galaxies}},
  \href{https://doi.org/10.1093/mnras/sts443}{\emph{Mon. Not. Roy. Astron.
  Soc.} {\bfseries 429} (2013) 1514}
  [\href{https://arxiv.org/abs/1206.5309}{{\ttfamily 1206.5309}}].

\bibitem{Xu:2012fw}
X.~Xu, A.J.~Cuesta, N.~Padmanabhan, D.J.~Eisenstein and C.K.~McBride,
  \emph{{Measuring $D_A$ and $H$ at $z=0.35$ from the SDSS DR7 LRGs using
  baryon acoustic oscillations}},
  \href{https://doi.org/10.1093/mnras/stt379}{\emph{Mon. Not. Roy. Astron.
  Soc.} {\bfseries 431} (2013) 2834}
  [\href{https://arxiv.org/abs/1206.6732}{{\ttfamily 1206.6732}}].

\bibitem{BOSS:2014hwf}
{\scshape BOSS} collaboration, \emph{{Baryon acoustic oscillations in the
  Ly\ensuremath{\alpha} forest of BOSS DR11 quasars}},
  \href{https://doi.org/10.1051/0004-6361/201423969}{\emph{Astron. Astrophys.}
  {\bfseries 574} (2015) A59}
  [\href{https://arxiv.org/abs/1404.1801}{{\ttfamily 1404.1801}}].

\bibitem{BOSS:2013igd}
{\scshape BOSS} collaboration, \emph{{Quasar-Lyman $\alpha$ Forest
  Cross-Correlation from BOSS DR11 : Baryon Acoustic Oscillations}},
  \href{https://doi.org/10.1088/1475-7516/2014/05/027}{\emph{JCAP} {\bfseries
  05} (2014) 027} [\href{https://arxiv.org/abs/1311.1767}{{\ttfamily
  1311.1767}}].

\bibitem{BOSS:2016wmc}
{\scshape BOSS} collaboration, \emph{{The clustering of galaxies in the
  completed SDSS-III Baryon Oscillation Spectroscopic Survey: cosmological
  analysis of the DR12 galaxy sample}},
  \href{https://doi.org/10.1093/mnras/stx721}{\emph{Mon. Not. Roy. Astron.
  Soc.} {\bfseries 470} (2017) 2617}
  [\href{https://arxiv.org/abs/1607.03155}{{\ttfamily 1607.03155}}].

\bibitem{Bourboux:2017cbm}
H.~du~Mas~des Bourboux et~al., \emph{{Baryon acoustic oscillations from the
  complete SDSS-III Ly$\alpha$-quasar cross-correlation function at $z=2.4$}},
  \href{https://doi.org/10.1051/0004-6361/201731731}{\emph{Astron. Astrophys.}
  {\bfseries 608} (2017) A130}
  [\href{https://arxiv.org/abs/1708.02225}{{\ttfamily 1708.02225}}].

\bibitem{Riess:2019cxk}
A.G.~Riess, S.~Casertano, W.~Yuan, L.M.~Macri and D.~Scolnic, \emph{{Large
  Magellanic Cloud Cepheid Standards Provide a 1\% Foundation for the
  Determination of the Hubble Constant and Stronger Evidence for Physics beyond
  $\Lambda$CDM}},
  \href{https://doi.org/10.3847/1538-4357/ab1422}{\emph{Astrophys. J.}
  {\bfseries 876} (2019) 85}
  [\href{https://arxiv.org/abs/1903.07603}{{\ttfamily 1903.07603}}].

\bibitem{Riess:2020sih}
A.G.~Riess, \emph{{The Expansion of the Universe is Faster than Expected}},
  \href{https://doi.org/10.1038/s42254-019-0137-0}{\emph{Nature Rev. Phys.}
  {\bfseries 2} (2019) 10} [\href{https://arxiv.org/abs/2001.03624}{{\ttfamily
  2001.03624}}].

\bibitem{Riess:2021jrx}
A.G.~Riess et~al., \emph{{A Comprehensive Measurement of the Local Value of the
  Hubble Constant with 1 km s$^{-1}$ Mpc$^{-1}$ Uncertainty from the Hubble
  Space Telescope and the SH0ES Team}},
  \href{https://doi.org/10.3847/2041-8213/ac5c5b}{\emph{Astrophys. J. Lett.}
  {\bfseries 934} (2022) L7}
  [\href{https://arxiv.org/abs/2112.04510}{{\ttfamily 2112.04510}}].

\bibitem{Freedman:2019jwv}
W.L.~Freedman et~al., \emph{{The Carnegie-Chicago Hubble Program. VIII. An
  Independent Determination of the Hubble Constant Based on the Tip of the Red
  Giant Branch}},  \href{https://arxiv.org/abs/1907.05922}{{\ttfamily
  1907.05922}}.

\bibitem{Freedman:2021ahq}
W.L.~Freedman, \emph{{Measurements of the Hubble Constant: Tensions in
  Perspective}},
  \href{https://doi.org/10.3847/1538-4357/ac0e95}{\emph{Astrophys. J.}
  {\bfseries 919} (2021) 16}
  [\href{https://arxiv.org/abs/2106.15656}{{\ttfamily 2106.15656}}].

\bibitem{Ade:2015xua}
{\scshape Planck} collaboration, \emph{{Planck 2015 results. XIII. Cosmological
  parameters}},
  \href{https://doi.org/10.1051/0004-6361/201525830}{\emph{Astron. Astrophys.}
  {\bfseries 594} (2016) A13}
  [\href{https://arxiv.org/abs/1502.01589}{{\ttfamily 1502.01589}}].

\bibitem{Ma:2010mr}
C.~Ma and T.-J.~Zhang, \emph{{Power of Observational Hubble Parameter Data: a
  Figure of Merit Exploration}},
  \href{https://doi.org/10.1088/0004-637X/730/2/74}{\emph{Astrophys. J.}
  {\bfseries 730} (2011) 74} [\href{https://arxiv.org/abs/1007.3787}{{\ttfamily
  1007.3787}}].

\bibitem{Wasserman:2001ng}
{\scshape PICA Group} collaboration, \emph{{Non-parametric inference in
  astrophysics}},  \href{https://arxiv.org/abs/astro-ph/0112050}{{\ttfamily
  astro-ph/0112050}}.

\bibitem{Sahni:2008xx}
V.~Sahni, A.~Shafieloo and A.A.~Starobinsky, \emph{{Two new diagnostics of dark
  energy}}, \href{https://doi.org/10.1103/PhysRevD.78.103502}{\emph{Phys. Rev.
  D} {\bfseries 78} (2008) 103502}
  [\href{https://arxiv.org/abs/0807.3548}{{\ttfamily 0807.3548}}].

\bibitem{Zunckel:2008ti}
C.~Zunckel and C.~Clarkson, \emph{{Consistency Tests for the Cosmological
  Constant}}, \href{https://doi.org/10.1103/PhysRevLett.101.181301}{\emph{Phys.
  Rev. Lett.} {\bfseries 101} (2008) 181301}
  [\href{https://arxiv.org/abs/0807.4304}{{\ttfamily 0807.4304}}].

\bibitem{Shafieloo:2009hi}
A.~Shafieloo and C.~Clarkson, \emph{{Model independent tests of the standard
  cosmological model}},
  \href{https://doi.org/10.1103/PhysRevD.81.083537}{\emph{Phys. Rev. D}
  {\bfseries 81} (2010) 083537}
  [\href{https://arxiv.org/abs/0911.4858}{{\ttfamily 0911.4858}}].

\bibitem{Clarkson:2007pz}
C.~Clarkson, B.~Bassett and T.H.-C.~Lu, \emph{{A general test of the Copernican
  Principle}},
  \href{https://doi.org/10.1103/PhysRevLett.101.011301}{\emph{Phys. Rev. Lett.}
  {\bfseries 101} (2008) 011301}
  [\href{https://arxiv.org/abs/0712.3457}{{\ttfamily 0712.3457}}].

\bibitem{Qi:2016wwb}
J.-Z.~Qi, M.-J.~Zhang and W.-B.~Liu, \emph{{Testing dark energy models with
  $H(z)$ data}},  \href{https://arxiv.org/abs/1606.00168}{{\ttfamily
  1606.00168}}.

\bibitem{Qi:2018pej}
J.-Z.~Qi, S.~Cao, M.~Biesiada, T.~Xu, Y.~Wu, S.~Zhang et~al., \emph{{What do
  parameterized $Om(z)$ diagnostics tell us in light of recent observations?}},
  \href{https://doi.org/10.1088/1674-4527/18/6/66}{\emph{Res. Astron.
  Astrophys.} {\bfseries 18} (2018) 066}
  [\href{https://arxiv.org/abs/1803.04109}{{\ttfamily 1803.04109}}].

\bibitem{Bengaly:2020neu}
C.A.P.~Bengaly, C.~Clarkson, M.~Kunz and R.~Maartens, \emph{{Null tests of the
  concordance model in the era of Euclid and the SKA}},
  \href{https://doi.org/10.1016/j.dark.2021.100856}{\emph{Phys. Dark Univ.}
  {\bfseries 33} (2021) 100856}
  [\href{https://arxiv.org/abs/2007.04879}{{\ttfamily 2007.04879}}].

\bibitem{OColgain:2021pyh}
E.~\'O~Colg\'ain and M.M.~Sheikh-Jabbari, \emph{{Elucidating cosmological model
  dependence with $H_0$}},
  \href{https://doi.org/10.1140/epjc/s10052-021-09708-2}{\emph{Eur. Phys. J. C}
  {\bfseries 81} (2021) 892}
  [\href{https://arxiv.org/abs/2101.08565}{{\ttfamily 2101.08565}}].

\bibitem{JMLR:v8:cawley07a}
G.C.~Cawley and N.L.C.~Talbot, \emph{Preventing over-fitting during model
  selection via bayesian regularisation of the hyper-parameters},
  {\emph{Journal of Machine Learning Research} {\bfseries 8} (2007) 841}.

\bibitem{10.1007/978-3-319-62416-7_14}
R.O.~Mohammed and G.C.~Cawley, \emph{Over-fitting in model selection with
  gaussian process regression},  in \emph{Machine Learning and Data Mining in
  Pattern Recognition}, P.~Perner, ed., (Cham), pp.~192--205, Springer
  International Publishing, 2017.

\bibitem{BeltranJimenez:2017tkd}
J.~Beltr\'an~Jim\'enez, L.~Heisenberg and T.~Koivisto, \emph{{Coincident
  General Relativity}},
  \href{https://doi.org/10.1103/PhysRevD.98.044048}{\emph{Phys. Rev. D}
  {\bfseries 98} (2018) 044048}
  [\href{https://arxiv.org/abs/1710.03116}{{\ttfamily 1710.03116}}].

\bibitem{Harko:2018gxr}
T.~Harko, T.S.~Koivisto, F.S.N.~Lobo, G.J.~Olmo and D.~Rubiera-Garcia,
  \emph{{Coupling matter in modified $Q$ gravity}},
  \href{https://doi.org/10.1103/PhysRevD.98.084043}{\emph{Phys. Rev. D}
  {\bfseries 98} (2018) 084043}
  [\href{https://arxiv.org/abs/1806.10437}{{\ttfamily 1806.10437}}].

\bibitem{Gakis:2019rdd}
V.~Gakis, M.~Kr\v{s}\v{s}\'ak, J.~Levi~Said and E.N.~Saridakis,
  \emph{{Conformal gravity and transformations in the symmetric teleparallel
  framework}}, \href{https://doi.org/10.1103/PhysRevD.101.064024}{\emph{Phys.
  Rev. D} {\bfseries 101} (2020) 064024}
  [\href{https://arxiv.org/abs/1908.05741}{{\ttfamily 1908.05741}}].

\bibitem{Soudi:2018dhv}
I.~Soudi, G.~Farrugia, V.~Gakis, J.~Levi~Said and E.N.~Saridakis,
  \emph{{Polarization of gravitational waves in symmetric teleparallel theories
  of gravity and their modifications}},
  \href{https://doi.org/10.1103/PhysRevD.100.044008}{\emph{Phys. Rev. D}
  {\bfseries 100} (2019) 044008}
  [\href{https://arxiv.org/abs/1810.08220}{{\ttfamily 1810.08220}}].

\end{thebibliography}\endgroup

\label{lastpage}

\end{document}